\documentclass[preprint,12pt]{elsarticle}
\usepackage{lmodern}

\usepackage{lineno,hyperref}
\usepackage{xcolor}
\usepackage{amssymb}
\usepackage{amsmath}
\usepackage{siunitx}
\usepackage{subfigure}
\usepackage{color, colortbl}
\usepackage{multirow}
\usepackage{float}
\usepackage{graphicx}
\usepackage{epsfig}
\usepackage{epstopdf}
\usepackage{caption}
\usepackage{booktabs,threeparttable}
\usepackage{bm}
\usepackage{lscape}
\usepackage{soul}

\usepackage{textcomp}

\usepackage{pifont}
\usepackage[normalem]{ulem}

\newcommand{\xmark}{\ding{55}}

\graphicspath{{./Figures/}}
\definecolor{Gray}{gray}{0.9}

\journal{Results in Engineering}

\begin{document}

\begin{frontmatter}



\title{Continuously Tunable Mid-Infrared Metalens Integrating Moiré and Pancharatnam-Berry Phases}

\author[FirstAuthorsAffiliation,SecondAuthorsAffiliation]{Keivan Kaboutari}
\author[FirstAuthorsAffiliation]{Muchuan Chen}
\author[FirstAuthorsAffiliation]{Xiu Liu}
\author[FirstAuthorsAffiliation]{Sheng Shen\corref{mycorrespondingauthor1}}
\cortext[mycorrespondingauthor1]{Corresponding author}
\ead{shengshe@andrew.cmu.edu}
\author[SecondAuthorsAffiliation]{Pedro Pinho}
\author[SecondAuthorsAffiliation]{Stanislav Maslovski\corref{mycorrespondingauthor2}}
\cortext[mycorrespondingauthor2]{Corresponding author}
\ead{stanislav.maslovski@ua.pt}
\address[FirstAuthorsAffiliation]{Department of Mechanical Engineering, Carnegie Mellon University, Pittsburgh, PA 15213, United States of America}
\address[SecondAuthorsAffiliation]{Instituto de Telecomunica\c{c}\~{o}es and Department of Electronics, Telecommunications and Informatics, University of Aveiro, 3810-193 Aveiro, Portugal}




\begin{abstract}
    A varifocal mid-infrared metalens designed for focusing applications is proposed to integrate both Pancharatnam-Berry (PB) and Moiré pattern-based mechanisms for rotational wavefront controll. The metalens employs two identical nanostructured discs featuring germanium ({\tt Ge}) meta-atoms on calcium fluoride ({\tt CaF$_{\rm 2}$}) substrates, operating as birefringent half-wave retarders with 99.95\% Polarization Conversion Ratio (PCR) and 83.65\% total transmission power at 2~$\mu\text{m}$ wavelength. Full-wave numerical simulations demonstrate continuous focal length tunability from 45~$\mu\text{m}$ to 12~$\mu\text{m}$ by rotating the second disk from 30$^\circ$ to 120$^\circ$, corresponding to the numerical aperture values ranging from 22.99\% to 68.67\%. The proposed metalens maintains reconfigurable focusing regardless of the incident light polarization, while concurrently enabling polarization-controlled operation via a phase-modulation mechanism that yields inverse responses under Right Circular Polarization (RCP) and Left Circular Polarization (LCP) illuminations.

\end{abstract}

\begin{graphicalabstract}
    \centering
    \includegraphics[width=1\linewidth,trim={0.0cm 0.0cm 0.0cm 0.0cm},clip=true]{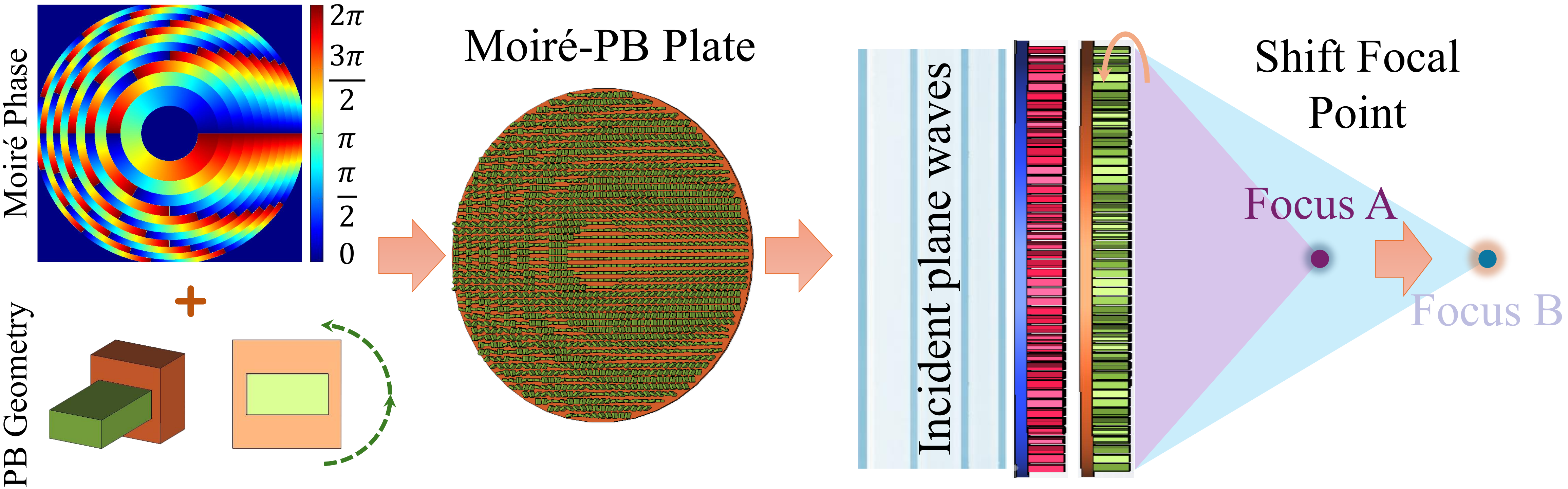}
\end{graphicalabstract}

\begin{highlights}
    \item Integrates Moiré and PB phases for mid-IR wavefront control.
    \item Continuous focal length tuning from 45 to 12~$\mu\text{m}$ via rotation.
    \item Features a single-mask architecture with two identical disks.
    \item Provides broadband potential using geometric phase meta-atoms.
\end{highlights}


\end{frontmatter}



\section{Introduction}
    In the Near-Infrared (NIR) spectral regime, metalenses represent a paradigm shift from conventional refractive optics by utilizing metamaterials and metasurfaces composed of subwavelength inclusions employed to realize arbitrary transmission phase profiles. Metasurfaces as planar optical components are particularly appealing for NIR systems, as they are compatible with semiconductor fabrication processes and can integrate complex optical functionalities into ultrathin platforms.
    
    Pancharatnam-Berry (PB) phase metalenses are a class of flat optics that control the phase of light through the geometric phase introduced by rotating anisotropic meta-atoms. Their performance comes from clever geometry rather than from the light’s travel distance. In fact, the PB-phase methodology provides polarization-encoded phase control, making it well-suited for broadband and polarization-dependent NIR applications. Yet another mechanism of flexible transmission phase control is achieved with so-called Moiré metalenses. The Moiré metalenses introduce an additional degree of control by overlaying rotated or laterally shifted Moiré patterned metasurface layers, enabling tunable focal-length properties. Integrating PB geometric phase modulation into Moiré-pattern wavefront control schemes may open new avenues for applying adaptive, multifunctional NIR metalenses in advanced optical and spectroscopic systems, including polarization-multiplexed IR systems.
    
    A varifocal Moiré zone plate using twin cubic-linear grids to enable continuous straightness measurement is proposed in~\cite{Burch_1977}. Relative grid translation continuously adjusts focal power while maintaining image focus. Utilizing phase-reversal grid structures markedly enhances diffraction image irradiance for photoelectric alignment. While offering strong potential for long-distance alignment and narrow-bandwidth spectroscopy, practical performance depends strictly on maintaining tight relative grid rotation tolerances and cam-compensated positioning mechanics. Bernet and Ritsch-Marte~\cite{bernetAdjustableRefractivePower2008c} demonstrate that mutual rotation of cascaded blazed diffractive Moiré optical elements enables continuous tuning of optical power and helical charge with high diffraction efficiency. Incorporating mathematical phase rounding eliminates secondary sector artifacts, maintaining over 85\% efficiency within a $\pm90^{\circ}$ rotation range. While highly adaptable for imaging and optical trapping, physical performance remains bounded by pixel resolution, axial Talbot limits, and rotation-dependent efficiency drop-offs.
    
    A study in~\cite{yilmazRotationallyTunablePolarizationinsensitive2019} demonstrates rotationally tunable, polarization-insensitive single and multifocal Moiré metasurfaces using {\tt TiO$_{\rm 2}$} nano-rods on {\tt SiO$_{\rm 2}$} substrate operating at 532~nm. Mutual rotation adjusts focal lengths nonlinearly from $11.4~\mu\text{m}$ to $4.2~\mu\text{m}$ with up to 55\% focusing efficiency. Regionally partitioned phase profiles enable dynamic multi-to-unifocal switching. While offering high efficiency and CMOS-compatible single-step lithography, performance relies on numerical FDTD modeling without physical fabrication validation. The research in~\cite{bernetDemonstrationFocustunableDiffractive2013} experimentally demonstrates focus-tunable diffractive Moiré lenses using 16-phase-level fused silica Diffractive Optical Elements (DOEs) blazed for 633~nm. Mutual rotation provides wide and precise refractive power tuning up to $\pm 26~\text{dpt}$ with diffraction-limited imaging performance. While integer Moiré DOEs (MDOEs) successfully eliminate sectoring with low contrast degradation, experimental efficiencies from 69\% to 86\% were limited by lithographic fabrication errors and axial positioning tolerances relative to the Talbot distance.
    
    Wen et al.~\cite{WENHelicityMultiplexedBroadband2015a} demonstrate a helicity-multiplexed metasurface hologram utilizing PB phase silver nanorods in a reflective tri-layer configuration. Reversing circular polarization helicity flips the phase profile sign, dynamically swapping two reconstructed off-axis images. Operating across 475-1,100~nm, the device achieves up to 59.2\% conversion efficiency and high image fidelity, offering strong potential for anti-counterfeiting and switchable display technologies. Research in~\cite{xuMechanicallyReprogrammablePancharatnam2022} demonstrates a mechanically reprogrammable PB metasurface for microwaves, utilizing motor-driven rotating meta-atoms to achieve twenty-eight phase levels. The platform successfully reconfigures metalensing, vortex beam generation, and holographic imaging on a single template with high efficiency and nonvolatility. Although mechanical gear assemblies limit modulation speed, the design avoids diode losses and provides continuous phase control for dynamic wavefront shaping. Yu Zou et al.~\cite{zouMoiretypePancharatnamBerry2025} demonstrate a Moiré-type PB phase element using cascaded liquid crystal plates to control polarization-rotating Bessel beams dynamically. Relative rotation tunes the polarization rotation rate and direction without changing spin states. Incorporating checkerboard phase encoding and back-propagation correction eliminates sector artifacts and diffraction gaps. Despite breakdown near extreme rotation angles, this approach enables highly integrated dynamic vector beam shaping for laser processing and microscopy.
    
    A reconfigurable terahertz metalens with dual-linear-focus phase distribution utilizing all-dielectric silicon nanopillars is proposed by Liu et al.~\cite{liuReconfigurableMetalensDuallinearfocus2023}. Stacking two orthogonal layers enables focal length tuning by stretching along individual axes rather than requiring uniform omnidirectional stretching. Combining propagation and PB phases provides polarization-controllable bifocal and varifocal performance. While numerical simulations show high linearity and diffraction-limited focusing, findings lack physical fabrication and experimental validation. Xu et al.~\cite{XUTerahertz4DBessel2025} demonstrate a 4D terahertz Bessel beam steering Moiré meta-device using three cascaded dielectric metasurfaces at 291~GHz. The methodology integrates polarization-dependent geometric phases and relative interlayer rotations to dynamically modulate the cone angle (governing non-diffracting distance), topological charge, and a linear phase gradient for 2D beam steering. Relative in-plane rotations and spin-decoupled phase modulation provide continuous control over non-diffracting propagation distance from 10 to 130~mm, $\pm40^{\circ}$ elevation angle, $360^{\circ}$ azimuthal angle, and $0^{\rm th}/1^{\rm st}$ beam order (topological charge). Experimental near-field communications and obstacle self-healing validate signal reliability, though overall efficiency remains limited by substrate reflections.
    
    An all-dielectric terahertz Moiré meta-device for dynamic Bessel beam generation at 0.9~THz is proposed in~\cite{nanGenerationBesselBeams2025}. By combining propagation and PB phases, relative layer rotation dynamically tunes topological charge, while overall rotation modulates linear polarization. Spatially non-uniform meta-atom layouts enable longitudinally rotating vector fields. Experimental terahertz focal-plane imaging confirms agreement with simulations, validating flexible multi-dimensional structured light manipulation. Xianglin Ye et al.~\cite{yeUltraBroadbandMoirePBDoublet2024} demonstrate an ultra-broadband Moiré-PB doublet lens using decentered liquid crystal phase plates with retardation compensation. Relative rotation yields a tunable phase grating, enabling real-time edge detection with resolution dynamically adjustable from 13.2 to 57.2~$\mu\text{m}$, alongside conventional and polarimetric imaging across 400 to 780~nm. High polarization conversion efficiency ($>$95\%) eliminates stray light, validating the device for low-contrast biological imaging and compact microscopic systems.
    
    Hui Li et al.~\cite{Li_2026_1} demonstrate a Polarization-multiplexed Cascaded Metasurface (PMCM) for adaptive 0.5~THz beam steering using all-silicon pillars. By combining spin-decoupled rectangular pillars and polarization-insensitive cylindrical pillars, in-plane layer rotations dynamically steer focused Gaussian ($l=0$) and vortex ($l=1$) beams along independent spatial trajectories. Experiments confirm beam displacement ($4.23\lambda$) and low channel crosstalk (average $<-15~\text{dB}$), validating a low-cost, multi-dimensional platform for compact THz systems. A reconfigurable all-silicon terahertz chiral metasurface combining Moiré phase engineering and spin-decoupled phase modulation at 0.5 THz is discussed in~\cite{Li_2026_2}. Relative layer rotation dynamically tunes broadband circular dichroism and focal lengths over a wide zoom ratio exceeding 4.6 times. An improved offset-phase design enables full-rotation continuous focusing and spin-selective transmission with low crosstalk. The voltage-free all-dielectric platform offers high spatial robustness for non-destructive testing and biomedical imaging.
    
    Despite major advances in both Moiré and PB-phase metasurfaces, their integration into a single, but simple platform remains fairly unexplored. Moiré metalenses provide continuous mechanical tunability through the mutual rotation of cascaded elements, enabling dynamic focal-length control without axial motion. In contrast, PB-phase metasurfaces provide phase modulation through the geometric rotation of anisotropic meta-atoms, which makes them attractive for broadband operation because the geometric phase can be, in principle, independent of the wavelength. Existing tunable metalenses, however, generally rely either on propagation-phase designs, which are prone to chromatic aberration, or on PB-phase elements that are not reconfigurable after fabrication. As a result, combining Moiré-based mechanical tunability with the broadband potential of PB-phase metasurfaces remains an important challenge in adaptive flat optics.
    
    Our work addresses this gap by introducing a mid-IR metalens based on hybrid Moiré and PB-phase concepts that integrates both mechanisms within a single but self-complementary design. The metalens described in this work preserves reconfigurable focusing for arbitrary incident polarization states by utilizing Moiré phase profiles synthesized using the half-wave plate PB geometric phase. While classical Moiré metalenses process all wavefronts identically and forfeit polarization control, typical PB metalenses are strictly spin-dependent. For any arbitrary incident polarization state, the metalens guarantees effective focal-spot generation simply by rotating the second disk in the proper direction to align with the dominant or targeted helicity component. Indeed, by appropriately configuring the relative rotation angle of the cascading layers, the proposed Moiré-PB metalens can be preset to focus a designated circular polarization state (either LCP or RCP). Additionally, the lens maintains a stable focal spot at the intended focal plane regardless of the incident linear polarization angle. However, this operational stability entails an efficiency trade-off: while a matched circularly polarized input achieves peak nominal efficiency, an LP input experiences an intrinsic $50\%$ efficiency reduction because the orthogonal, unfocused spin component is dispersed into the background. On the other hand, it enables polarization-dependent control through a phase-modulation mechanism that generates inverse responses under RCP and LCP illumination. This dual functionality provides a versatile platform that is robust to arbitrary inputs while remaining tunable via the helicity of the incident wave.
    
    Such a metalens offers three main advantages: ($I$) Continuous mechanical tunability through the mutual rotation of complementary phase disks, enabling focal-length adjustment from infinity to about 17~$\mu\text{m}$ without axial displacement; ($II$) Broadband potential derived from the PB-phase mechanism, in which the geometric phase may rely mainly on the rotation angles of the meta-atoms rather than on resonant dispersion; and ($III$) Operation with high Polarization Conversion Ratio (PCR) and total transmission power enabled by optimized {\tt Ge} half-wave-plate meta-atoms. This hybrid approach provides a versatile platform for adaptive imaging, beam steering, and other applications that require both reconfigurability and broadband performance.
    
    It is worth noting that the hybrid Moiré/PB-phase concept introduced here may also provide an effective approach for controlling wavefront aberrations and improving focusing performance of reconfigurable metalenses. Since this method superposes spatially varying birefringent phase distributions onto the Moiré patterns, it creates a separate degree of control over the effective phase profile across the metalens aperture. Combining the anisotropic phase response of birefringent materials with the periodic Moiré-type modulation enables multi-level phase quantization without introducing artifacts related to abrupt phase jumps on the lens aperture. Its main advantages are: ($I$) Continuous phase scaling, which reduces phase quantization errors and improves diffraction efficiency; ($II$) Greater design flexibility for simultaneous control of focal length, inherent polarization conversion, and potential for aberration corrections; ($III$) Reduced sensitivity to fabrication tolerances; and ($IV$) The possibility of compact broadband operation through birefringent meta-atoms engineered for a desired spectral response.
    
    Compared to the studies by Yu Zou et al.~\cite{zouMoiretypePancharatnamBerry2025}, Tong Nan et al.~\cite{nanGenerationBesselBeams2025}, and Xianglin Ye et al.~\cite{yeUltraBroadbandMoirePBDoublet2024}, the primary novelty of this work lies in integrating PB geometric phase with Moiré pattern rotations specifically for continuous varifocal imaging rather than for generating polarization-rotating Bessel beams or adjusting THz Bessel beam topological charges and polarization. Additionally, Yu Zou et al. require two distinct phase distributions incorporating a checkerboard-encoded grid pattern on the first element to synthesize their Moiré profile, and rely on phase-conjugate back-propagation to compensate for the 10~mm interlayer separation gap. Tong Nan et al. require two distinct metasurface masks (meta1 with circular pillars and meta2 with rectangular pillars) to decouple phase and polarization under strictly $x$-polarized incidence. Furthermore, Xianglin Ye et al. rely on physical spatial decentering of optical axes and a complex multi-twisted liquid crystal architecture to accomplish broadband retardation compensation. As an example, Haifeng Xu et al.~\cite{XUTerahertz4DBessel2025} rely on three non-identical metasurfaces with multi-layer rotational coordination to achieve 4D Bessel beam steering for near-field communications.
    
    To conclude, the major practical advantage of the proposed approach is its single-mask, self-complementary architecture. While the prior works successfully mixed propagation and geometric phases for dynamic beam manipulation, they rely on complex, non-identical, or liquid-crystal architectures targeted at specialized beam shaping. In contrast, our methodology introduces a streamlined, mid-infrared varifocal metalens constructed from two identical {\tt Ge} nanobars on {\tt CaF$_{\rm 2}$} disks. Since the engineered anisotropic {\tt Ge} meta-atoms operate as half-wave retarders, two structurally identical metasurfaces fabricated from a single lithographic mask automatically synthesize the required complementary Moiré phase distributions upon mutual rotation. The hybrid Moiré-PB phase integration realized in our design achieves continuous mechanical focal tuning (from 68~$\mu\text{m}$ to 17~$\mu\text{m}$) and remarkably high PCR ($>$99.95\%) at 2~$\mu\text{m}$ while remaining adjustable to varying incident polarization states.
   
\section{Near‑Infrared Wavefront Control and Methodology of Focal Length Modulation}
    In what follows, we introduce a wavefront control methodology that utilizes birefringent response of metamaterials. To quantify phase retardation induced by birefringent unit cells, a half-wave condition is formulated as a function of both the wavelength and the material thickness along the direction of wave propagation. Next, a polarization-optimized meta-atom geometry must be chosen. For this purpose, we select a rectangular prism-shaped meta-atom made of germanium ({\tt Ge}) and model its electromagnetic response in a full-wave simulator. Then, simplified analytical models are used to assess the resulting geometry. We ensure that the meta-atom exhibits robust half-wave behavior and delivers the required phase delays.
    
    Once a suitable meta-atom is selected, Moiré phase patterning is employed to create spatially varying supercells. These supercells can produce prescribed wavefront gradients while maintaining polarization selectivity. Next, the PB-phase framework is applied, which translates in-plane rotations of the meta-atoms into geometric phase profiles using the Moiré phase profiles as the underlying basis. As a result, we achieve polarization-engineered control over the focal length with a possibility to compensate for optical aberrations. Finally, all of these components are brought together into a complete metalens disk structure. In this final integration, the selected meta-atoms, Moiré superlattices, and PB‑phase distributions are co-optimized. This co-optimization enables focal-length modulation in the near-infrared while respecting bandwidth constraints.
    \subsection{Birefringence-Induced Phase Retardation}
        Birefringent materials are optically anisotropic media in which the refractive index depends on the polarization and propagation direction of light. In these materials, a single incident beam naturally splits into two orthogonally polarized eigenmodes with different phase velocities, i.e., ordinary and extraordinary waves.
        
        Thus, birefringence refers to the existence of at least two primary refractive indices, whose difference is represented by $\Delta n=n_{\rm e}-n_{\rm o}$ ($\Delta n$ is greater than zero for positive uniaxial birefringent media). Here, $n_{\rm e}$ is the extraordinary index, and $n_{\rm o}$ is the ordinary index. In uniaxial and biaxial crystals, as well as in artificially structured media, these indices correspond to perpendicular optical axes. As light propagates through these materials, polarization components that are aligned parallel and perpendicular to a given axis experience different phase delays. Due to the differing refractive indices for two orthogonal polarization components, a slab of thickness $h$ creates a relative phase shift (phase retardation) $\delta=2\pi\Delta n h/\lambda$ between them (here, $\lambda$ is the operating wavelength), enabling functionalities such as half- and quarter-wave plates. Birefringent materials typically exhibit linear eigenpolarizations that propagate independently, enabling precise control of polarization while maintaining high transmission. These characteristics can be exploited in polarizers, retarders, and polarization-sensitive metasurfaces.
        
        In the design of infrared metalenses, birefringent meta-atoms, such as anisotropic dielectric nanofins or nanoblocks, function as localized waveplates. The optical axes and phase delay of these waveplates can be customized through geometric modifications, enabling control over both optical and geometric phases, including the PB-phase. By rotating these local anisotropic elements, it is possible to create a spatially varying phase profile that corresponds to a specific polarization state while simultaneously manipulating the polarization itself. This approach enables advanced functionalities, such as spin-dependent focusing and vector-beam generation. Furthermore, the independent phase channels provided by the birefringent meta-atoms for two orthogonal polarization states can be employed to allow for dual or multiplexed operation, which leads to polarization-controlled focusing profiles for various inputs applied to the same metalens.
    \subsection{Polarization-Sensitive Meta-Atoms: Unit Cell Geometry and Numerical-Analytical Study}
\begin{figure}
            \begin{center}
                \subfigure[]{\label{Fig: Fig01.a}
                \includegraphics[height=0.3\linewidth,trim={0.0cm 0.0cm 0.0cm 0.0cm},clip=true]{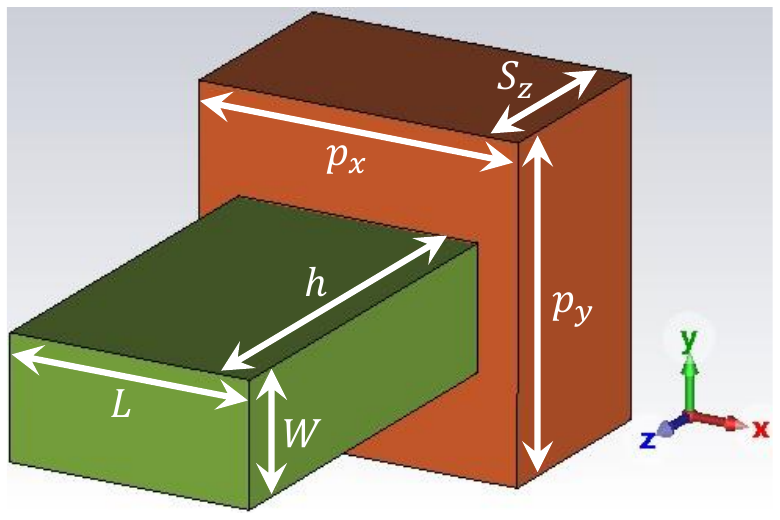}}
                \subfigure[]{\label{Fig: Fig01.b}
                \includegraphics[height=0.3\linewidth,trim={0.0cm 0.0cm 0.0cm 0.0cm},clip=true]{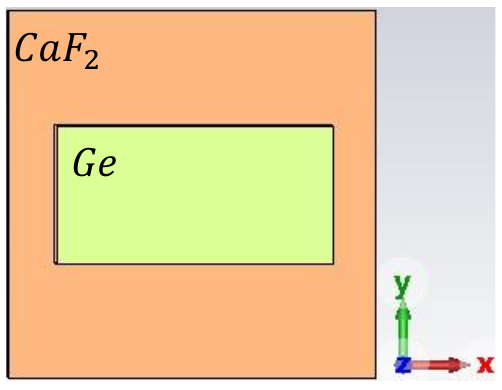}}
            \end{center}
            \vspace*{-5mm}
            \caption{The unit-cell structure employed in the design of the proposed integrated Moiré and PB-phase metalens. (a) Three‑dimensional view and (b) Top view of the utilized unit cell.}
            \label{Fig: Fig01}
        \end{figure}

        Germanium, {\tt Ge}, is an appropriate material for birefrigent nanobars deposited on a calcium fluoride substrate ({\tt CaF$_{\rm 2}$}). Respectively, these materials have been selected for numerical analysis in SIMULIA CST Studio Suite, Fig.~\ref{Fig: Fig01}. The meta-atom is excited using a Floquet port with two modes: Mode 1, polarized along the $y$-axis [Transverse Electric (TE) mode], and Mode 2, polarized along the $x$-axis [Transverse Magnetic (TM) mode]. The boundary conditions are set as follows: Bloch-periodic along the $x$ and $y$ axes, and Perfectly Matched Layer (PML) with added space along the $z$-axis, which is the main wave propagating direction.
        
        To accurately assess phase retardation, two reference planes are positioned directly above and below the meta-atom (if the $z$-axis is vertical). Although the metalens designs that follow the PB-phase concept typically operate only with one preselected incident polarization state (such as RCP or LCP), here we are interested in developing a hybrid Moiré/PB-phase methodology that functions in a way that reconfigurable focusing is achievable for LP or EP light, as well as for RCP and LCP waves.Therefore, $x$- and $y$-directed LP simulations are conducted specifically to verify the fulfillment of phase retardation conditions. This verification ensures that efficient polarization conversion occurs when incident light interacts with the rotated meta-atoms.
        
        The complex transmission coefficients for the two main modes, $t_{xx}$ and $t_{yy}$, under the above-mentioned Bloch-periodic scenario must satisfy Eq.~\ref{Eq: Eq01} to exhibit birefringence-induced half-wave plate properties for linearly polarized light:
        \begin{equation}
            \angle t_{xx}-\angle t_{yy}\approx\pm\pi,
            \label{Eq: Eq01}
        \end{equation}
        where $\angle z \equiv \arg{z}$.
        
        On the other hand, the modal propagation phase difference in an analytical context (e.g.\ under effective medium approximation) for a birefringent structure is defined using the following equation:
        \begin{equation}
            \delta=2\pi\Delta n h/\lambda\approx\pm\pi,
            \label{Eq: Eq02}
        \end{equation}
        where $\Delta n$ represents the difference in effective refractive index caused by the birefringent properties of the meta-atoms. The parameter $h$ denotes the thickness or height of the meta-atoms. This height is measured along the direction of propagation.
        
        The refractive index behavior of {\tt CaF$_{\rm 2}$}, {\tt Ge}, and air with respect to the operating wavelength at 2~$\mu\text{m}$ is adopted from the sources of Refs.~\cite{RefractiveIndexInfo,Li_1980,Amotchkina_2020,Mathar_2007}. Respectively, in the unit cell of Fig.~\ref{Fig: Fig01}, the refractive indices of the designed meta-atom at 2~$\mu\text{m}$ are 1.42 for {\tt CaF$_{\rm 2}$}, 4.04 for crystalline {\tt Ge} (with an extinction coefficient of $k\approx 0$), and 1 for air. It should be noted that {\tt Ge} and air do not exhibit intrinsic birefringence. Instead, birefringence is achieved by using a biaxial or anisotropic {\tt Ge}  nanobar geometry, such as a rectangular prism with uneven sides.
        
\begin{figure}
            \begin{center}
                \subfigure[]{\label{Fig: Fig02.a}
                \includegraphics[width=0.6\linewidth,trim={0.0cm 0.0cm 0.0cm 0.0cm},clip=true]{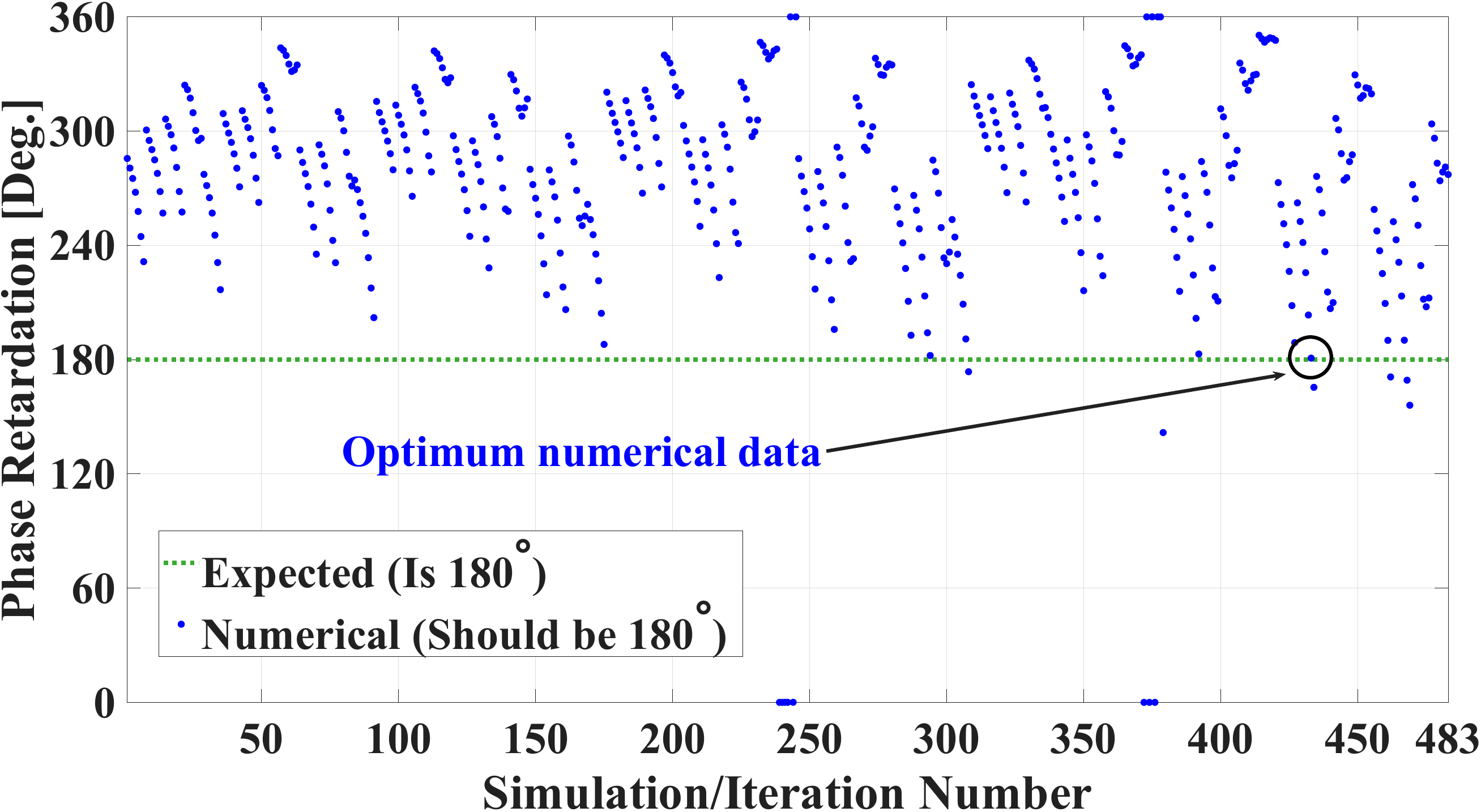}}
                \vspace*{-2mm}
                \subfigure[]{\label{Fig: Fig02.b}
                \includegraphics[height=0.25\linewidth,trim={0.0cm 0.0cm 0.0cm 0.0cm},clip=true]{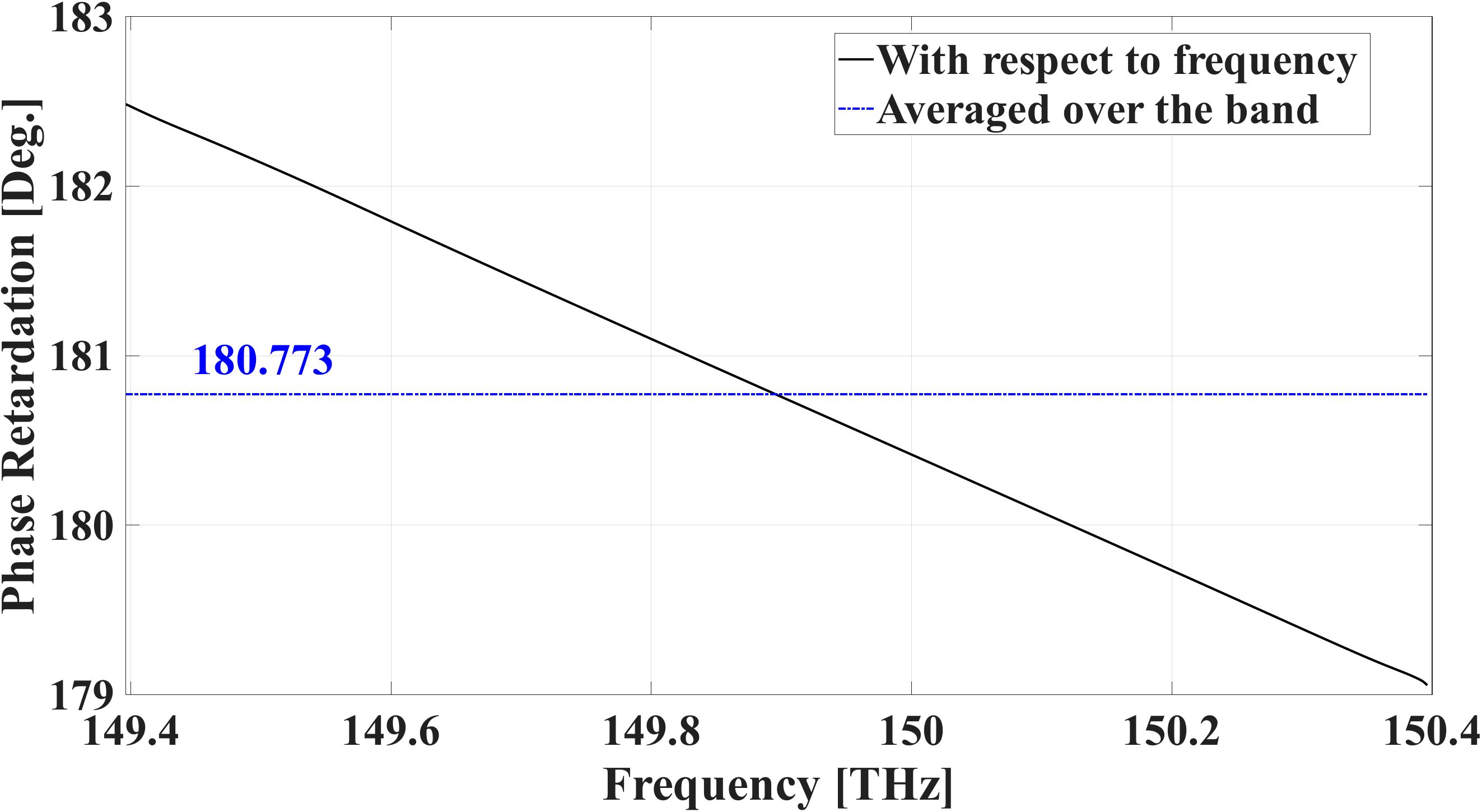}}
                \subfigure[]{\label{Fig: Fig02.c}
                \includegraphics[height=0.25\linewidth,trim={0.0cm 0.0cm 0.0cm 0.0cm},clip=true]{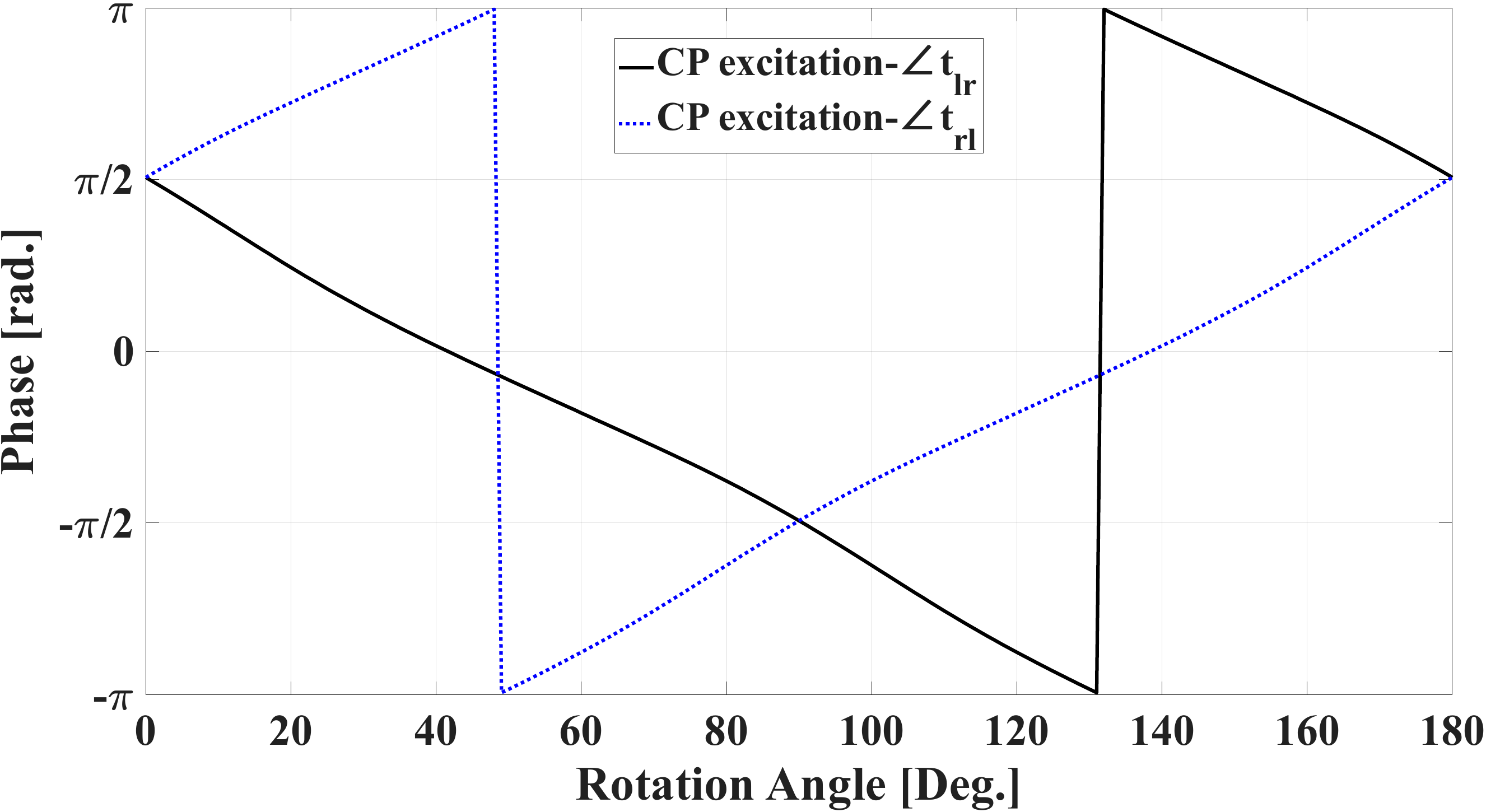}}
                
                \subfigure[]{\label{Fig: Fig02.d}
                \includegraphics[width=0.65\linewidth,trim={0.0cm 0.0cm 0.0cm 0.0cm},clip=true]{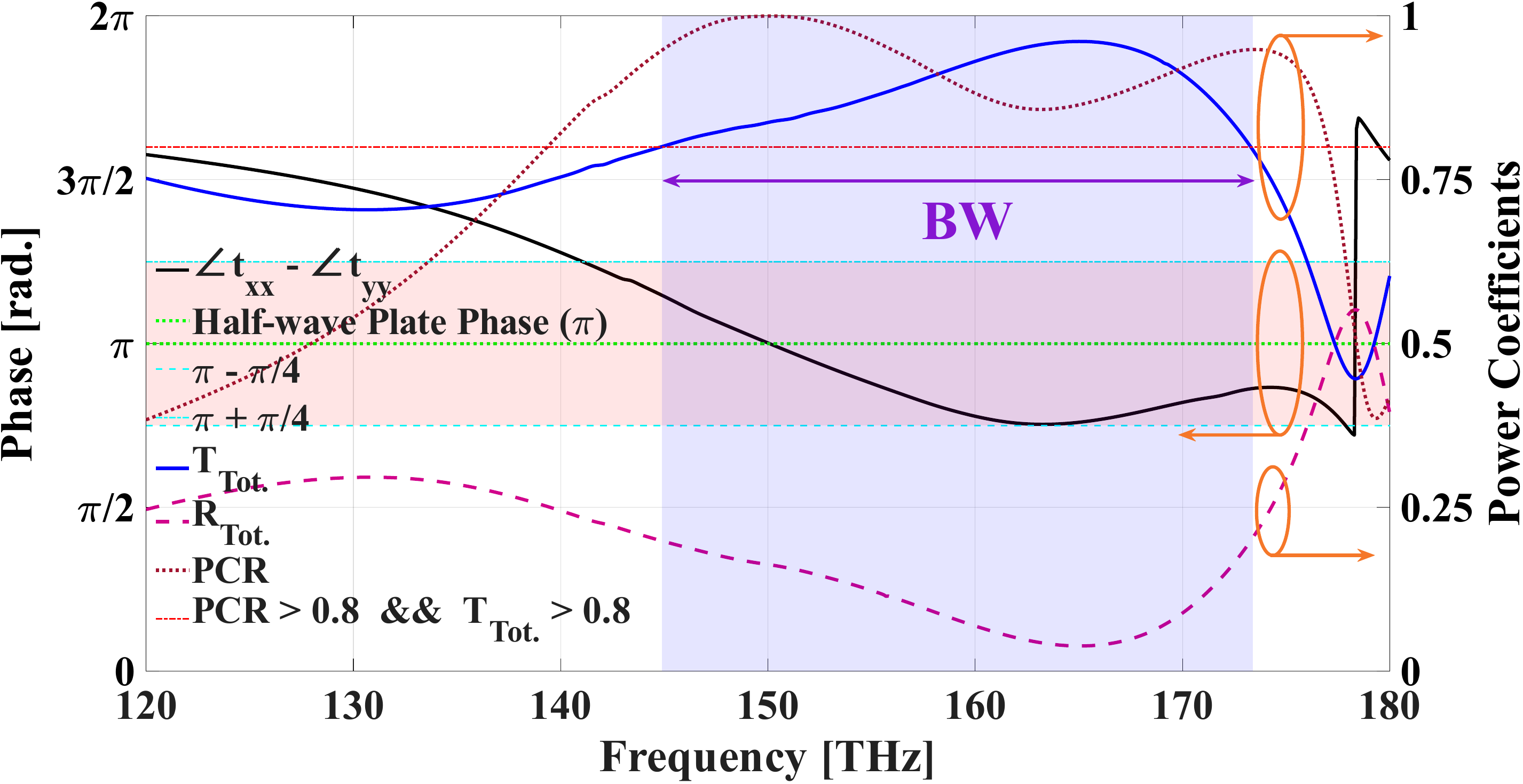}}
            \end{center}
            \vspace*{-7mm}
            \caption{Numerical studies on the birefringent and half-wave plate phase properties of the proposed {\tt Ge}-based meta-atom. (a) Phase retardation of the meta-atom while varying the parameters $p$, $L$, $W$, and $h$, (b) Phase retardation of the selected meta-atom with respect to frequency, (c) Phase retardation of the meta-atom with respect to the rotation angle while excited by circularly polarized illuminations at a wavelength of 2~$\mu\text{m}$, and (d) Frequency-dependent characteristics of transmission, reflection, PCR, and half-wave plate phase retardation behavior of the proposed meta-atom.}
            \label{Fig: Fig02}
        \end{figure}

        To determine a suitable material for the nanobar and the dimensions of the meta-atom, a parametric study was conducted using SIMULIA CST Studio Suite. In this study, the substrate thickness was set to 0.25~$\mu\text{m}$. This thickness imposes a constant phase shift on all incident waves. Therefore, it does not affect the variation of the metalens focal point. The parameters $p = p_x = p_y$, $L$, $W$, and $h$ of the meta-atom were varied under the constraints $p<\lambda/2$, $p>L>W$, and $p>\sqrt{L^{\rm 2}+W^{\rm 2}}$. By adjusting these parameters, the conditions in Eqs.~\ref{Eq: Eq01} and~\ref{Eq: Eq02} were targeted. Figure~\ref{Fig: Fig02} shows the numerical phase retardation results for the meta-atom, whose properties vary across different simulations.
        
        Based on the numerical phase simulation results in Fig.~\ref{Fig: Fig02.a}, we have found that a meta-atom with the {\tt Ge} nanobar geometry corresponding to the 433$^{\rm rd}$ iteration shows the closest match to the expected 180$^\circ$ phase retardation. Its geometrical parameters are $p=0.7$~$\mu\text{m}$, $L=0.55$~$\mu\text{m}$, $W=0.25$~$\mu\text{m}$, and $h=0.75$~$\mu\text{m}$. The obtained phase retardation is 180.77$^\circ$. For this selected case, $\Delta n$ is approximately 1.34 (Eq.~\ref{Eq: Eq02}). This value indicates a strong birefringent behavior of the {\tt Ge} nanobar with such parameters.
        
        Figure~\ref{Fig: Fig02.b} shows the numerical phase retardation and its mean value for the selected meta-atom. The results confirm its half-wave-plate behavior. Figure~\ref{Fig: Fig02.c} shows the phase variation of the metalens under RCP and LCP excitation with respect to nanopillar rotation angles, which covers a phase range of 0$^\circ$ to 360$^\circ$ upon rotation from 0$^\circ$ to 180$^\circ$, making it ideal for focusing applications. Full-wave simulations further show that the unit cell operates in a highly transmissive regime near the design wavelength of 2~$\mu\text{m}$ ($\sim$149.896~THz). To determine the operational bandwidth, we adopt the Rayleigh criterion for phase tolerance while requiring the PCR to remain above 80\%, as indicated by the blue shaded region in Fig.~\ref{Fig: Fig02.d}. The Rayleigh criterion is widely employed to define the acceptable phase error for maintaining high-efficiency broadband cross-polarization conversion in optical systems~\cite{BornWolf1999}. For an ideal half-wave plate, the phase retardation should remain within the Rayleigh tolerance of $\pi \pm \pi/2$, which is equivalent to $\lambda \pm \lambda/4$. In this work, however, a more stringent criterion is adopted to ensure robust lens performance. Specifically, the allowable phase retardation is restricted to $\pi \pm \pi/4$, corresponding to $\lambda \pm \lambda/8$, as highlighted by the red shaded region in Fig.~\ref{Fig: Fig02.d}. Consequently, the operational Bandwidth (BW) of the structure is defined as the intersection of these two shaded areas, spanning from 144.90~THz to 173.40~THz. The Fractional Bandwidth (FBW) is 17.91\%, indicating a fairly small/narrow FBW (i.e., high quality factor), which is generally beneficial for applications requiring highly precise and focused performance. At the targeted operating wavelength of 2~$\mu\text{m}$, the structure achieves a 99.95\% PCR, 83.65\% total transmission power, and 16.33\% total reflection power. These results confirm that the meta-atom behaves as a low-loss and weakly reflecting birefringent element. It is therefore suitable for efficient phase-only wavefront shaping in the proposed metalens.
    \subsection{\label{sec:phaseeng} Moiré Phase Engineering for Wavefront Shaping}
        Moiré lenses are phase-only optical elements, also known as {\em kinoforms}. They generate a continuous or discretized phase profile for wavefront shaping. This approach is particularly useful for focusing applications that avoid unnecessary amplitude modulation. A Moiré metalens consists of two cascaded metasurfaces with complex conjugate or complementary phase profiles. One metasurface is rotated by an angle $\Theta$ relative to the other. This mutual rotation produces a combined phase distribution, making the focal length tunable.
        
        The Transmission Functions (TFs) of the two Moiré metasurfaces are defined as follows:
        \begin{equation}
            T_{\rm 1}(r,\theta)=e^{i\phi(r,\theta)},
            \label{Eq: Eq03}
        \end{equation}
        \begin{equation}
            T_{\rm 2}(r,\theta)=e^{-i\phi(r,\theta)},
            \label{Eq: Eq04}
        \end{equation}
        where $\phi(r,\theta)$ denotes the phase distributions generated by the metasurfaces as functions of $r$ and $\theta$. Here, $r$ and $\theta$ represent the polar coordinates.
        
        Taking into account the phase-complementary nature of the Moiré lenses, we additionally enforce the relation $\phi(r,\theta)=-\phi(r,-\theta)$. This means that a planar DOE with this property produces a conjugate phase profile when flipped, because flipping replaces the coordinates $(r,\theta)$ with $(r,-\theta)$. In this case, one can take a pair of identical DOEs, turn one upside down, and place it face-to-face on top of another. This will yield the desired Moiré metalens configuration. When two such DOEs are positioned as just described, the Transmission Function (TF) of the combined metalens can be expressed as follows:
        \begin{multline}
            T_{\rm Joint}(r,\theta;\Theta)=T_{\rm 1}(r,\theta)T_{\rm 2}(r,\theta-\Theta)\\=e^{i\phi(r,\theta)}e^{-i\phi(r,\theta-\Theta)}=e^{i(\phi(r,\theta)-\phi(r,\theta-\Theta))},
            \label{Eq: Eq05}
        \end{multline}
        where the $\Theta > 0$ represents the rotation angle of the second disk in the Clockwise (CW) direction. For $\Theta < 0$, however, the rotation direction changes to Counterclockwise (CCW). 
        
        In practice, the phase profile in Eq.~\ref{Eq: Eq05} is approximated assuming $\phi(r,\theta)=\phi_{\rm r}(r)\phi_{\rm \theta}(\theta)$ and employing power series expansions in $r$ and $\Theta$:
        \begin{equation}
            T_{\rm Joint}(r,\theta;\Theta)=e^{i\phi_{r}(r)\Theta}
            \label{Eq: Eq06}
        \end{equation}
        where $\phi_{\rm \theta}(\theta)=\theta$, $\phi_{\rm r}(r)=ar^{\rm 2}$,  and $a$ is the linear scaling factor representing the optical power or Power Factor (PF). In optical systems, the power factor indicates the refractive power of a lens, which is the reciprocal of its focal length $f$, i.e., $PF\equiv 1/f$. The linear scaling factor for a Moiré metalens must remain constant and satisfy the condition $a < 1/(2pR_{\rm Disk})$, where $R_{\rm Disk}$ denotes the radius of the Moiré disk. On the other hand, the relationship between the rotation angle and the focal length for a Moiré metalens is given by $a = \pi/(\lambda f |\Theta|)$.
        
        It is well known that constructing a Moiré lens based on Eqs.~\ref{Eq: Eq03}, ~\ref{Eq: Eq04}, and ~\ref{Eq: Eq06} leads to a sectoring effect. This phenomenon reduces the focusing performance and overall efficiency of the lens. The problem arises because the phase profiles of the two DOEs do not vary continuously around the origin after a full $2\pi$ rotation in the $\theta$ angle. Instead, phase discontinuities appear that are not simple integer multiples of $2\pi$. When the polar angle $\theta$ is bounded to the interval $(-\pi;\pi]$, these discontinuities occur along the radial line at $\theta=\pi$, i.e. along the path extending from the center toward the left edge of each DOE. To eliminate these discontinuities and improve focusing efficiency, the phase profiles of the disks forming the Moiré lens can be slightly modified as follows:
        \begin{equation}
            T_{\rm 1}(r,\theta)=e^{i\cdot{\tt round}(ar^{\rm 2})\theta}
            \label{Eq: Eq07}
        \end{equation}
        \begin{equation}
            T_{\rm 2}(r,\theta)=e^{-i\cdot{\tt round}(ar^{\rm 2})\theta}
            \label{Eq: Eq08}
        \end{equation}
        where the {\tt round($\cdot$)} operation indicates rounding the argument to the nearest integer. Such a modification is equivalent to introducing circular zones analogous to the zones on a Fresnel zone plate.
        
        Finally, the transmission phase profiles of the disks are restricted to the interval $[0,2\pi)$ using the ${\tt Confine}(x) \equiv x \mod 2\pi$ phase wrap function, which wraps and maps any input angle into this interval:
        \begin{equation}
            T_{\rm 1}(r,\theta)=e^{i\cdot{\tt Confine}({\tt round}(ar^{\rm 2})\theta)},
            \label{Eq: Eq09}
        \end{equation}
        \begin{equation}
            T_{\rm 2}(r,\theta)=e^{-i\cdot{\tt Confine}({\tt round}(ar^{\rm 2})\theta)}
            \label{Eq: Eq10}.
        \end{equation}
        This process removes redundant phase values that are physically equivalent and yield identical optical responses. As a result, the final TF for the designed Moiré lens can be represented as follows:
        \begin{equation}
            T_{\rm Joint}(r,\theta;\Theta)=e^{i\cdot{\tt Confine}({\tt round}(ar^{\rm 2})\Theta)}.
            \label{Eq: Eq11}
        \end{equation}
        Based on Eqs.~\ref{Eq: Eq09} and~\ref{Eq: Eq10}, one can conclude that creating a Moiré lens requires two DOE disks that generate the transmission phase distributions ${\tt Confine}({\tt round}(ar^{\rm 2})\theta)$ and $-{\tt Confine}({\tt round}(ar^{\rm 2})\theta)$, respectively. In addition, there is a physical limit on the maximum allowable axial distance between the two DOEs. To ensure that light emanating from an element of the first DOE (a ``pixel'') properly passes through the corresponding ``pixel'' of the second DOE, the separation between the two DOEs must be less than half of the Talbot length, i.e., $T_{\rm Talbot}=2p^{\rm 2}/\lambda$.
    \subsection{Pancharatnam-Berry Phase and Meta-atom Orientation Profile Mapping}
        Metasurfaces with varying geometries, such as PB-phase lenses, are planar arrays of subwavelength nanostructures. They have become powerful platforms for wavefront manipulation. The PB-phase mechanism provides a unique approach to wavefront shaping as compared to typical metasurfaces that control propagation phase. In this method, polarized light acquires the so-called geometric phase after passing through anisotropic optical elements. PB-phase metalenses use the link between photonic angular momentum (photonic spin) and local geometric rotation to generate spatially varying phase profiles. This enables subwavelength control of amplitude, phase, and polarization while maintaining high diffraction efficiency.
        
        Thus the main advantage of PB-phase designs is in their geometrically controlled local phase modulation. This effect arises from the geometric phase $\Psi_{\rm PB}(r,\theta)=2\sigma\Phi(r,\theta)$, which depends on the rotation angle $\Phi$ of the anisotropic meta-atoms at the polar position $(r,\theta)$. Here, $\sigma=\pm1$ represents the photonic spin of the incident wave, with $+1$ for RCP and $-1$ for LCP. However, efficient PB-phase focusing requires each meta-atom to function as a half-wave plate with polarization conversion efficiency close to unity. In practice, this requirement demands precise electromagnetic design and optimization of the unit cell geometry, such as the {\tt Ge} nanobar length, width, and height, through coupled eigenmode and full-wave simulations. A target phase profile for a PB-phase metalens used in focusing applications requires parabolic phase distribution, as described in Eq.~\ref{Eq: Eq12}.
        \begin{equation}
            \Psi_{\rm PB}(r,\theta,\lambda)=-\frac{2\pi}{\lambda}(\sqrt{r^{\rm 2}+f^{\rm 2}}-f)
            \label{Eq: Eq12}
        \end{equation}
        As a result, the phase mapping (i.e. phase profile to meta-atom rotation profile mapping) of the PB-phase metalens is conducted using the in-plane orientation angles of the meta-atoms:
        \begin{equation}
            \Phi(r,\theta)=\frac{\Psi_{\rm PB}(r,\theta,\lambda)}{2\sigma}=\frac{-\frac{2\pi}{\lambda}(\sqrt{r^{\rm 2}+f^{\rm 2}}-f)}{2\sigma}=-\frac{\pi}{\sigma\lambda}(\sqrt{r^{\rm 2}+f^{\rm 2}}-f).
            \label{Eq: Eq13}
        \end{equation}
    \subsection{From Spin to Phase: Jones Vector Analysis of PB-Phase Polarization Conversion}
        The polarization manipulation characteristics of an elementary cell of a PB-phase metalens operating as described above can be analyzed using the Jones vector framework. In this study, the harmonic time dependence is assumed in the form of $e^{-i\omega t}$, consistent with optical conventions. The incident wave is assumed to propagate along the positive $z$-direction. For a rotated anisotropic meta-atom characterized by transmission coefficients $t_{xx}$ and $t_{yy}$ in its principal axes, the Jones matrix is written as
        \begin{equation}
            J(\Phi(r,\theta))=R(-\Phi(r,\theta))
            \begin{bmatrix}
                t_{xx} & 0 \\
                0 & t_{yy}
            \end{bmatrix}
            R(\Phi(r,\theta)),
            \label{Eq: Eq14}
        \end{equation}
        in which $R(\Phi(r,\theta))$ denotes the rotation matrix:
        \begin{equation*}
            R(\Phi)=
            \begin{bmatrix}
                \cos\Phi & -\sin\Phi \\
                \sin\Phi & \quad\cos\Phi
            \end{bmatrix}.
        \end{equation*}
        
        Since each meta-atom operates similarly to a half-wave plate, the transmission amplitudes along the principal axes satisfy $t_{xx}\approx -t_{yy}$ with nearly equal magnitudes. Under this condition, the structure efficiently manipulates the handedness of the polarization. It also introduces a geometric phase determined by the meta-atom rotation.
        
        To evaluate the performance of the proposed structure and confirm its polarization conversion capability, we analytically examine the excitation of the device using RCP, LCP, and LP light. Because our structure consists of two cascaded disks, polarization conversion occurs twice. The polarization states of the incident RCP and LCP lights are defined as follows:
        \begin{equation}
            \mathbf{E}^{\rm R/L}_{\rm I}=\frac{1}{\sqrt{2}}
            \begin{bmatrix}
                1 \\
                \pm i
            \end{bmatrix}.
            \label{Eq: Eq15}
        \end{equation}
        As a result, the light transmitted through the first disk in the forward direction (at a given point on the lens aperture) can be expressed as:
        \begin{equation}
            \mathbf{E}^{\rm L/R}_{\rm Out_{\rm 1}}=J(\Phi_{\rm 1}(r,\theta))\mathbf{E}^{\rm R/L}_{\rm I}=\frac{t_{xx}+t_{yy}}{2}
            \begin{bmatrix}
                1 \\
                \pm i
            \end{bmatrix}+\frac{t_{xx}-t_{yy}}
            {2}e^{\mp i2\Phi_{\rm 1}(r,\theta)}
            \begin{bmatrix}
                1 \\
                \mp i
            \end{bmatrix},
            \label{Eq: Eq16}
        \end{equation}
        in which the first term reduces to zero because the first disk satisfies the half-wave condition. Consequently, only the field component with the flipped polarization state remains. This field then propagates across the gap between the two disks without polarization manipulation. It subsequently illuminates the second disk. Therefore, the light transmitted through the second disk (at the same location on the lens aperture) in the forward direction is given by
        \begin{equation}
            \mathbf{E}^{\rm R/L}_{\rm Out_{\rm 2}}=J(\Phi_{\rm 2}(r,\theta))\mathbf{E}^{\rm L/R}_{\rm Out_{\rm 1}}\approx t_{xx}^2 e^{\mp i2\Phi_{\rm 1}(r,\theta)}e^{\pm i2\Phi_{\rm 2}(r,\theta)}
            \begin{bmatrix}
                1 \\
                \pm i
            \end{bmatrix},
            \label{Eq: Eq17}
        \end{equation}
        where the component with the same polarization as the one incident on the second disk vanishes and is omitted because the second disk also satisfies the half-wave condition.
        
        Analysis of Eqs.~\ref{Eq: Eq14}-\ref{Eq: Eq17} reveals an important consequence of the polarization-flipping nature of the disks. Whenever one of the disks is illuminated by light with a given polarization (e.g.~RCP), polarization flipping phenomenon (e.g.~RCP is converted to LCP) leads to an effective negation of the required phase action of the other disk. This means that, under a given illumination, the transmission phase profiles of the first and second disks will have complementary properties, even with identical position-dependent rotation profiles of their {\tt Ge} nanopillars! In other words, polarization-converted transmitted waves effectively introduce complementary phase behavior for the second disk, although it is structurally identical to the first one. Therefore, applying the PB-phase concept when realizing the required {\em complementary} Moiré transmission phase profiles of cascaded disks results in structurally {\em identical design and orientation} of the two disks.
        
        Consequently, to connect the PB-phase metalens formalism with the Moiré approach outlined in Sec.~\ref{sec:phaseeng}, we expand Eq.~\ref{Eq: Eq13} in Taylor series considering the ratio $r/f$ as a small parameter while keeping terms up to the second order in this parameter. Since in our reconfigurable lens the total phase retardation must be split between the two Moiré patterned disks, the profiles of the meta-atom rotation angle in Disks $I$ and $II$ (relative to the $x$-axis in the local coordinate system of each disc) are obtained as, respectively,
        \begin{equation}
            \Phi_{\rm 1}(r,\theta)=\Phi_{\rm 2}(r,\theta) = \frac{{\tt Confine}({\tt round}(ar^{\rm 2})\theta)}{2\sigma}.
            \label{Eq: Eq18}
        \end{equation}
        
        Interestingly, the proposed structure enables tunable focal length creation for both RCP and LCP illuminations. This capability exists despite the structure relying on polarization-coded PB-phase disks. However, for a given rotation angle $\Theta$, the required rotation direction of the second disk depends on the incident polarization.
        
        Specifically, to realize focusing behavior under RCP excitation that is similar in action to a convex lens, the second disk should be rotated CW by angle $0 < \Theta < \pi$. Under an LCP excitation, a CCW rotation is mandated. If the rotation direction is reversed, i.e.\ when $ -\pi < \Theta < 0$, the operation of the lens is reversed too (e.g.\ an effectively convex lens becomes concave). As a result, the metalens performance at arbitrary excitation is dictated by the superposition of its response to the constituent circular components.
        
        The proposed hybrid Moiré-PB architecture introduces a rotation-controlled mechanism to unlock polarization-selective functionality. In particular, since an LP wave can be decomposed into equal-magnitude RCP and LCP components: LP = $\frac{1}{\sqrt{2}}$(RCP + LCP), the hybrid Moiré-PB metasurface simultaneously imparts conjugate transmission phase profiles to each. Under EP illumination, also, the incident wave can be decomposed into an asymmetric combination of RCP and LCP components, expressed generally as $\mathbf{E}_{\rm EP} = \alpha \mathbf{E}_{\rm RCP} + \beta e^{j\Delta\phi} \mathbf{E}_{\rm LCP}$, where $\alpha$ and $\beta$ represent the unequal amplitude coefficients ($\alpha^2 + \beta^2 = 1$) and $\Delta\phi$ denotes their relative phase difference. Consequently, the hybrid Moiré-PB metasurface imparts conjugate phase profiles to opposing helicities, while only one specific spin component undergoes convergence to the target focal point with respect to the second disk rotation direction. In contrast, the counterpart helicity diverges into the background. Since the peak focusing efficiency occurs under pure circular excitation of the responsive helicity (where either $\alpha=1$ or $\beta=1$ for EP), any linear/elliptical state distributing power across both spin states inevitably routes a fraction of the total incident energy into the divergent channel. Consequently, the operational focusing efficiency under linear excitation remains fundamentally bounded by a maximum theoretical limit of 50\% relative to the total available power.
        
        Nevertheless, considering the behavior just discussed, the proposed metalens is anyway expected to form a focal spot under LP/EP illumination, albeit with a 50\% loss in power. Interestingly, both CW and CCW rotations of the second disk for LP illumination will yield the same focal length, determined by the rotation angle magnitude $|\Theta| < \pi$. However, the opposite directions of rotation will lead to focal spots formed by light of opposite helicity.
    \subsection{From Rotation Angle to Focal Length: Positive vs. Negative Mutual Rotation}
        As was mentioned before, the focal length for various rotation angles $\Theta$ can be calculated analytically using the equation
        \begin{equation}
            f=\frac{\pi}{a\lambda|\Theta|}.
        \label{Eq: Eq19}
        \end{equation}
        The mutual rotation angle of the second disk with respect to the first disk, which is the meaning of angle $\Theta$, controls the effective focal length of the Moiré-PB metalens. It does so by tuning the net quadratic phase term on the combined aperture. At zero mutual rotation, the conjugate phase profiles cancel each other, and the focal length tends to infinity. Increasing the rotation angle progressively increases the optical power and moves the focal spot toward the metalens.
        
        As established earlier, the focusing or defocusing function of the proposed metalens is governed by the rotation direction when excited with RCP, LCP, EP, or LP light. For RCP light excitation, the rotation must follow the CW direction and not exceed 180$^\circ$. This allows the structure to function as a convex lens producing a real focus. Under LCP illumination, a similarly limited rotation must be applied in the CCW direction. This also enables convex lens behavior and formation of a real focus.
        
        For RCP excitation with CCW rotation and LCP excitation with CW rotation, the signs of the quadratic terms in the metalens transmission phase profile are reversed. Therefore, in these cases the metalens has a virtual focus on the object side rather than a real focus on the image side. This corresponds to a concave (defocusing) operation of the metalens. Finally, note that under EP or LP illuminations, one of the circular components always forms a real focus, the distance to which is still given by Eq.~\ref{Eq: Eq19}.
                
        Moreover, when rotation exceeds $180^\circ$ CCW or CW, i.e.\ when $|\Theta| > \pi$, the periodicity of the Moiré phase pattern implies that such a rotation is equivalent to oppositely directed rotation by an angle $|\Theta'| = 2\pi-|\Theta| < \pi$. In other words, rotations exceeding $180^\circ$ do not create new categories of phase profiles. Instead, they map onto the same set of focal powers already accessed by smaller rotations in opposite directions. For this reason, we constrain the rotations to $|\Theta| \leq \pi$.
\section{Implementation of Hybrid Moiré-PB-Phase Metalens and Related Phase Profiles}
\begin{figure}
        \begin{center}
            \subfigure[]{\label{Fig: Fig03.a}
            \includegraphics[width=0.48\linewidth,trim={0.0cm 0.0cm 0.0cm 0.0cm},clip=true]{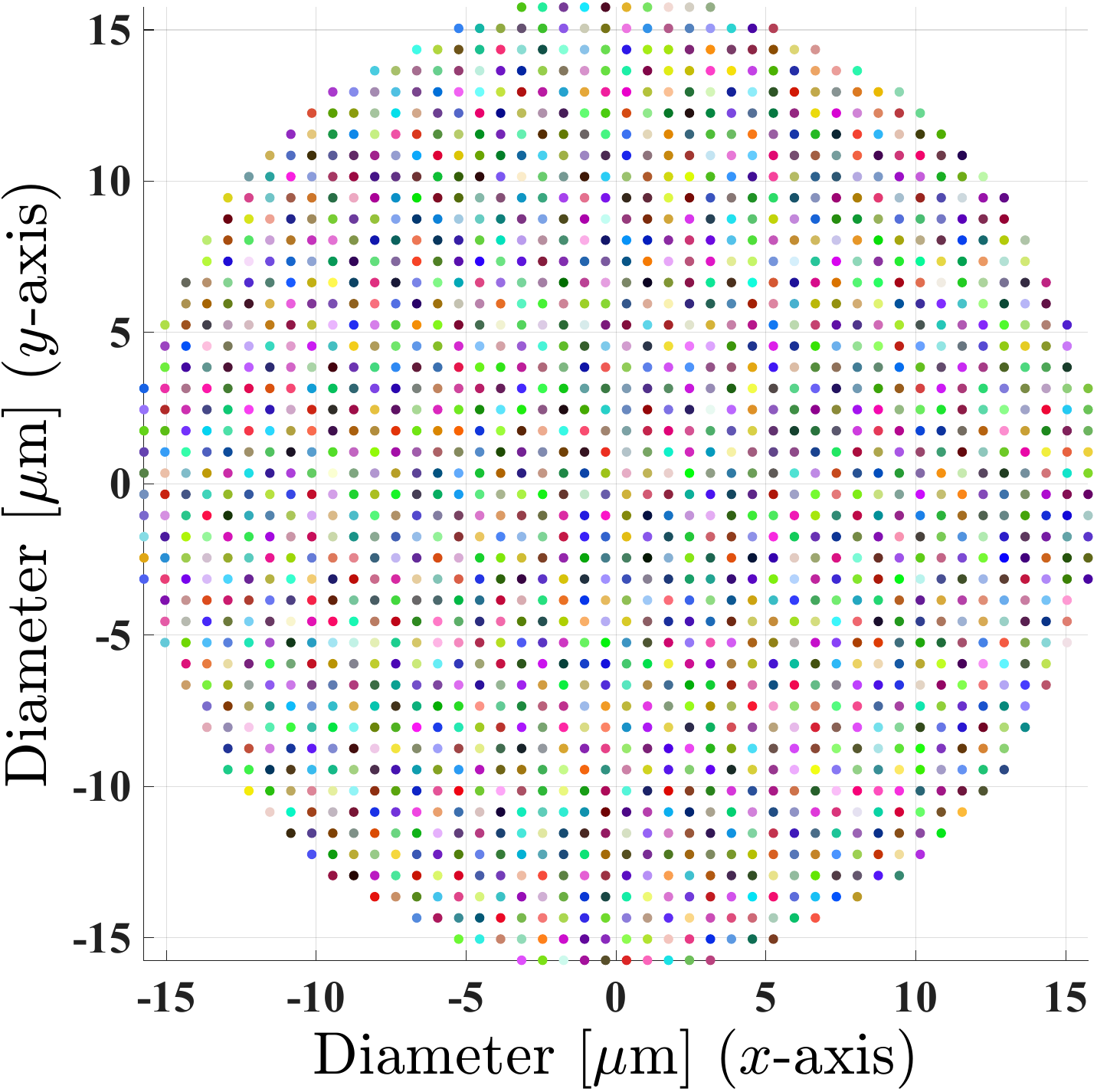}}
            \subfigure[]{\label{Fig: Fig03.b}
            \includegraphics[width=0.48\linewidth,trim={0.0cm 0.0cm 0.0cm 0.0cm},clip=true]{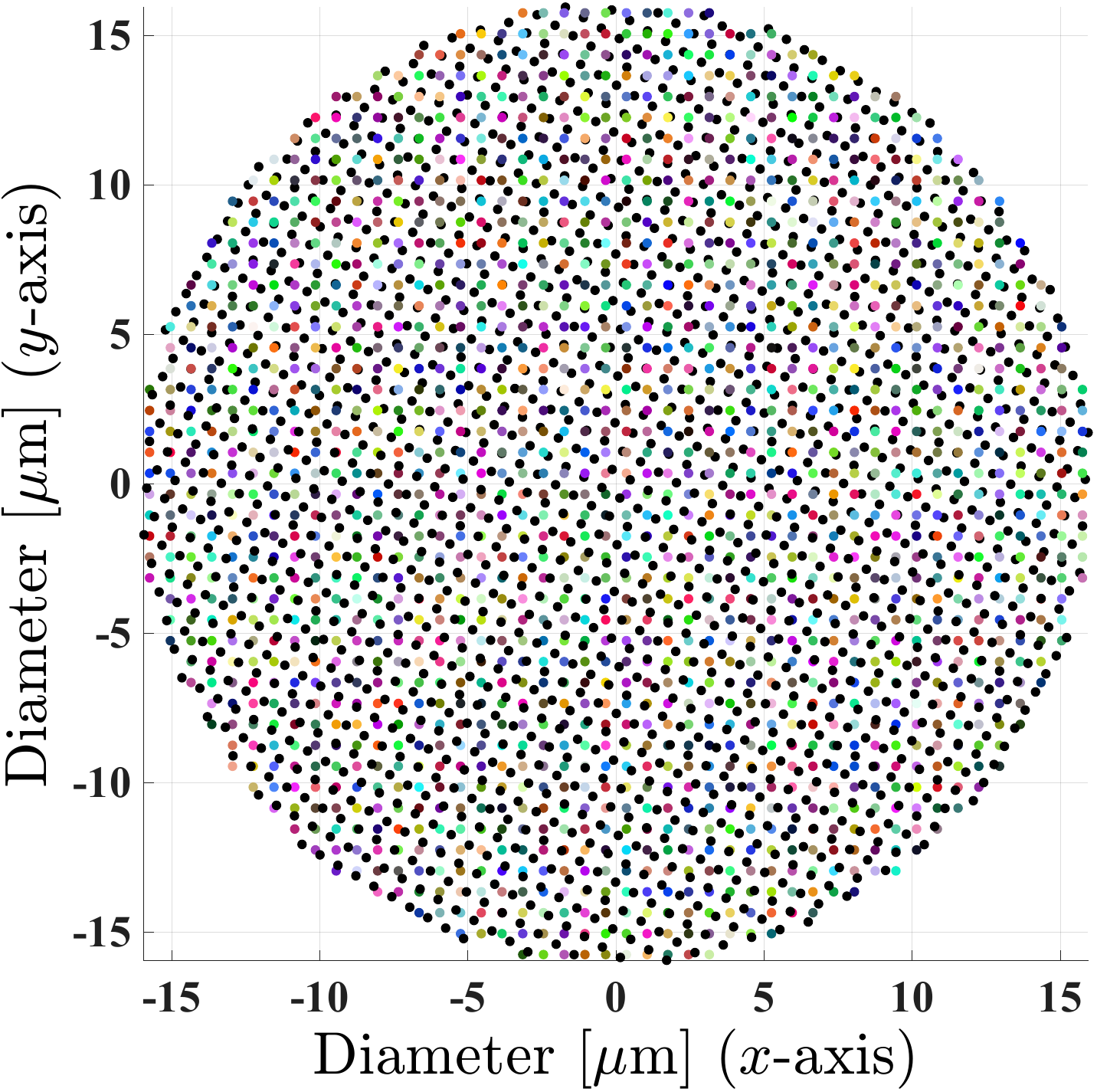}}

            \vspace*{-2mm}
            \subfigure[]{\label{Fig: Fig03.e}
            \includegraphics[width=0.48\linewidth,trim={1.6cm 0.6cm 4.0cm 2.35cm},clip=true]{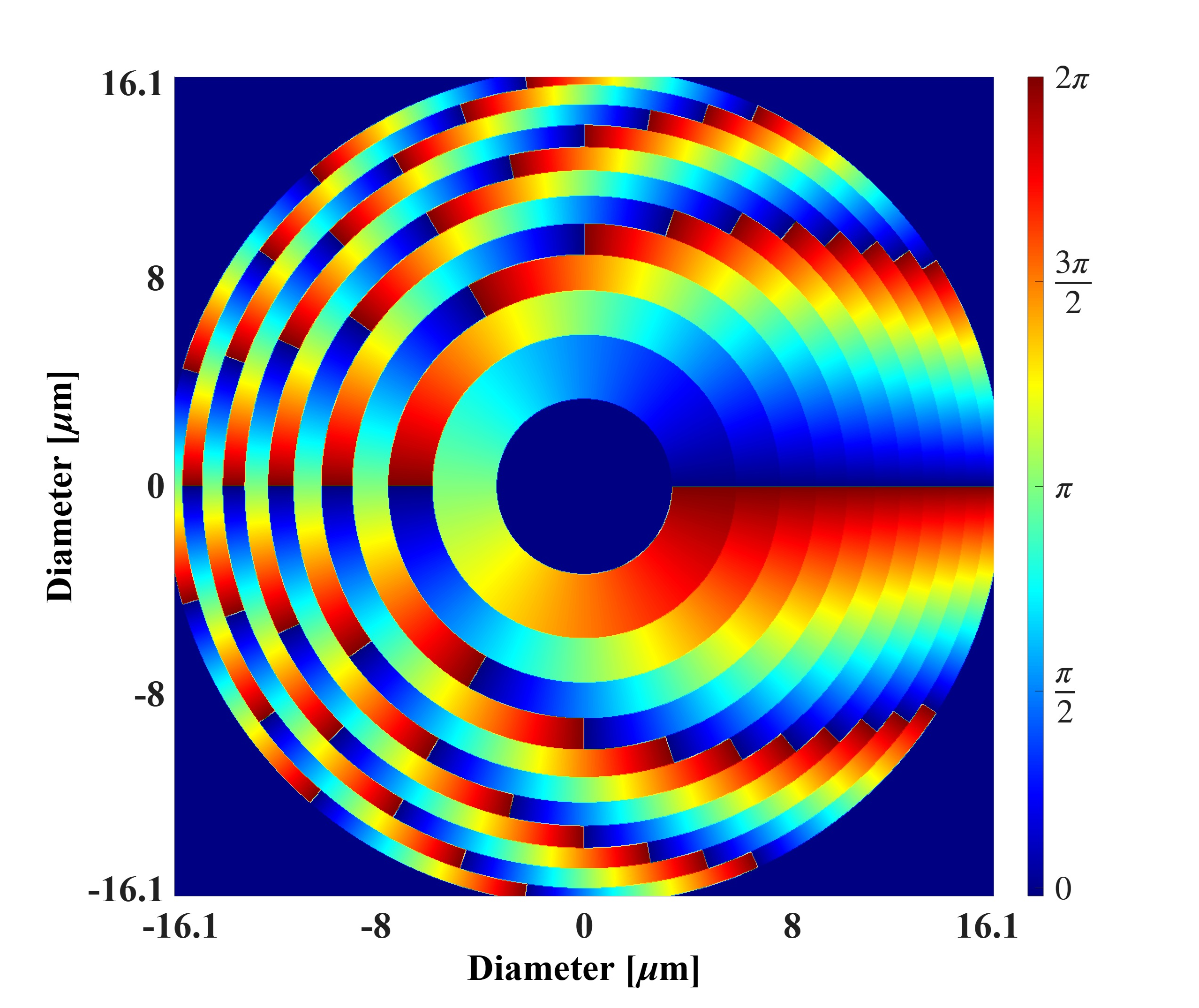}}
        \end{center}
        \vspace*{-5mm}
        \caption{Localization of unit cells in Cartesian coordinates and phase distribution of Moiré-PB disks. (a) Distribution of meta-atoms on each disk in Cartesian coordinates, (b) Localization of meta-atoms when the second disk is rotated at an arbitrary angle, (c) Moiré phase distribution for Disks $I$ and $II$.}
        \label{Fig: Fig03}
    \end{figure}

    In the studies that are described next, a hybrid reconfigurable metalens consisting of two Moiré-PB disks is considered. A three-dimensional full-wave electromagnetic model of the metalens structure with realistic material parameters is implemented in ANSYS Lumerical. The model incorporates a total of 1664 meta-atoms distributed over each disk in a square lattice. This arrangement maintains a fixed lattice constant along the $x$ and $y$ axes, as illustrated in Fig.~\ref{Fig: Fig03.a}. The diameter of the disk is such that 46 unit cells fit along the disk diameters collinear with the $x$ and $y$ axes. An aperture matching the metalens diameter is introduced at the first disk to suppress incident wave propagation outside the lens perimeter.
    
    Considering the meta-atom lattice period $p=0.7~\mu\text{m}$, the cascaded disks have a diameter of 32.2~$\mu\text{m}$ and an interlayer space of $g_{\rm 0}=T_{\rm Talbot}/3=0.163\ \mu\text{m}$, respectively. With reference to Eqs.~\ref{Eq: Eq09} and \ref{Eq: Eq10}, the Moiré-PB disks must exhibit phase profiles $\phi(r,\theta) = {\tt Confine}({\tt round}(ar^{\rm 2})\theta)$. Furthermore, for the designed Moiré-PB metalens, the maximum attainable linear scaling factor is $a = 4.437\times10^{10}$~m$^{-2}$. From this condition, a value of $a = 4.4\times10^{10}$~m$^{-2}$ has been selected for implementing the phase profiles in practice. The required Moiré-PB phase profiles (interpolated to improve picture clarity (see Supplementary Material Section 1, Fig.~1)) for the hybrid metalens are shown in Fig.~\ref{Fig: Fig03.e}.
        
\begin{figure}
        \begin{center}
            \subfigure[]{\label{Fig: Fig04.a}
            \includegraphics[width=0.48\linewidth,trim={1.5cm 0.55cm 3.95cm 2.4cm},clip=true]{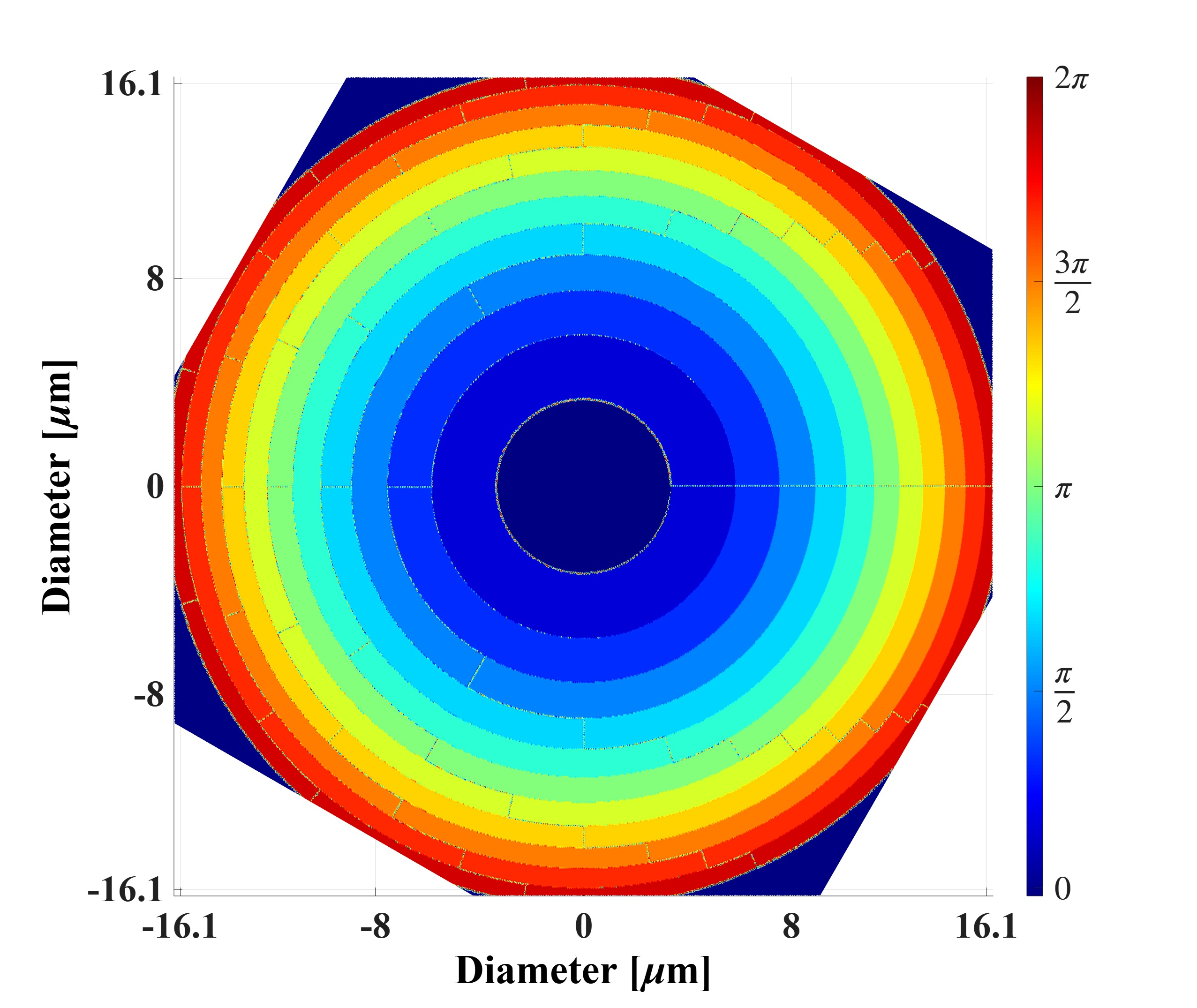}}
            \subfigure[]{\label{Fig: Fig04.b}
            \includegraphics[width=0.48\linewidth,trim={1.5cm 0.55cm 3.95cm 2.4cm},clip=true]{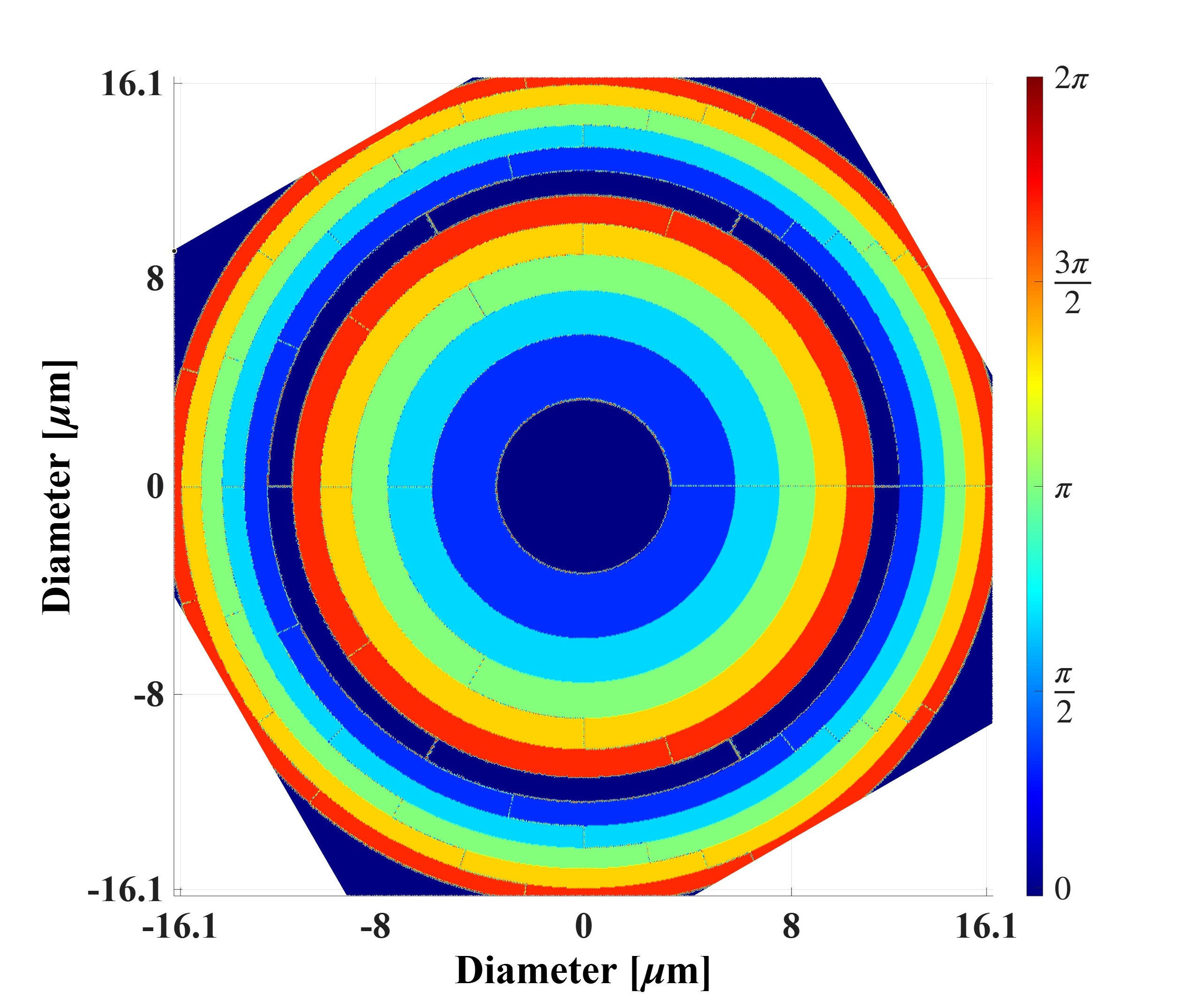}}
            
            \vspace*{-2mm}
            \subfigure[]{\label{Fig: Fig04.c}
            \includegraphics[width=0.48\linewidth,trim={1.5cm 0.55cm 3.95cm 2.4cm},clip=true]{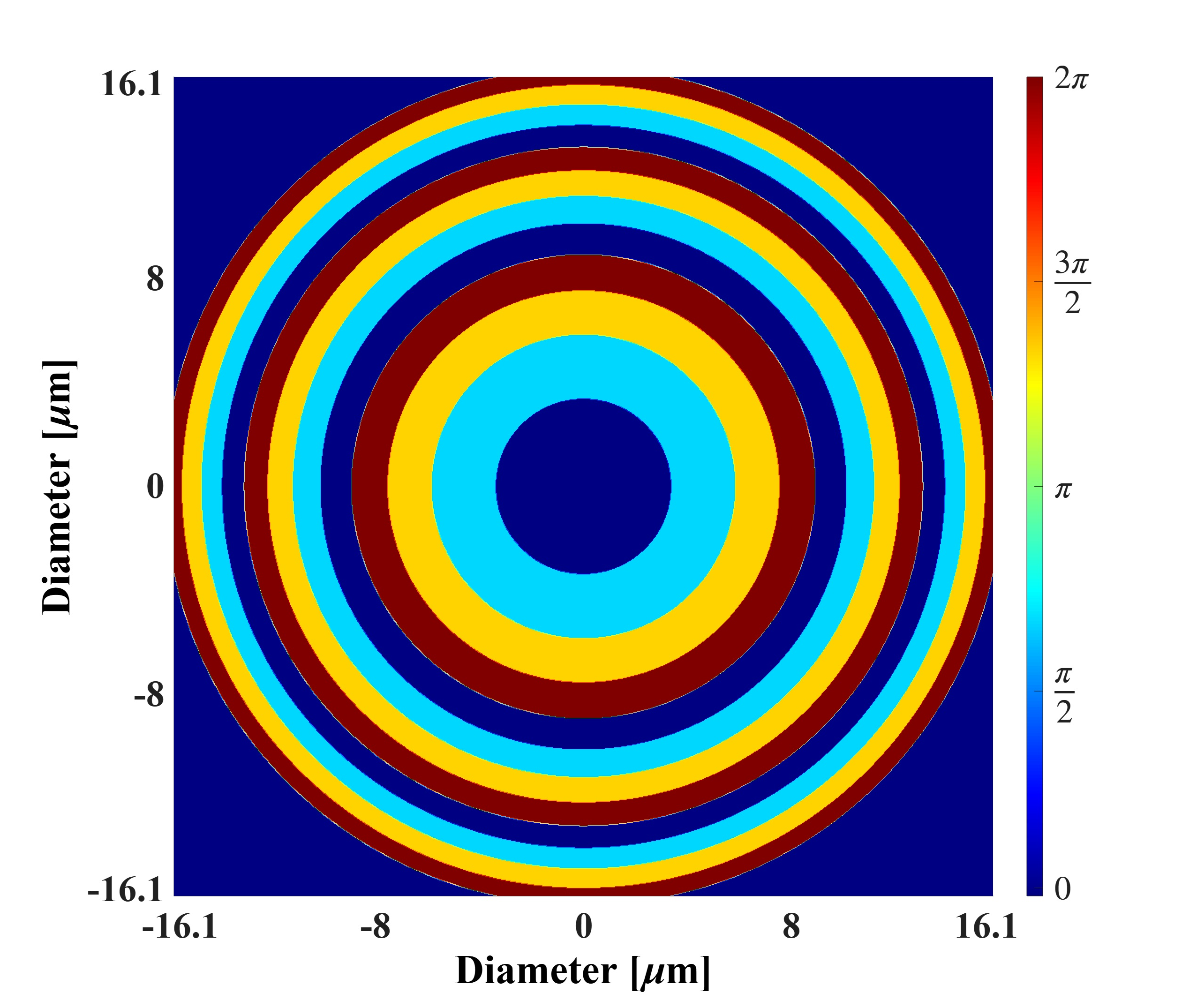}}
            \subfigure[]{\label{Fig: Fig04.d}
            \includegraphics[width=0.48\linewidth,trim={1.5cm 0.55cm 3.95cm 2.4cm},clip=true]{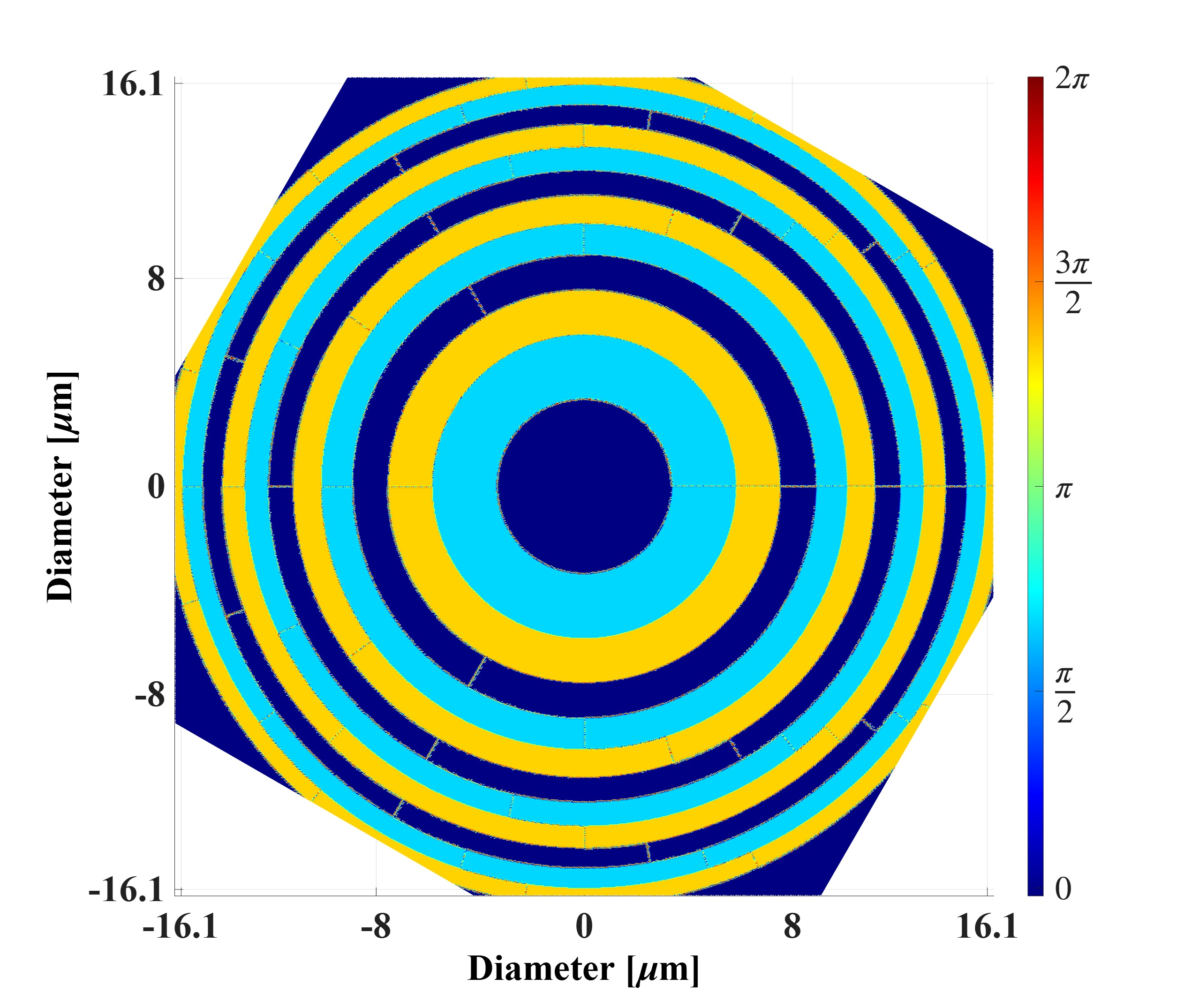}}
        \end{center}
        \vspace*{-5mm}
        \caption{Phase distribution of the Moiré system in Cartesian coordinates when the rotation angles for Disk $II$ are set to: (a) 30$^\circ$, (b) 60$^\circ$, (c) 90$^\circ$,  and (d) 120$^\circ$.}
        \label{Fig: Fig04}
    \end{figure}

    The phase profile performance of the Moiré-PB system is evaluated by rotating Disk $II$ through a series of discrete angles. The corresponding phase distribution and analytical focal response are examined in Fig.~\ref{Fig: Fig04} and Table~\ref{Tab: Table01}. At intermediate twist angles, namely $[30^\circ, 60^\circ, 90^\circ, 120^\circ]$, the relative rotation between the two diffractive disks reshapes the interference pattern. This reshaping produces concentric phase rings with progressively varying gradients. Figure~\ref{Fig: Fig04} demonstrates continuous modulation of the generated phase profile across the aperture (see Supplementary Material Section 1, Fig.~2). These engineered phase distributions yield a tunable focal length that decreases approximately monotonically. The focal length goes from an effectively infinite value at $30^\circ$ to shorter focal distances as the rotation angle increases. This behavior is consistent with the general dependence of the Moiré lens optical power on the twist angle between the disks.
    
    \begin{table}[!htbp]
        \centering
        \caption{Analytically predicted focal response ($f_{\rm A.}$) as a function of the rotation angle of Moiré-PB Disk $II$.}
        \bigskip
        \setlength{\tabcolsep}{3pt}
        \begin{tabular}{c|c|c|c|c}
            \toprule
            \toprule
            \rowcolor{Gray}
             Rotation angle [$^\circ$] & 30 & 60 & 90 & 120 \\
            \hline
             $f_{\rm A.}$ [$\mu\text{m}$] & 68.18 & 34.09 & 22.73 & 17.05 \\
            \toprule
            \toprule
        \end{tabular}
        \begin{minipage}{7.5cm}
        \vspace{0.05cm}
        \end{minipage}
        \label{Tab: Table01}
    \end{table}
    
    A metric for evaluating the proposed metalens is the Numerical Aperture (\textit{NA}). This is a wavelength-dependent dimensionless number that quantifies the angular range over which a metalens can accept or emit electromagnetic waves. Given the metalens dimensions and its focal length tunability ranging from $68.18$ to $17.05~\mu\text{m}$, the \textit{NA} is computed to range from 22.99\% to 68.67\% using Eq.~\ref{Eq: Eq20}:
    \begin{equation}
        \textit{NA}
        =n\displaystyle\frac{\frac{D}{2}}{\sqrt{\left(\frac{D}{2}\right)^{\rm 2}+f^{\rm 2}}},
        \label{Eq: Eq20}
    \end{equation}
    where $n$ is the refractive index of the medium in which the metalens operates, with $n=1.00027304$.
    \begin{figure}
        \begin{center}
            \subfigure[]{\label{Fig: Fig05.a}
            \includegraphics[height=0.4\linewidth,trim={0.0cm 0.0cm 0.0cm 0.0cm},clip=true]{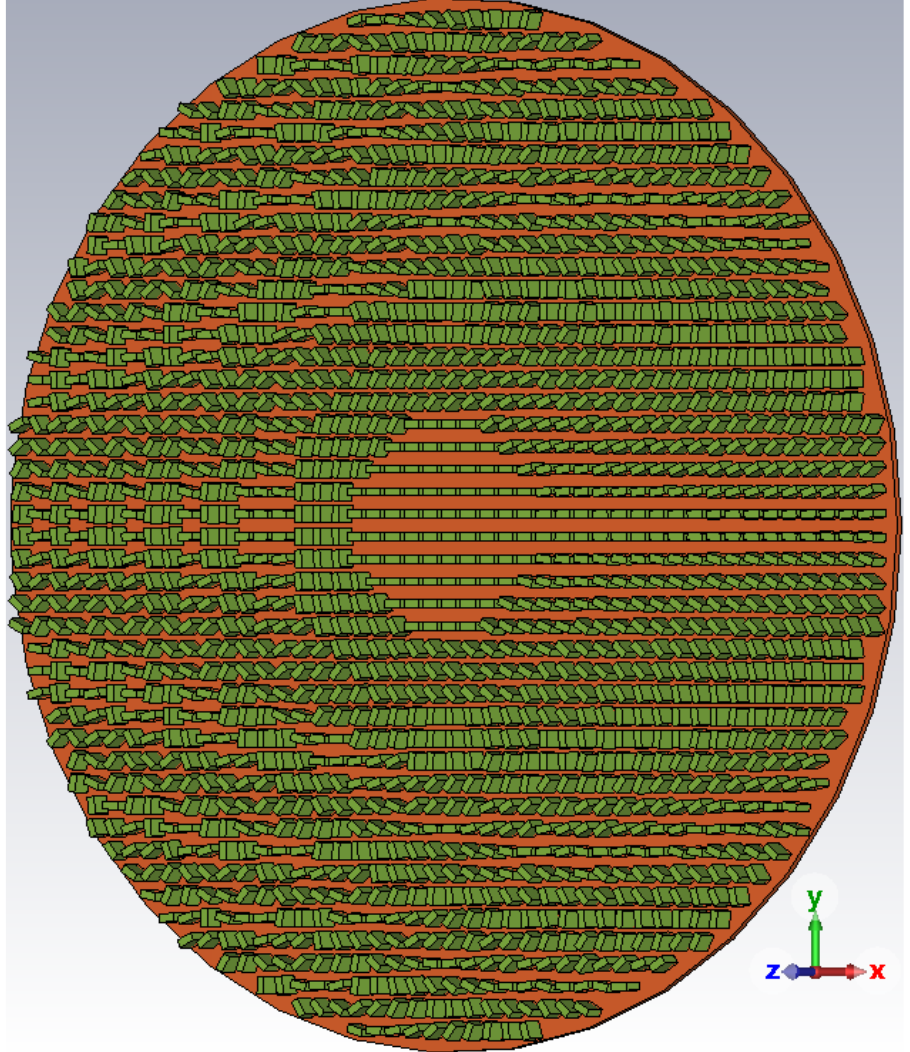}}
            \subfigure[]{\label{Fig: Fig05.b}
            \includegraphics[height=0.4\linewidth,trim={0.0 0.0cm 0.0cm 0.0cm},clip=true]{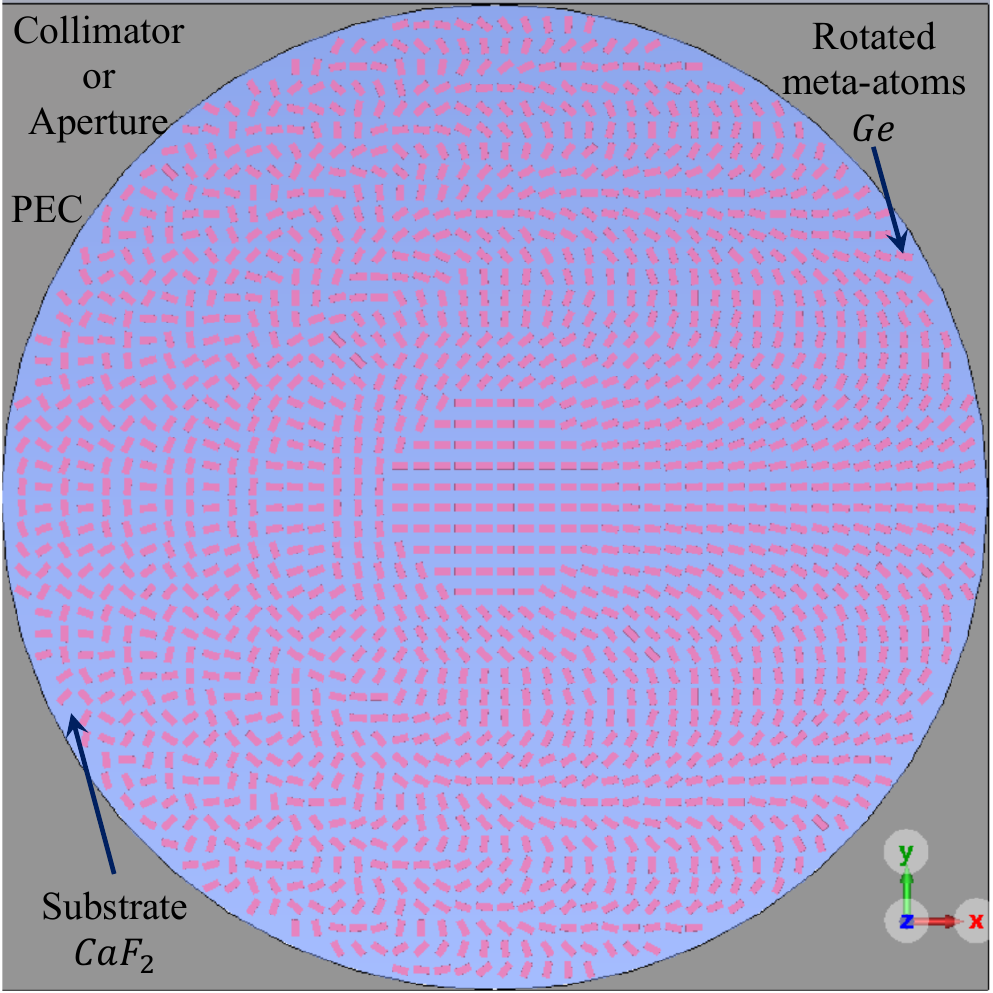}}

            \vspace*{-2mm}
            \subfigure[]{\label{Fig: Fig05.c}
            \includegraphics[height=0.07\linewidth,trim={0.0cm 0.0cm 0.0cm 0.0cm},clip=true]{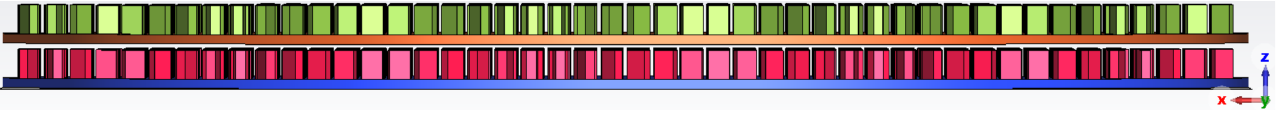}}
        \end{center}
        \vspace*{-5mm}
        \caption{Constructed Moiré-PB metalens configuration: (a) 3D view of the upper disk (Disk $I$), (b) Top-view of the bottom disk (Disk $II$) in companion with collimator/aperture, and (c) Side view of the structure.}
        \label{Fig: Fig05}
    \end{figure}
    
    The Fresnel number ($F_{N}$) is another dimensionless parameter that determines whether wave propagation is in the near-field (Fresnel) or far-field (Fraunhofer) regime. Specifically, {$F_{N}\ll 1$} corresponds to the Fraunhofer (far-field) diffraction regime, {$F_{N}\approx 1$} represents the transition region between near- and far-field diffraction, ${F_{N}>1}$ indicates the Fresnel (near-field) diffraction regime, and $F_{N}\gg 1$ corresponds to strong near-field focusing, where the near-field behavior of the optical wave dominates diffraction effects. Given the metalens operating wavelength, dimensions, and its focal length tunability, the $F_{N}$ is computed to range from 1.90 to 7.60 employing Eq.~\ref{Eq: Eq21}:
    \begin{equation}
        F_{N}=\frac{D^{\rm 2}}{4\lambda f}.
        \label{Eq: Eq21}
    \end{equation}
\section{Full-wave Simulation Results and Discussion}
    Figure~\ref{Fig: Fig05} summarizes the full-wave simulation model used to investigate the proposed Moiré-PB metalens. The model highlights both the in-plane phase engineering and the out-of-plane structural layout of the device. The 3D and top views of Disk $II$ and Disk $I$ show the spatial distribution and local rotation of the PB meta-atoms over the {\tt CaF$_{\rm 2}$} substrates. The construction illustrates how the virtually complementary patterns on the two disks jointly synthesize the required Moiré phase profile for tunable focusing. These complementary patterns arise from the phase-sign inversion introduced by one of the disks, due to its half-wave-plate properties. The side view further reveals the multilayer stack and the spacing between the patterned {\tt Ge} meta-atoms. Additionally, Fig.~\ref{Fig: Fig05.b} depicts the collimator, which defines the illumination aperture and is modeled as a Perfect Electric Conductor (PEC).
    
\begin{figure*}[ht!]
        \begin{center}
            \subfigure[]{\label{Fig: Fig06.a}
            \includegraphics[height=0.23\linewidth,trim={0.0cm 0.0cm 0.0cm 0.0cm},clip=true]{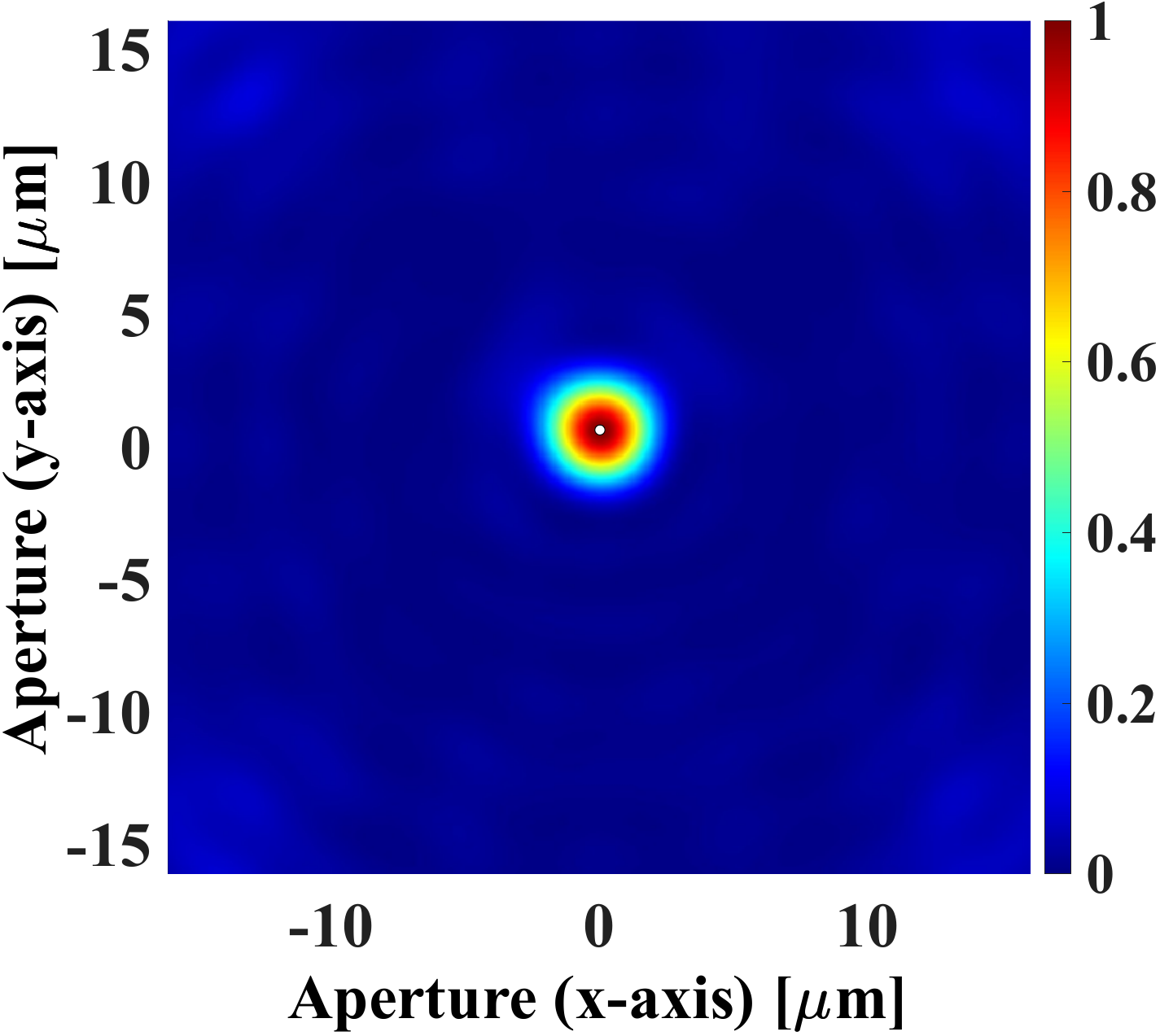}}
            \subfigure[]{\label{Fig: Fig06.b}
            \includegraphics[height=0.23\linewidth,trim={0.0 0.0cm 0.0cm 0.0cm},clip=true]{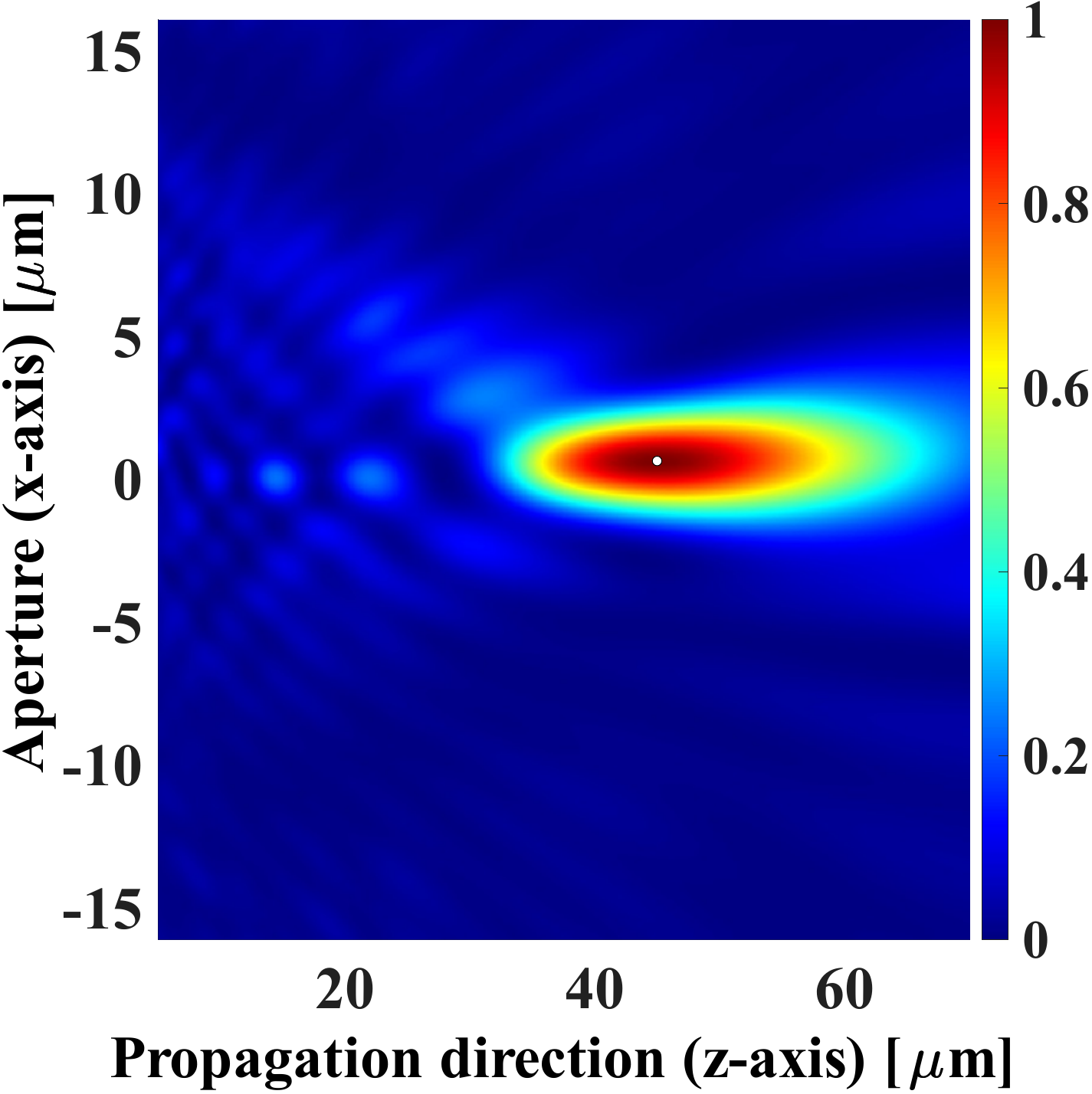}}
            \subfigure[]{\label{Fig: Fig06.c}
            \includegraphics[height=0.23\linewidth,trim={0.0cm 0.0cm 0.0cm 0.0cm},clip=true]{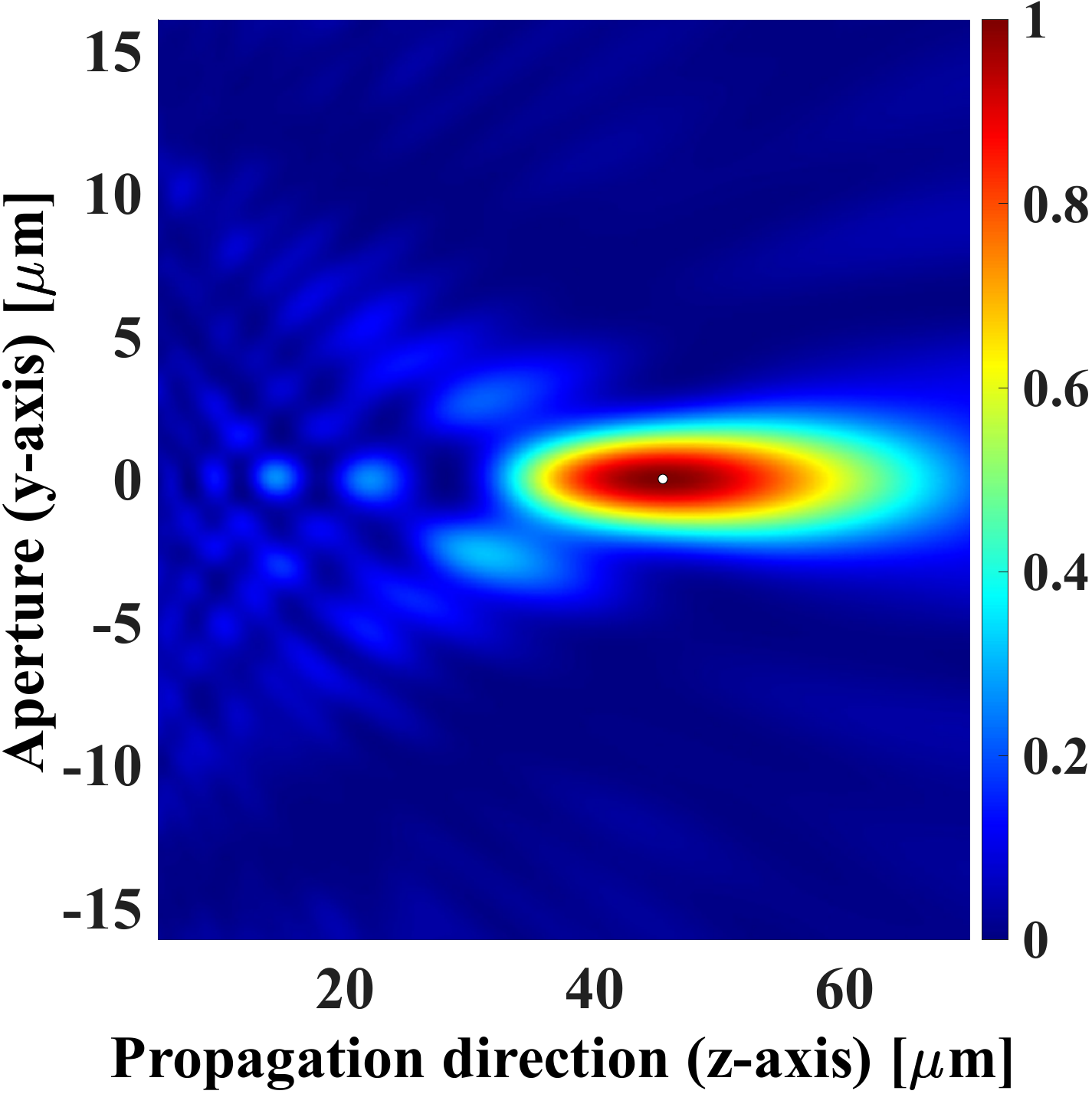}}
            \vspace*{-2mm}
            \subfigure[]{\label{Fig: Fig06.d}
            \includegraphics[height=0.23\linewidth,trim={0.0cm 0.0cm 0.0cm 0.0cm},clip=true]{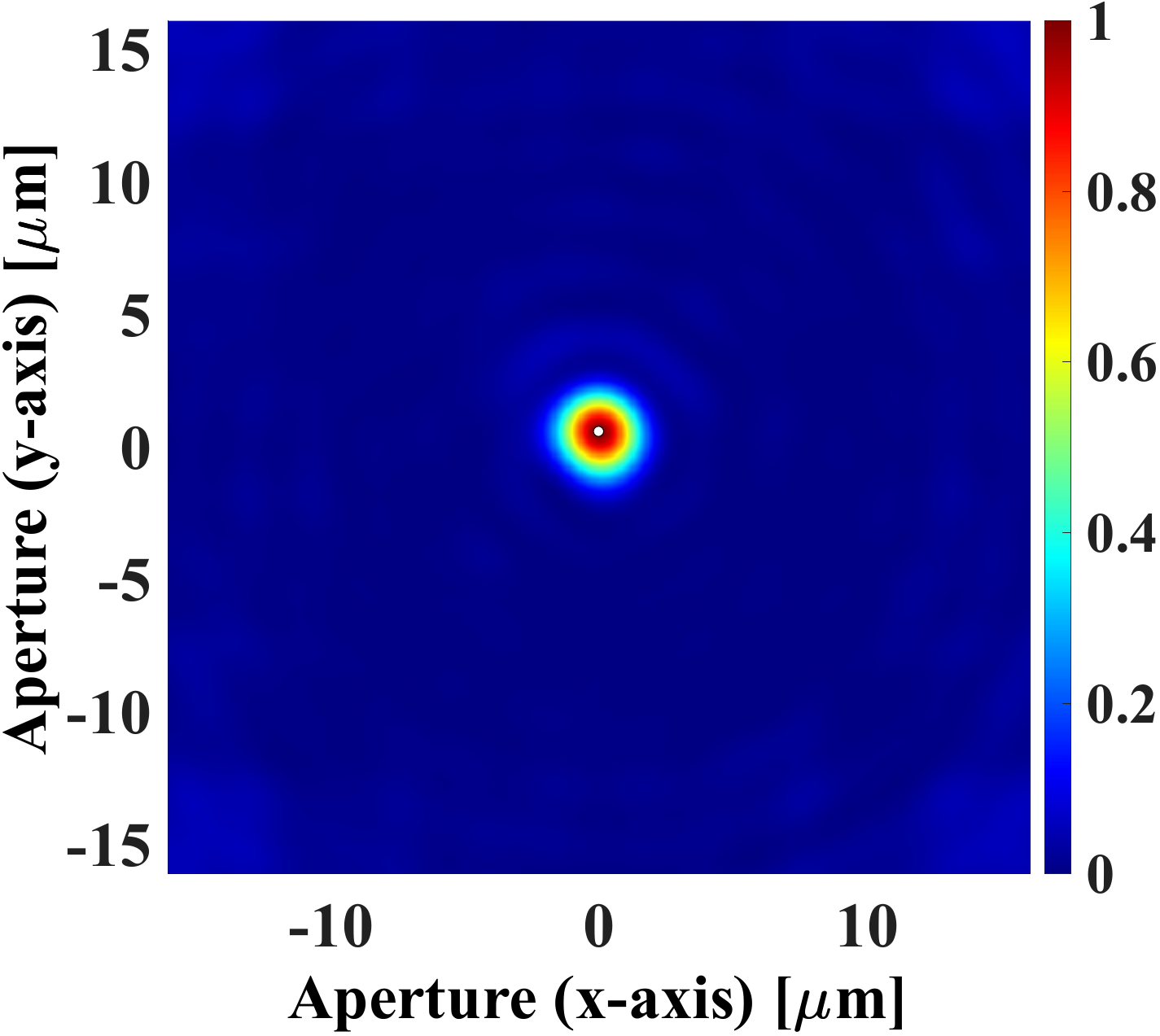}}
            \subfigure[]{\label{Fig: Fig06.e}
            \includegraphics[height=0.23\linewidth,trim={0.0cm 0.0cm 0.0cm 0.0cm},clip=true]{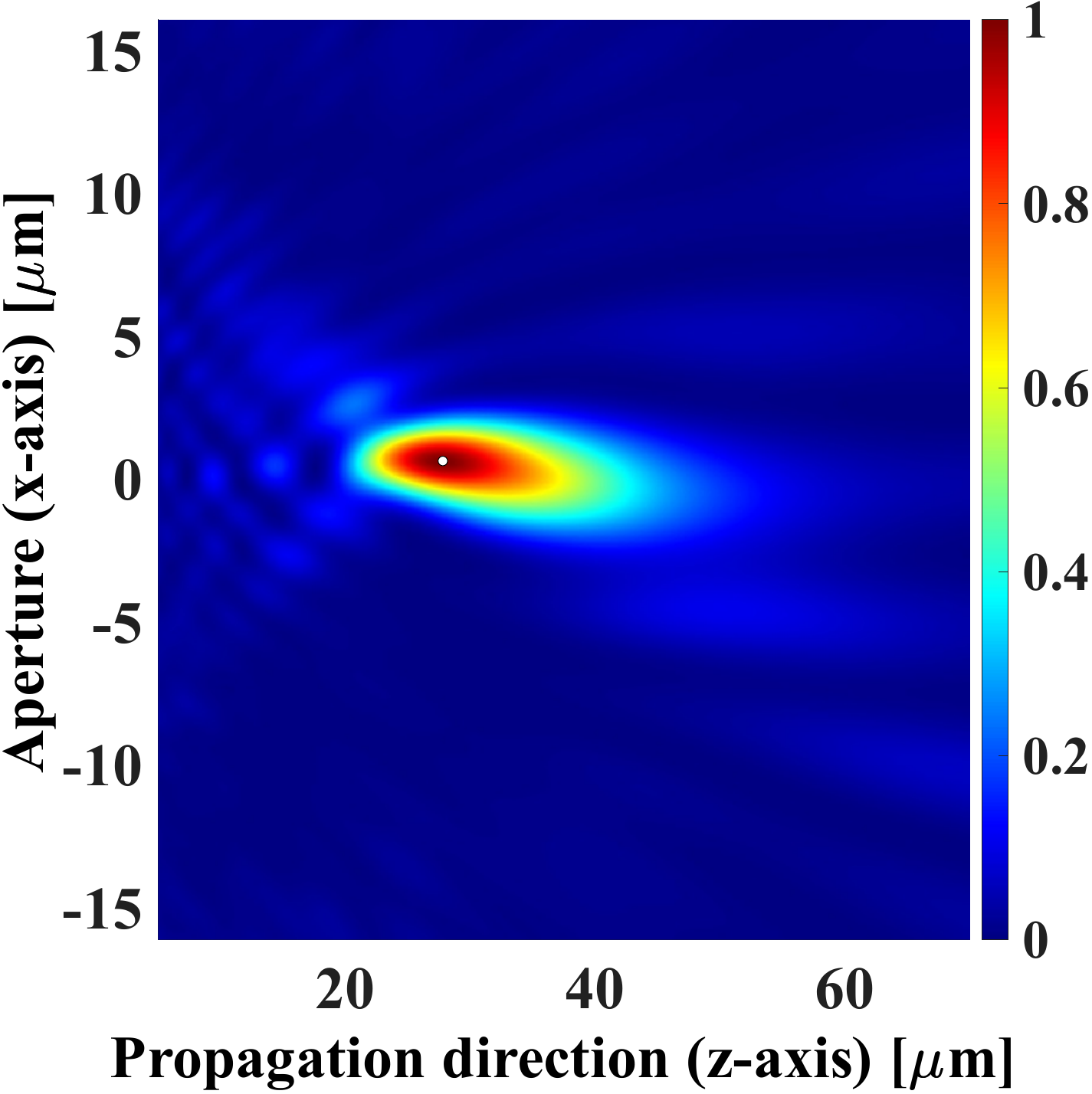}}
            \subfigure[]{\label{Fig: Fig06.f}
            \includegraphics[height=0.23\linewidth,trim={0.0cm 0.0cm 0.0cm 0.0cm},clip=true]{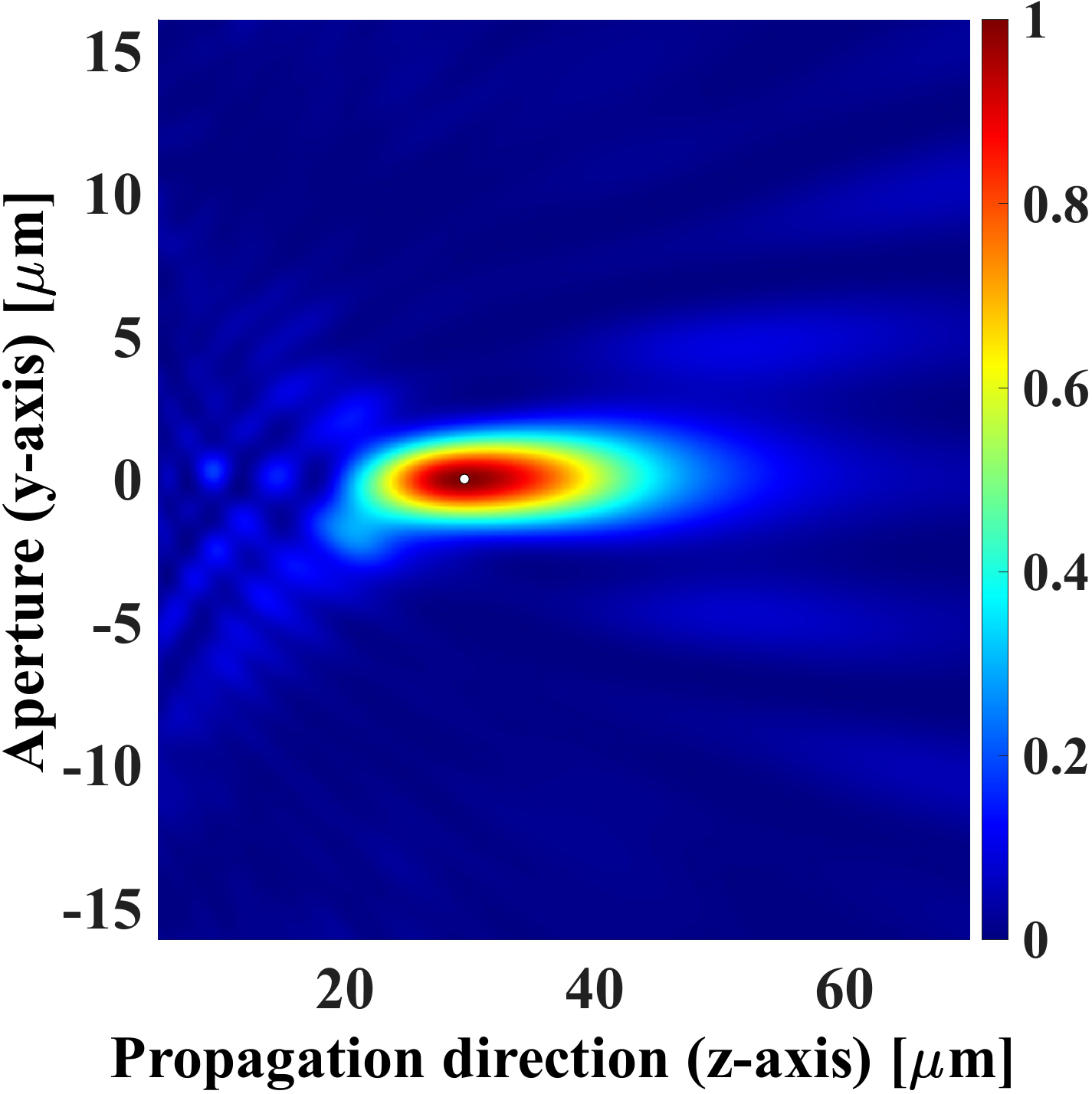}}
            \vspace*{-2mm}
            \subfigure[]{\label{Fig: Fig06.g}
            \includegraphics[height=0.23\linewidth,trim={0.0cm 0.0cm 0.0cm 0.0cm},clip=true]{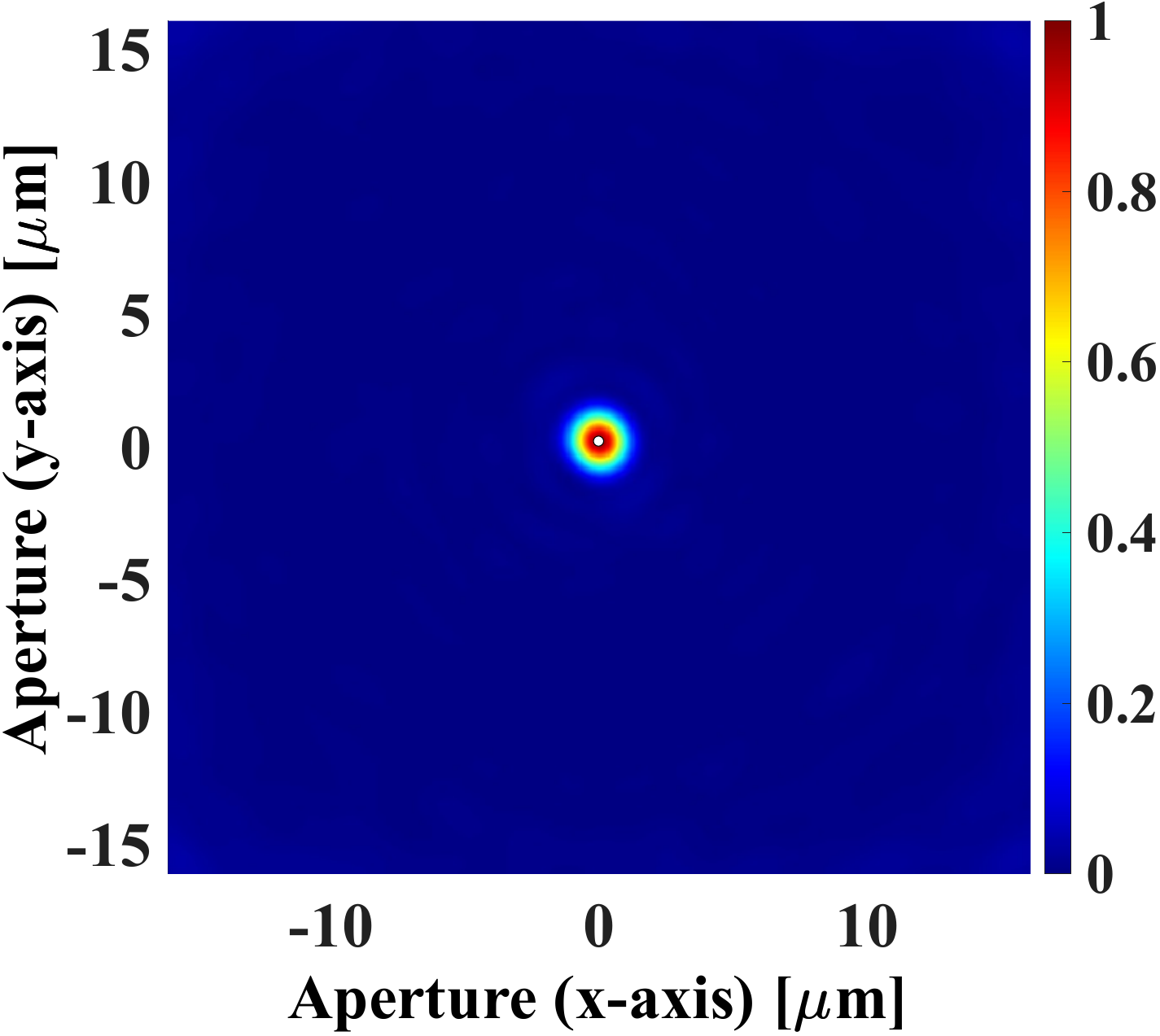}}
            \subfigure[]{\label{Fig: Fig06.h}
            \includegraphics[height=0.23\linewidth,trim={0.0cm 0.0cm 0.0cm 0.0cm},clip=true]{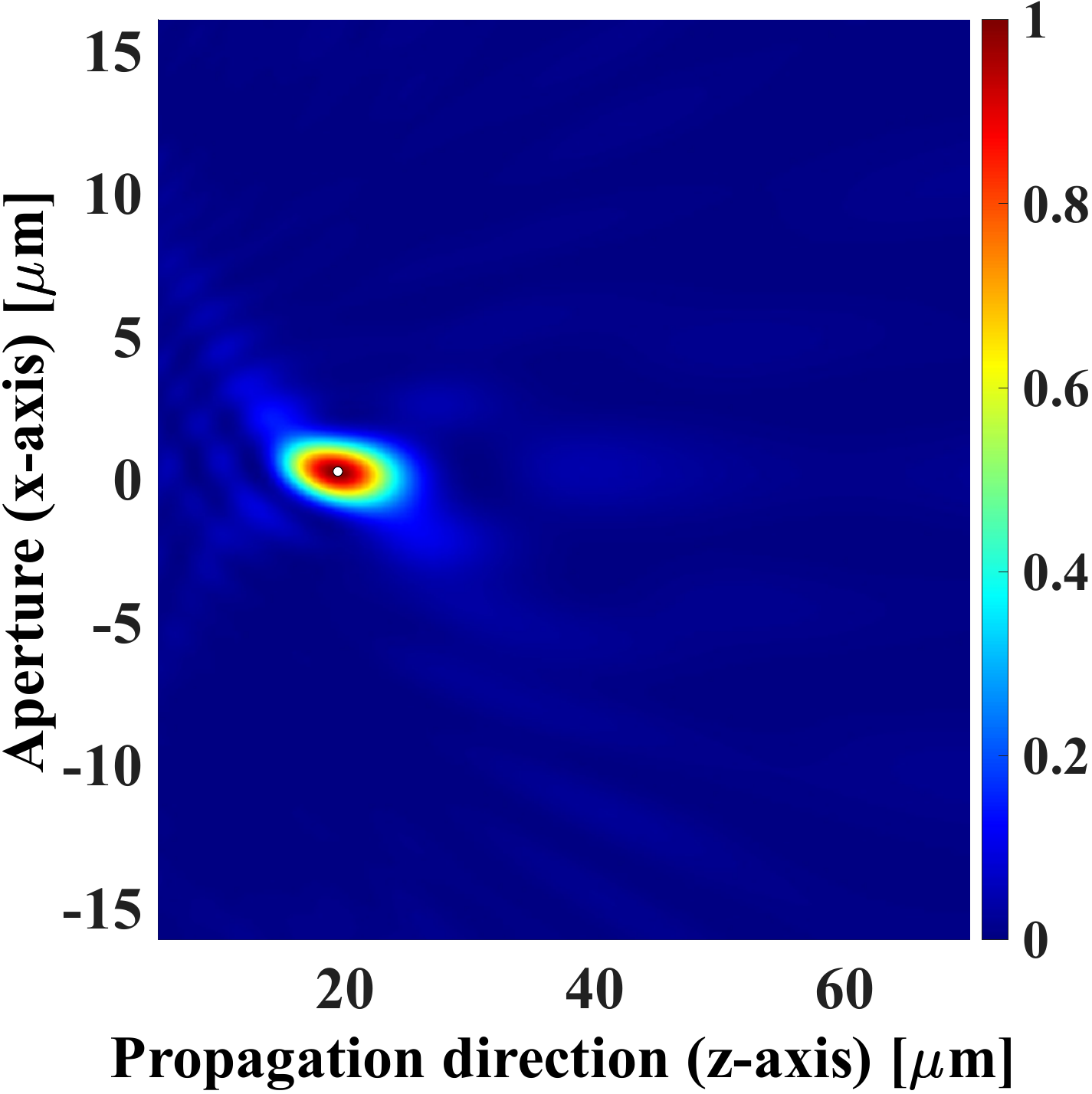}}
            \subfigure[]{\label{Fig: Fig06.i}
            \includegraphics[height=0.23\linewidth,trim={0.0cm 0.0cm 0.0cm 0.0cm},clip=true]{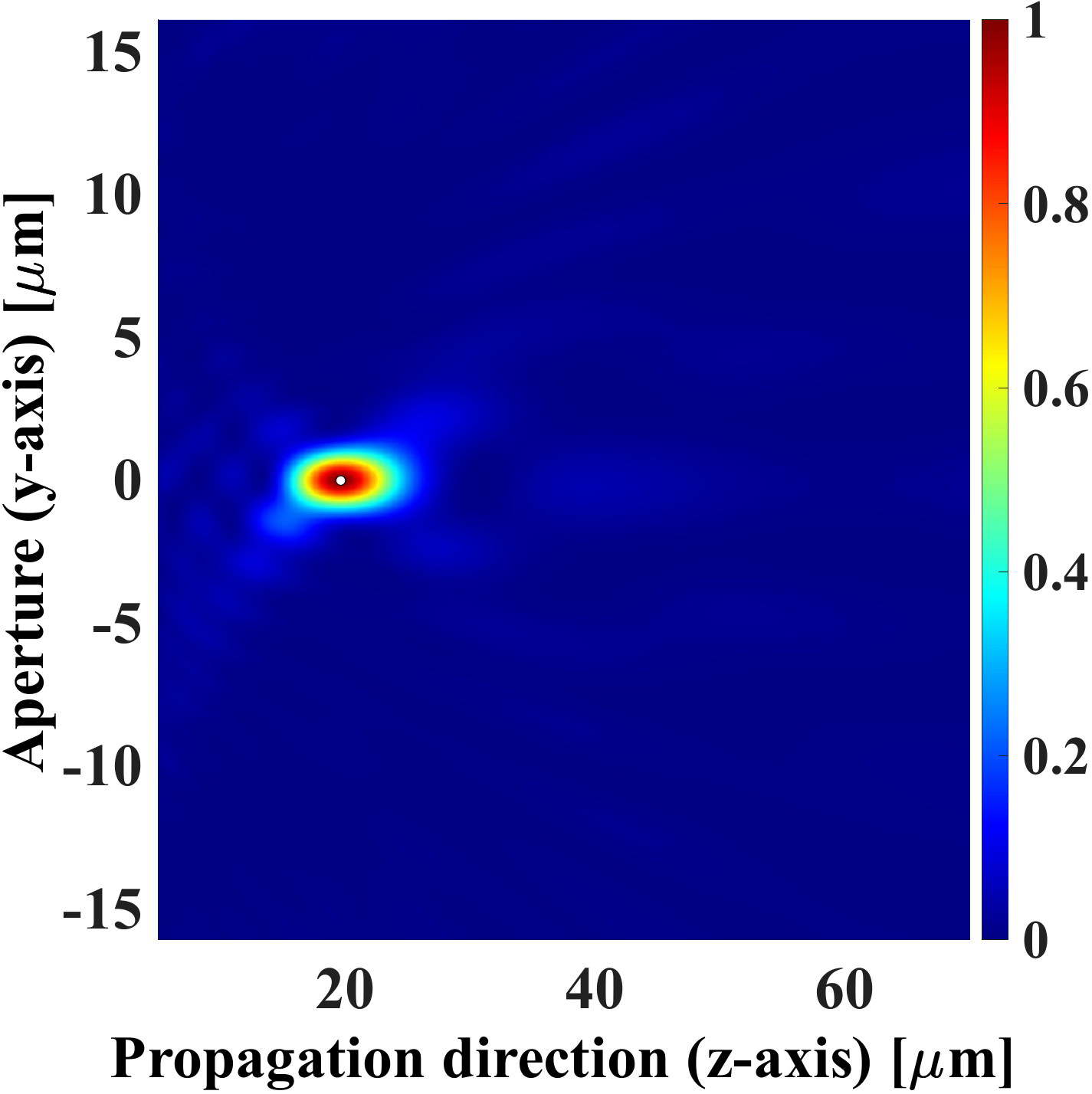}}
            \vspace*{-2mm}
            \subfigure[]{\label{Fig: Fig06.j}
            \includegraphics[height=0.23\linewidth,trim={0.0cm 0.0cm 0.0cm 0.0cm},clip=true]{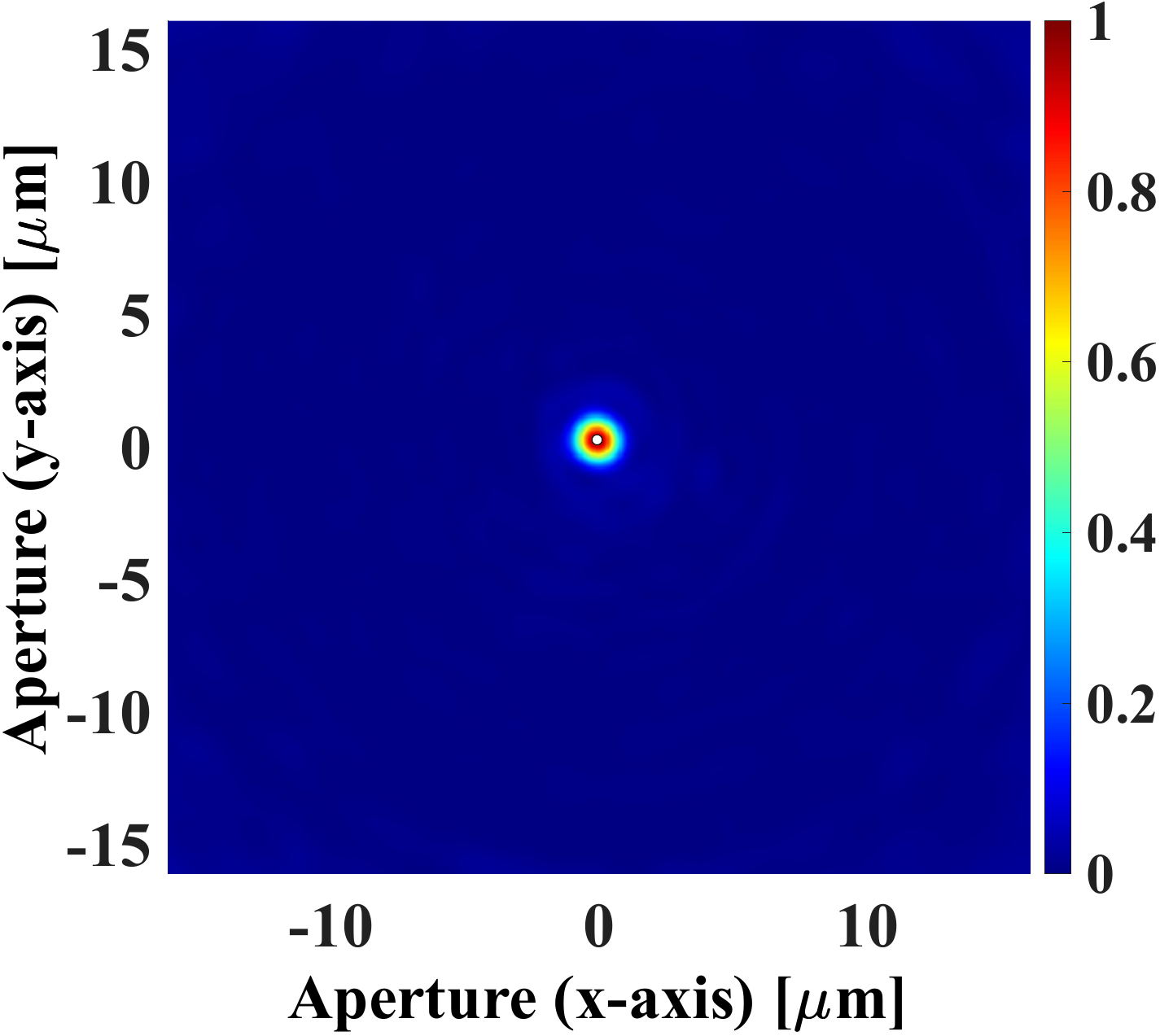}}
            \subfigure[]{\label{Fig: Fig06.k}
            \includegraphics[height=0.23\linewidth,trim={0.0cm 0.0cm 0.0cm 0.0cm},clip=true]{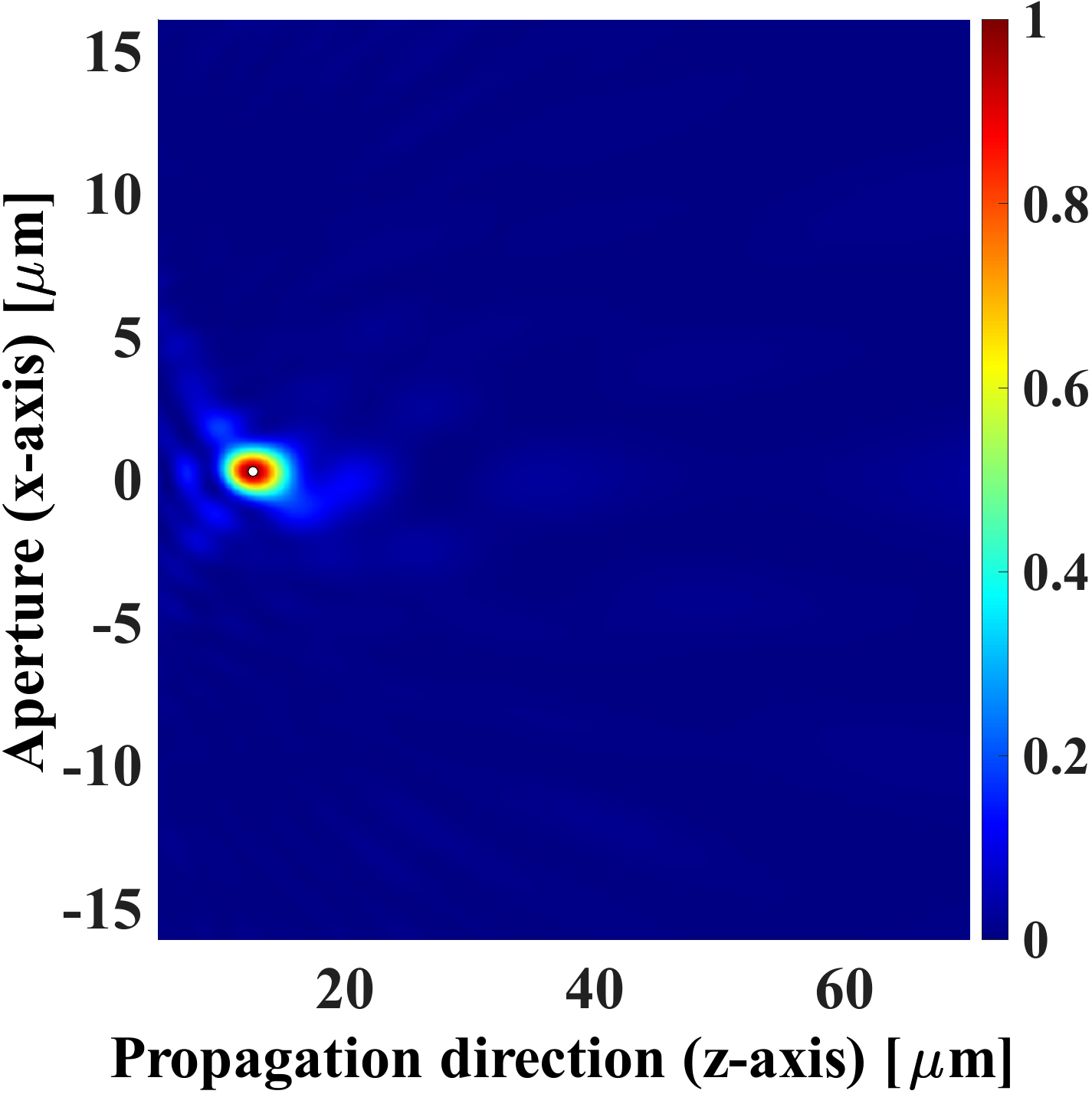}}
            \subfigure[]{\label{Fig: Fig06.l}
            \includegraphics[height=0.23\linewidth,trim={0.0cm 0.0cm 0.0cm 0.0cm},clip=true]{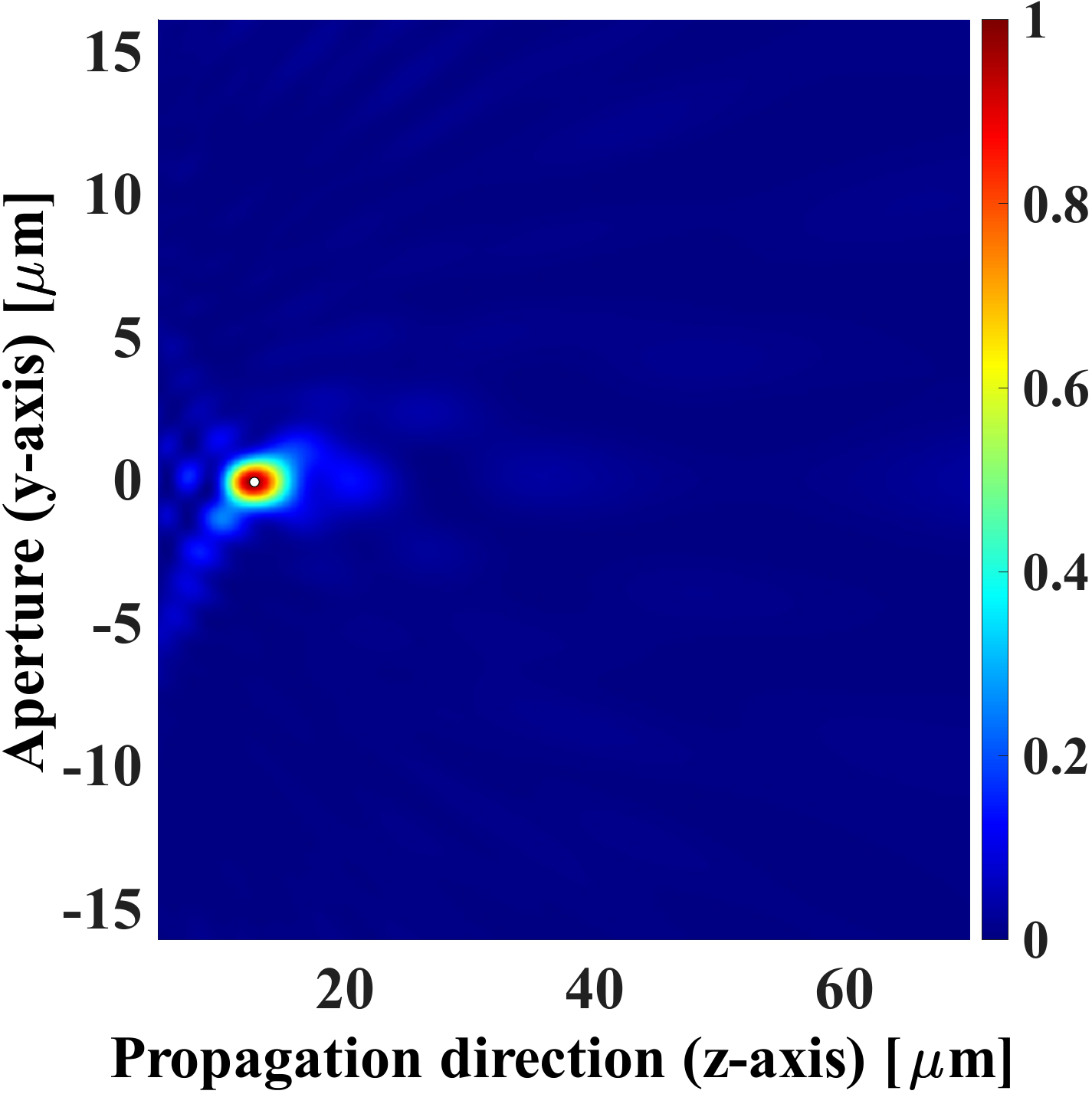}}
        \end{center}
        \vspace*{-5mm}
        \caption{Normalized $E$-field distributions in the $xy$-, $zx$-, and $yz$-plane under RCP illumination with CW rotation. The rotation angles of (a-c) 30$^\circ$, (d-f) 60$^\circ$, (g-i) 90$^\circ$, and (j-l) 120$^\circ$.}
        \label{Fig: Fig06}
    \end{figure*}

    The focusing operation of the proposed Moiré-PB metalens depends on the incident light’s polarization state, as was discussed earlier. Figure~\ref{Fig: Fig06} illustrates the focusing performance under RCP illumination. To this end, the field distributions in the $xy$-, $zx$-, and $yz$-planes are presented. Specifically, Figs.~\ref{Fig: Fig06.a}-\ref{Fig: Fig06.c},~\ref{Fig: Fig06.d}-\ref{Fig: Fig06.f},~\ref{Fig: Fig06.g}-\ref{Fig: Fig06.i}, and ~\ref{Fig: Fig06.j}-\ref{Fig: Fig06.l} depict the focusing states when the second disk is rotated CW by $30^\circ$, $60^\circ$, $90^\circ$, and $120^\circ$, respectively. The $xy$-plane slices are taken at the axial position where the $E$-field intensity along the propagation direction reaches its maximum. The results show that as the rotation angle of the second disk decreases, the focal spots become increasingly diffuse and lose sharpness (i.e., become blurred). This degradation occurs because the lens aperture size relative to the operating wavelength is insufficient at smaller rotation angles of Disk $II$. Consequently, diffraction effects become more apparent, thereby blurring the focus.
    
\begin{figure*}[htb]
        \begin{center}
            \subfigure[]{\label{Fig: Fig07.a}
            \includegraphics[height=0.3\linewidth,trim={0.0cm 0.0cm 0.0cm 0.0cm},clip=true]{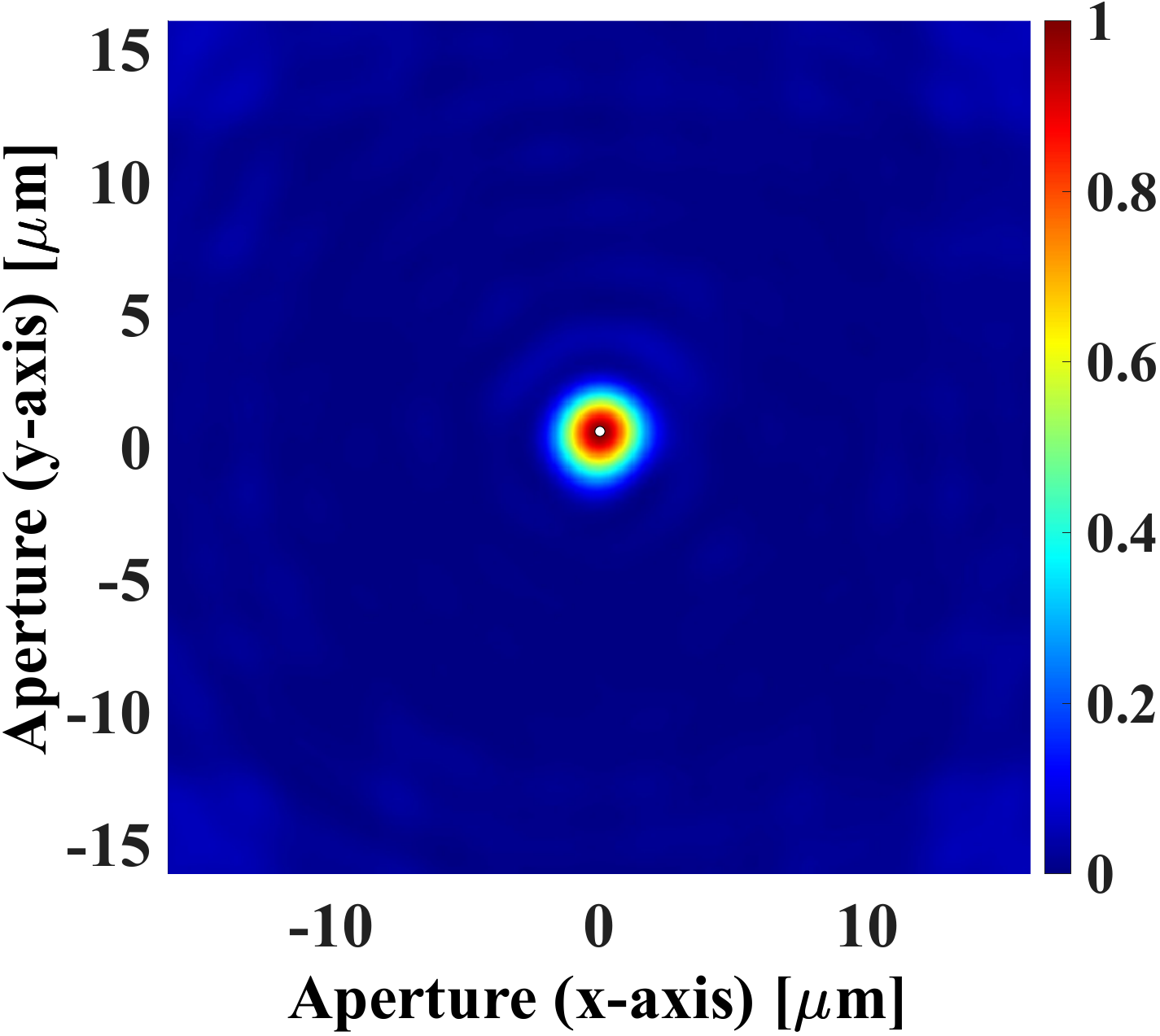}}
            \subfigure[]{\label{Fig: Fig07.b}
            \includegraphics[height=0.3\linewidth,trim={0.0 0.0cm 0.0cm 0.0cm},clip=true]{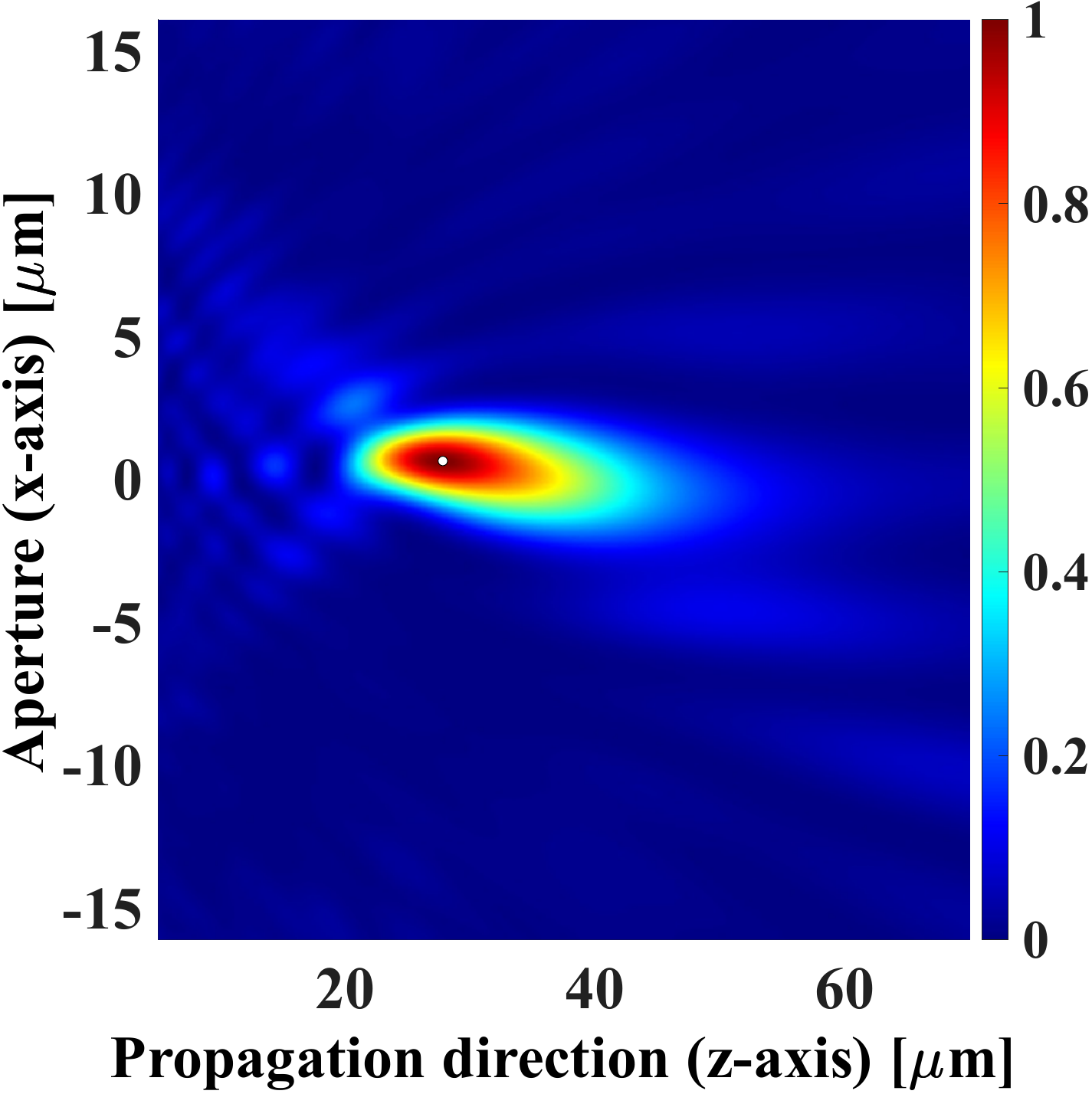}}
            \subfigure[]{\label{Fig: Fig07.c}
            \includegraphics[height=0.3\linewidth,trim={0.0cm 0.0cm 0.0cm 0.0cm},clip=true]{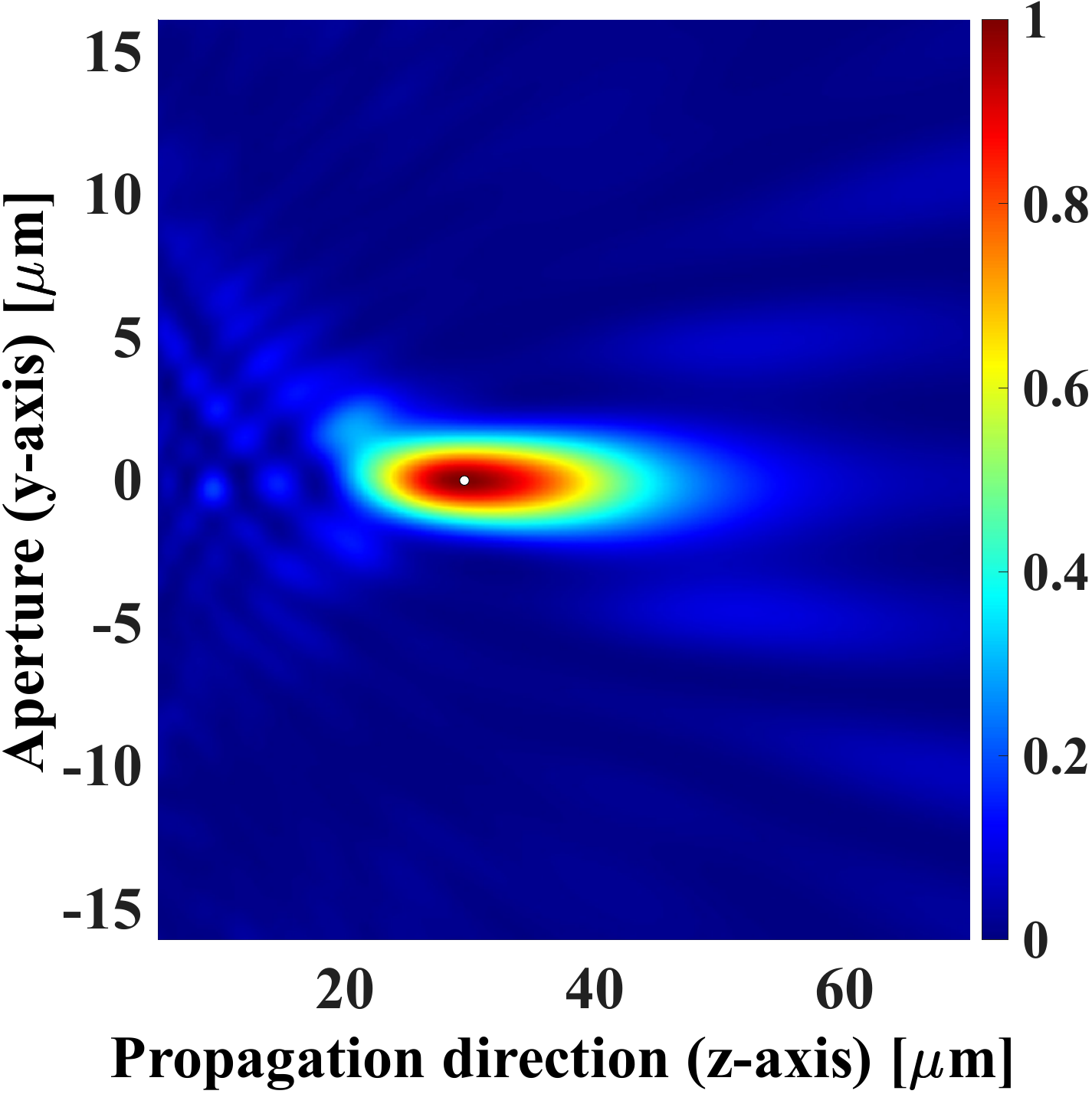}}
        \end{center}
        \vspace*{-5mm}
        \caption{Normalized $E$-field distributions for rotation angles of 60$^\circ$ in the $xy$-, $zx$-, and $yz$-plane under LCP illumination with CCW rotation.}
        \label{Fig: Fig07}
    \end{figure*}

    As a proof of concept, a $60^\circ$ rotation angle is selected for the second disk under both LCP and LP excitations. This choice demonstrates the metalens's focusing performance. Figure~\ref{Fig: Fig07} presents the results for LCP illumination with a CCW rotation of Disk $II$. The observed focusing behavior is identical to that obtained for the 60$^\circ$ CW rotation under RCP excitation, confirming the expected symmetry.
    \begin{table*}[htbp]
        \centering
        \caption{Comparison between the analytically predicted ($f_{\rm A.}$), numerically retrieved ($f_{\rm N.}$) focal lengths, and analytical $\textit{NA}_{\rm A.}$ and $F_{\rm N_{A.}}$ as a function of the rotation angle of Moiré-PB Disk $II$.}
        \vspace*{2mm}
        \setlength{\tabcolsep}{3pt}
        \begin{tabular}{c|c|c|c|c|c|c|c|c|c}
            \toprule
            \toprule
            \rowcolor{Gray}
             Rotation angle [$^\circ$] & 30 & 60 & 60 & 60 & 60 & 60 & 60 & 90 & 120 \\
            \hline
             Disk $II$ rotation direction & CW & CW & CCW & CW & CCW & CW & CCW & CW & CW \\
            \hline
            \rowcolor{Gray}
             Polarization & RCP & RCP & LCP & x-LP & x-LP & y-LP & y-LP & RCP & RCP \\
             \hline
             $f_{\rm A.}$ [$\mu\text{m}$] & 68.18 & 34.09 & 34.09 & 34.09 & 34.09 & 34.09 & 34.09 & 22.73 & 17.05 \\
            \hline
            \rowcolor{Gray}
             $f_{\rm N.}$ [$\mu\text{m}$] & 45.41 & 29.52 & 29.52 & 29.41 & 29.41 & 29.52 & 29.52 & 19.62 & 12.71 \\
            \hline
             $\textit{NA}_{\rm A.}$ [\%] & 22.99 & 42.72 & 42.72 & 2.72 & 42.72 & 2.72 & 42.72 & 57.82 & 68.67 \\
            \hline
            \rowcolor{Gray}
             $F_{\rm N_{A.}}$ & 1.90 & 3.80 & 3.80 & 3.80 & 3.80 & 3.80 & 3.80 & 5.70 & 7.60 \\
            \hline
             $\Delta f$ [\%] & 33.41 & 13.41 & 13.41 & 13.75 & 13.75 & 13.41 & 13.41 & 13.67 & 25.42 \\
            \toprule
            \toprule
        \end{tabular}
        \begin{minipage}{7.5cm}
        \vspace{0.05cm}
        \end{minipage}
        \label{Tab: Table02}
    \end{table*}
    
\begin{figure*}
        \begin{center}
            \subfigure[]{\label{Fig: Fig08.a}
            \includegraphics[height=0.3\linewidth,trim={0.0cm 0.0cm 0.0cm 0.0cm},clip=true]{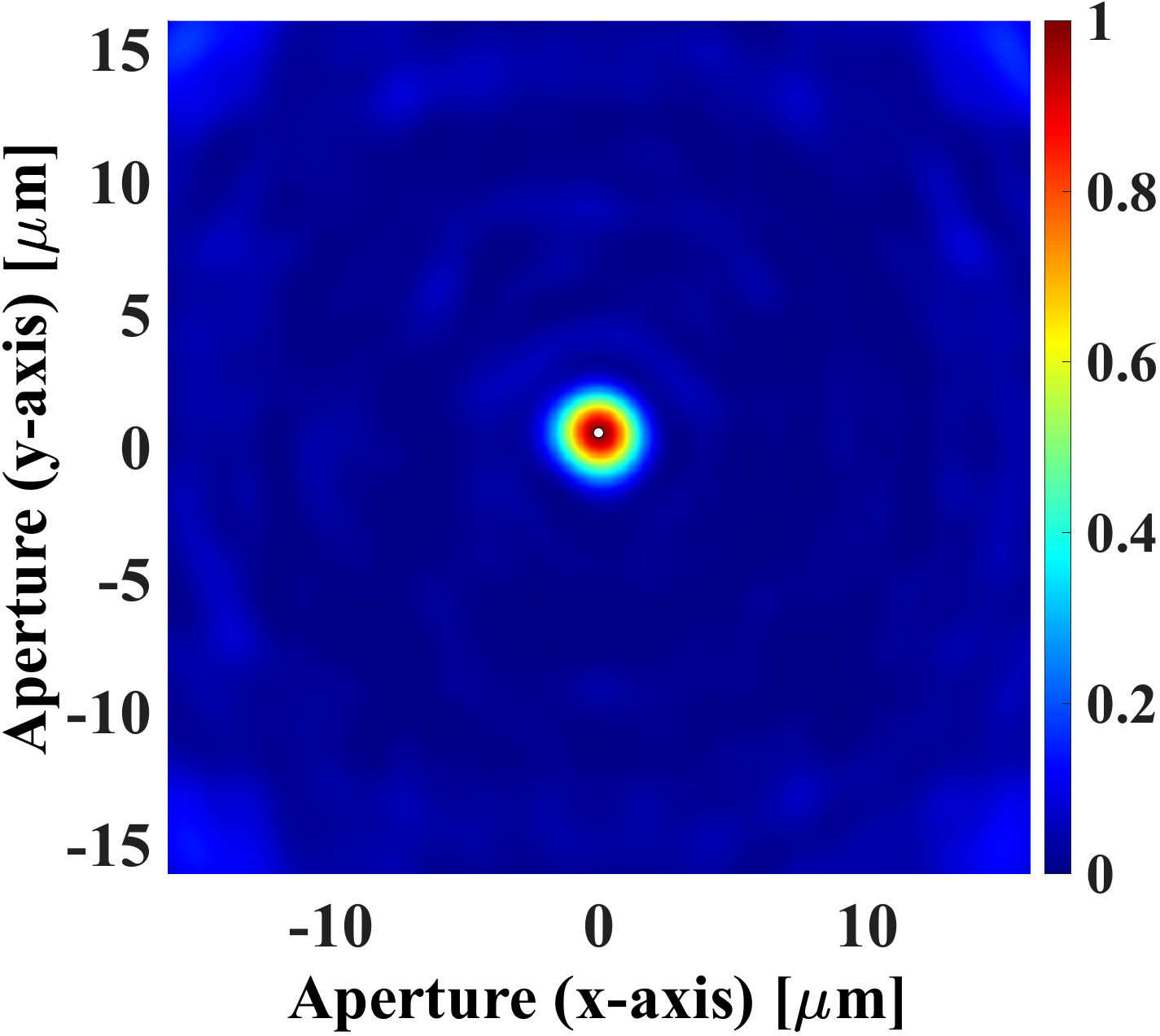}}
            \subfigure[]{\label{Fig: Fig08.b}
            \includegraphics[height=0.3\linewidth,trim={0.0 0.0cm 0.0cm 0.0cm},clip=true]{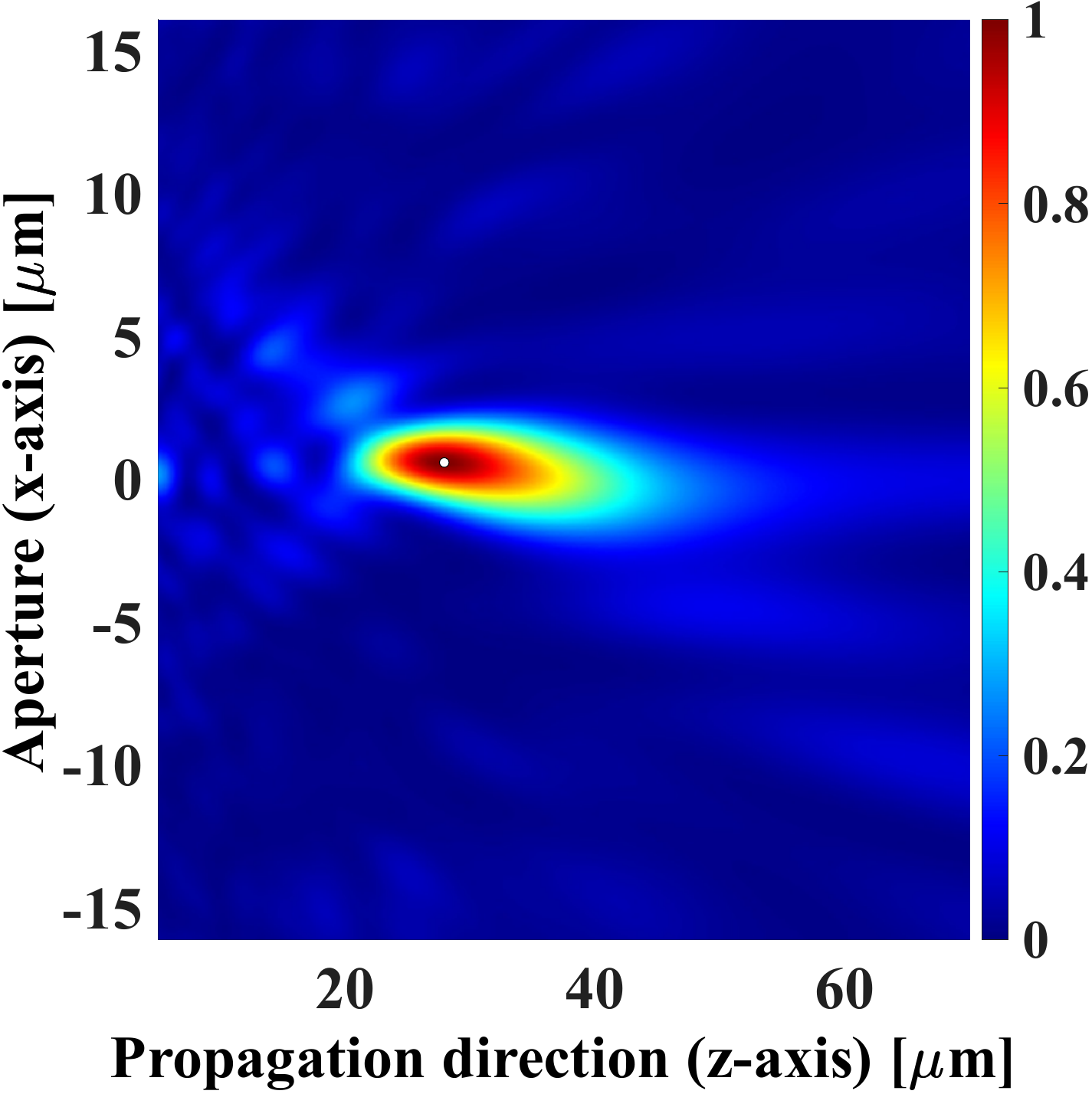}}
            \subfigure[]{\label{Fig: Fig08.c}
            \includegraphics[height=0.3\linewidth,trim={0.0cm 0.0cm 0.0cm 0.0cm},clip=true]{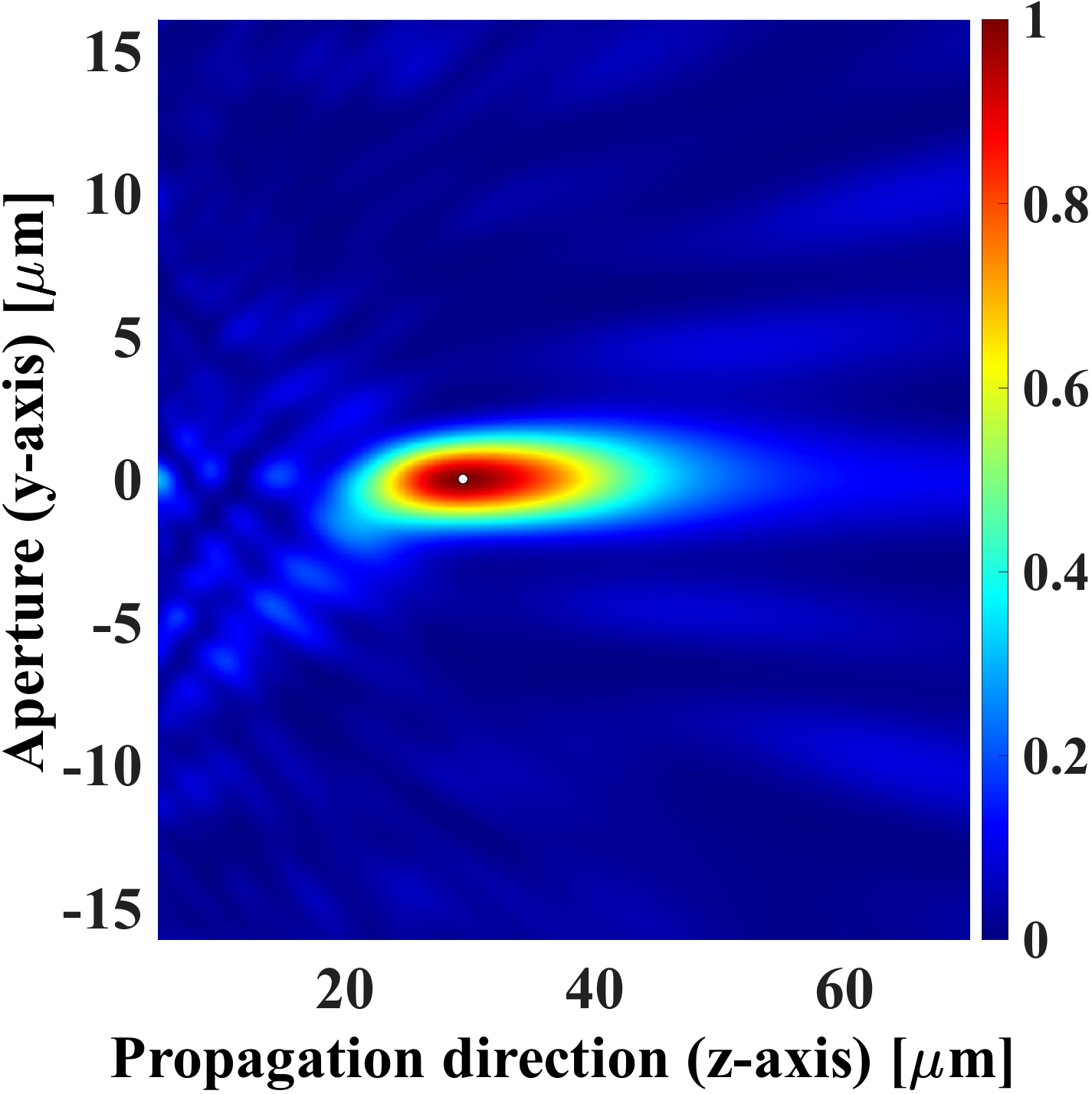}}
            \vspace*{-2mm}
            \subfigure[]{\label{Fig: Fig08.d}
            \includegraphics[height=0.3\linewidth,trim={0.0cm 0.0cm 0.0cm 0.0cm},clip=true]{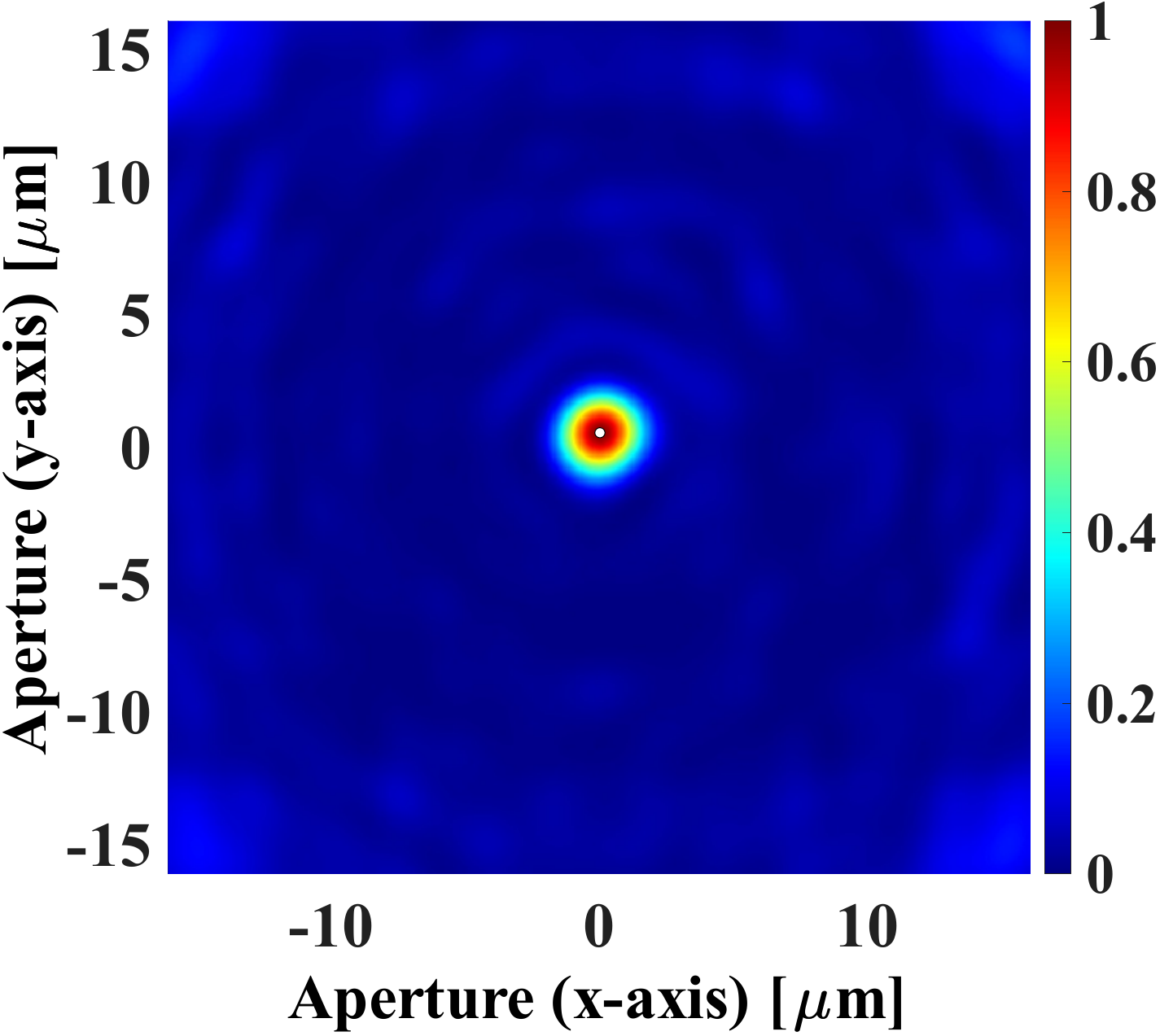}}
            \subfigure[]{\label{Fig: Fig08.e}
            \includegraphics[height=0.3\linewidth,trim={0.0cm 0.0cm 0.0cm 0.0cm},clip=true]{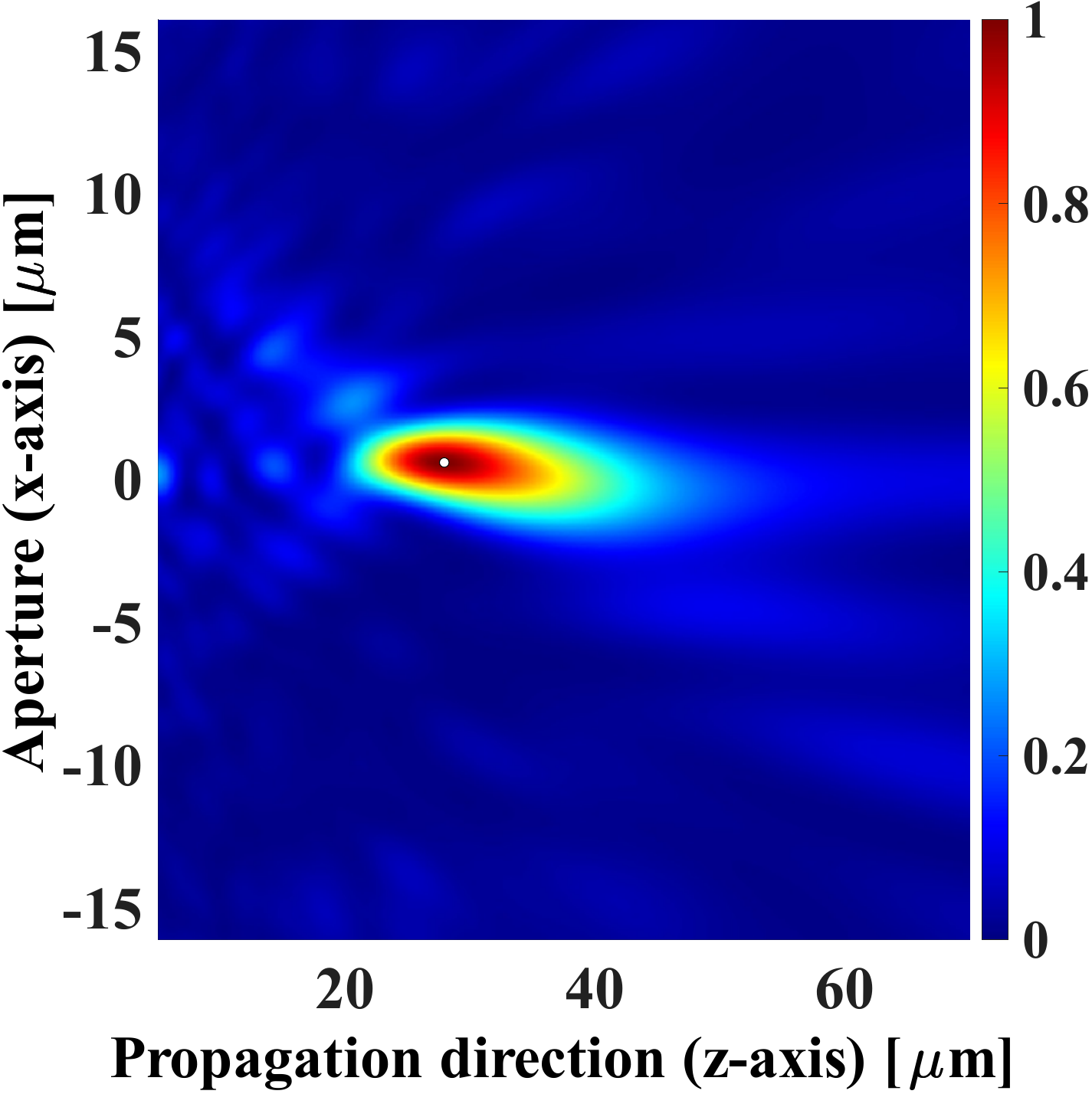}}
            \subfigure[]{\label{Fig: Fig08.f}
            \includegraphics[height=0.3\linewidth,trim={0.0cm 0.0cm 0.0cm 0.0cm},clip=true]{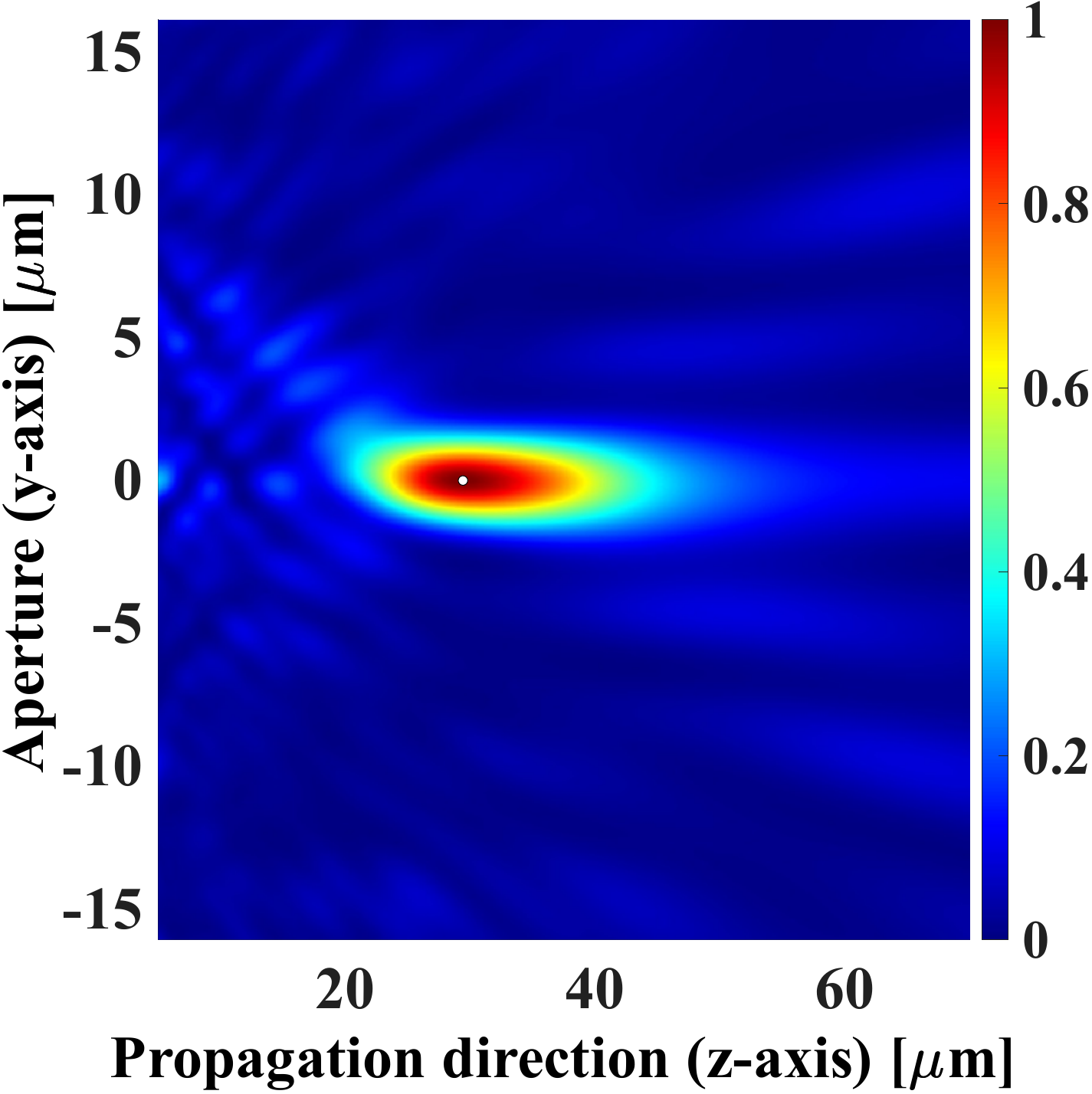}}
        \end{center}
        \vspace*{-5mm}
        \caption{Normalized $E$-field distributions for rotation angles of 60$^\circ$ in the $xy$-, $zx$-, and $yz$-plane under x-LP illumination with CW and CCW rotation. (a-c) CW rotation and (d-f) CWW rotation.}
        \label{Fig: Fig08}
    \end{figure*}

\begin{figure*}
        \begin{center}
            \subfigure[]{\label{Fig: Fig09.a}
            \includegraphics[height=0.3\linewidth,trim={0.0cm 0.0cm 0.0cm 0.0cm},clip=true]{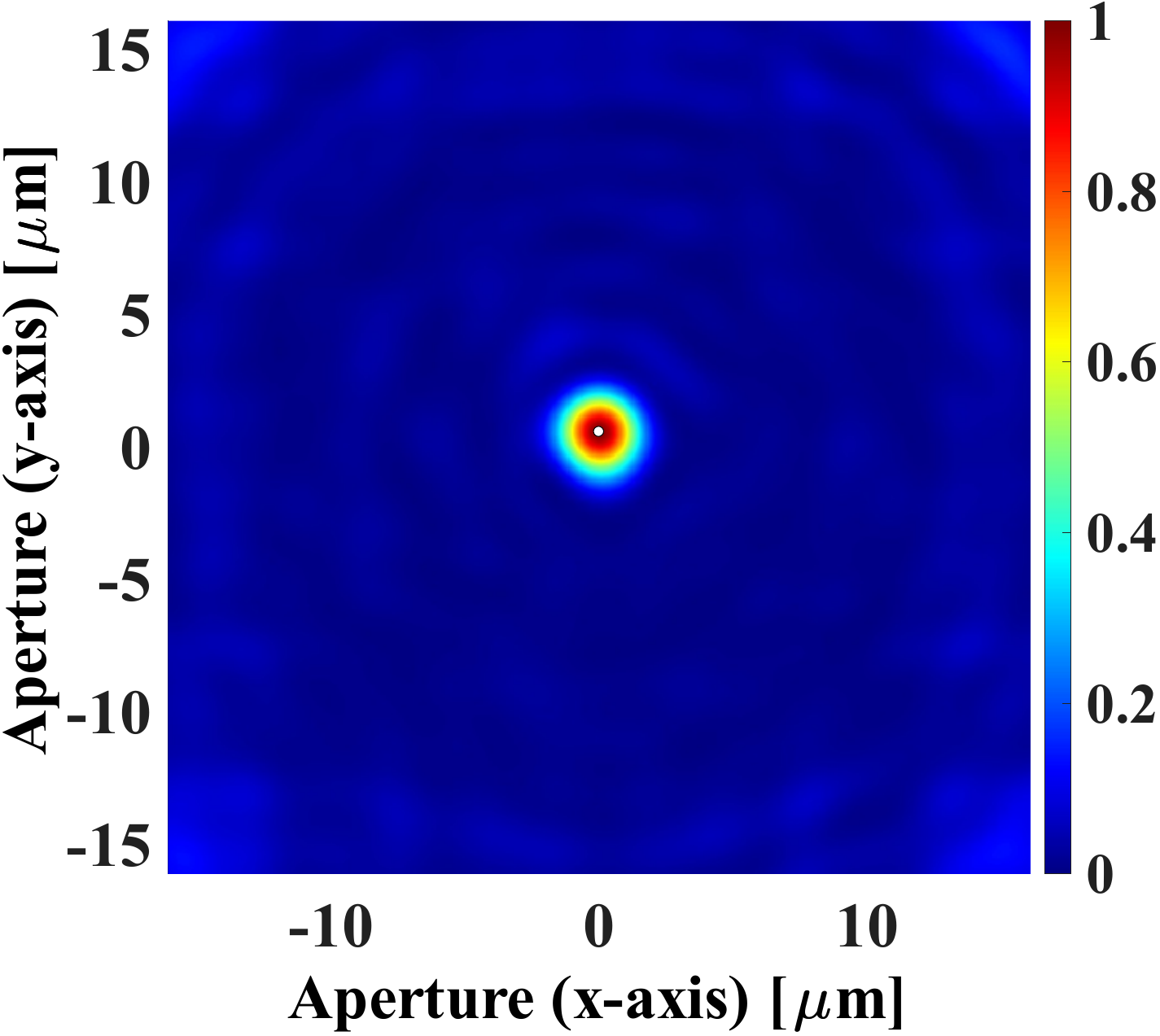}}
            \subfigure[]{\label{Fig: Fig09.b}
            \includegraphics[height=0.3\linewidth,trim={0.0 0.0cm 0.0cm 0.0cm},clip=true]{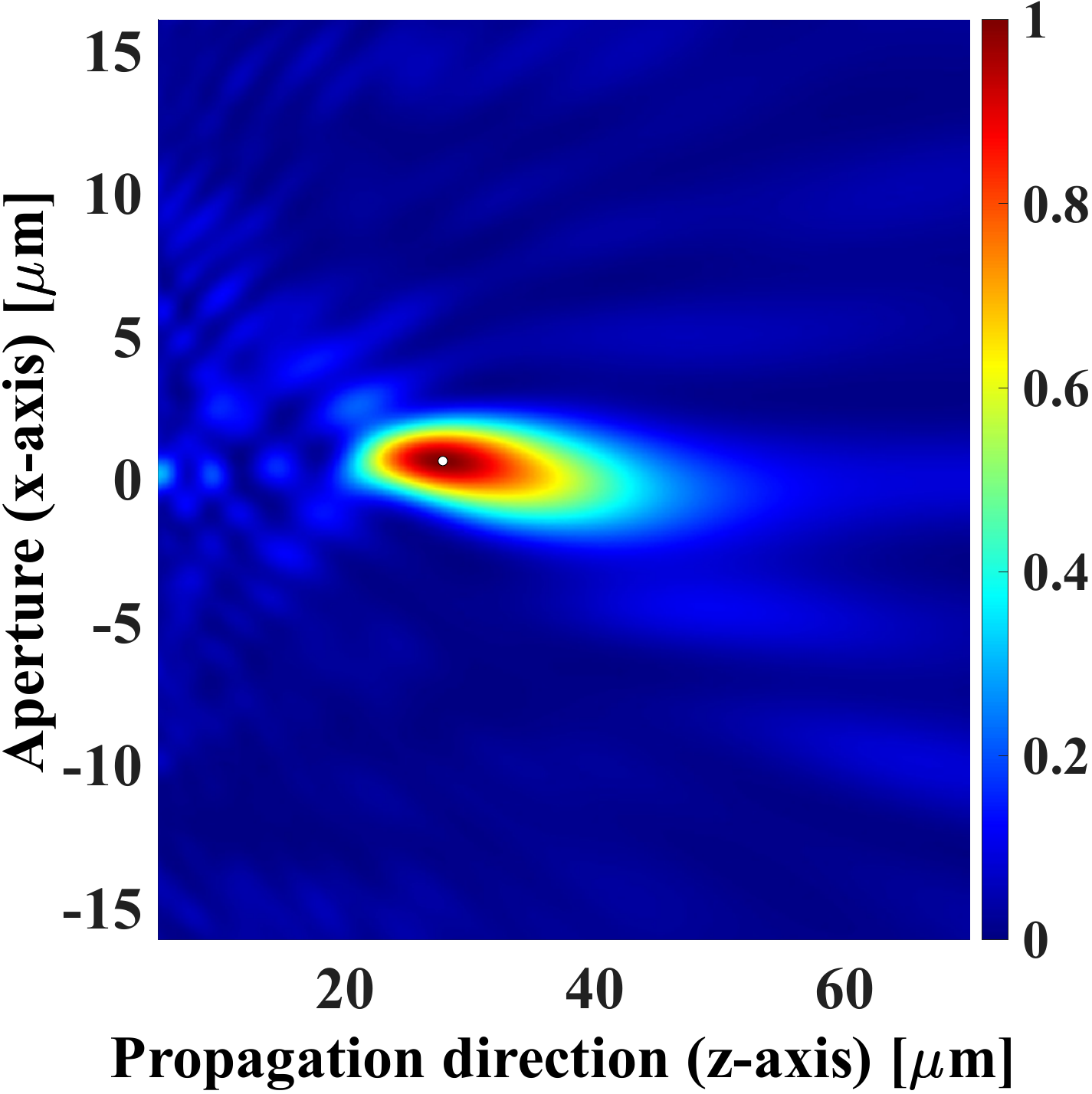}}
            \subfigure[]{\label{Fig: Fig09.c}
            \includegraphics[height=0.3\linewidth,trim={0.0cm 0.0cm 0.0cm 0.0cm},clip=true]{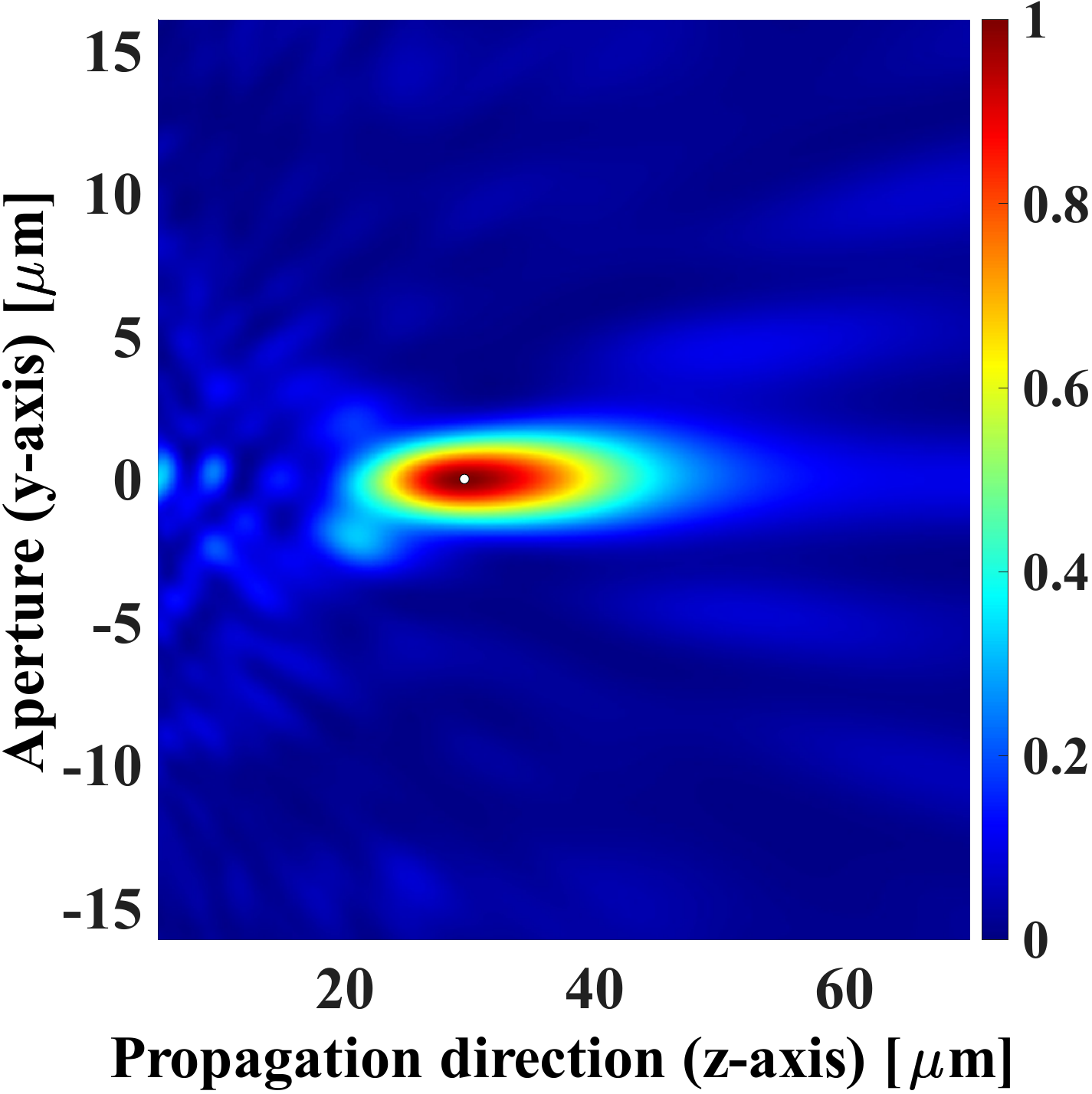}}
            \vspace*{-2mm}
            \subfigure[]{\label{Fig: Fig09.d}
            \includegraphics[height=0.3\linewidth,trim={0.0cm 0.0cm 0.0cm 0.0cm},clip=true]{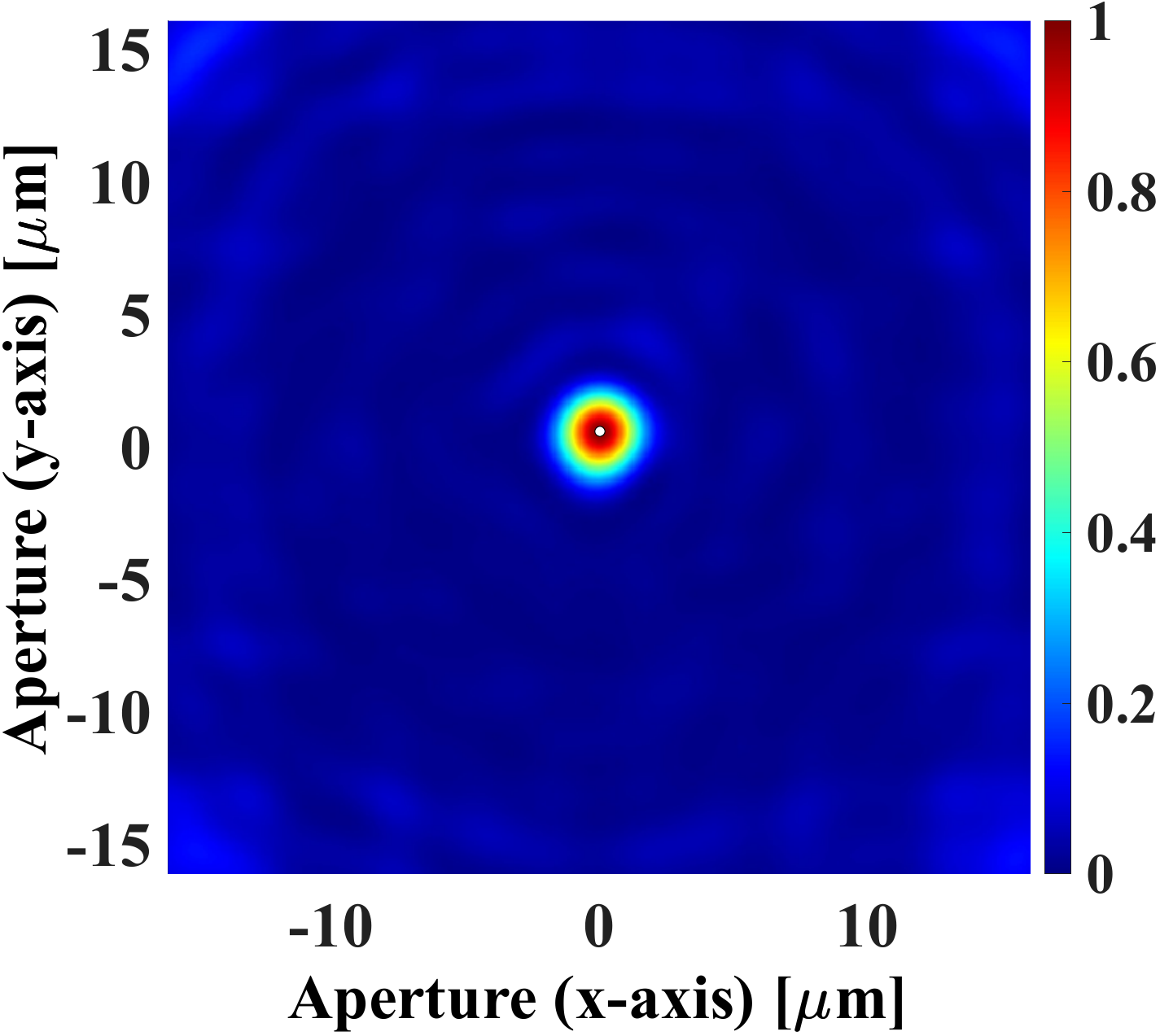}}
            \subfigure[]{\label{Fig: Fig09.e}
            \includegraphics[height=0.3\linewidth,trim={0.0cm 0.0cm 0.0cm 0.0cm},clip=true]{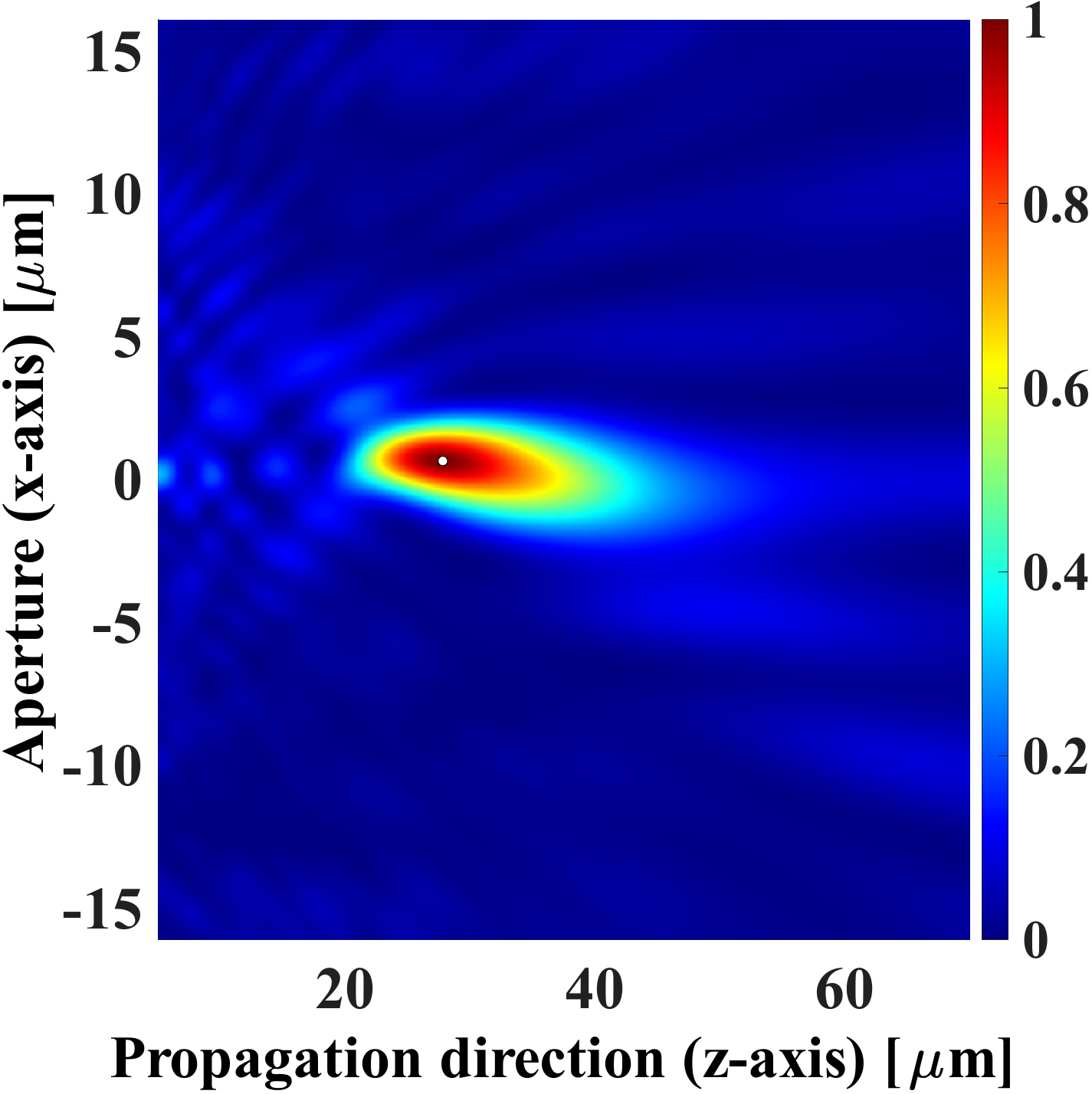}}
            \subfigure[]{\label{Fig: Fig09.f}
            \includegraphics[height=0.3\linewidth,trim={0.0cm 0.0cm 0.0cm 0.0cm},clip=true]{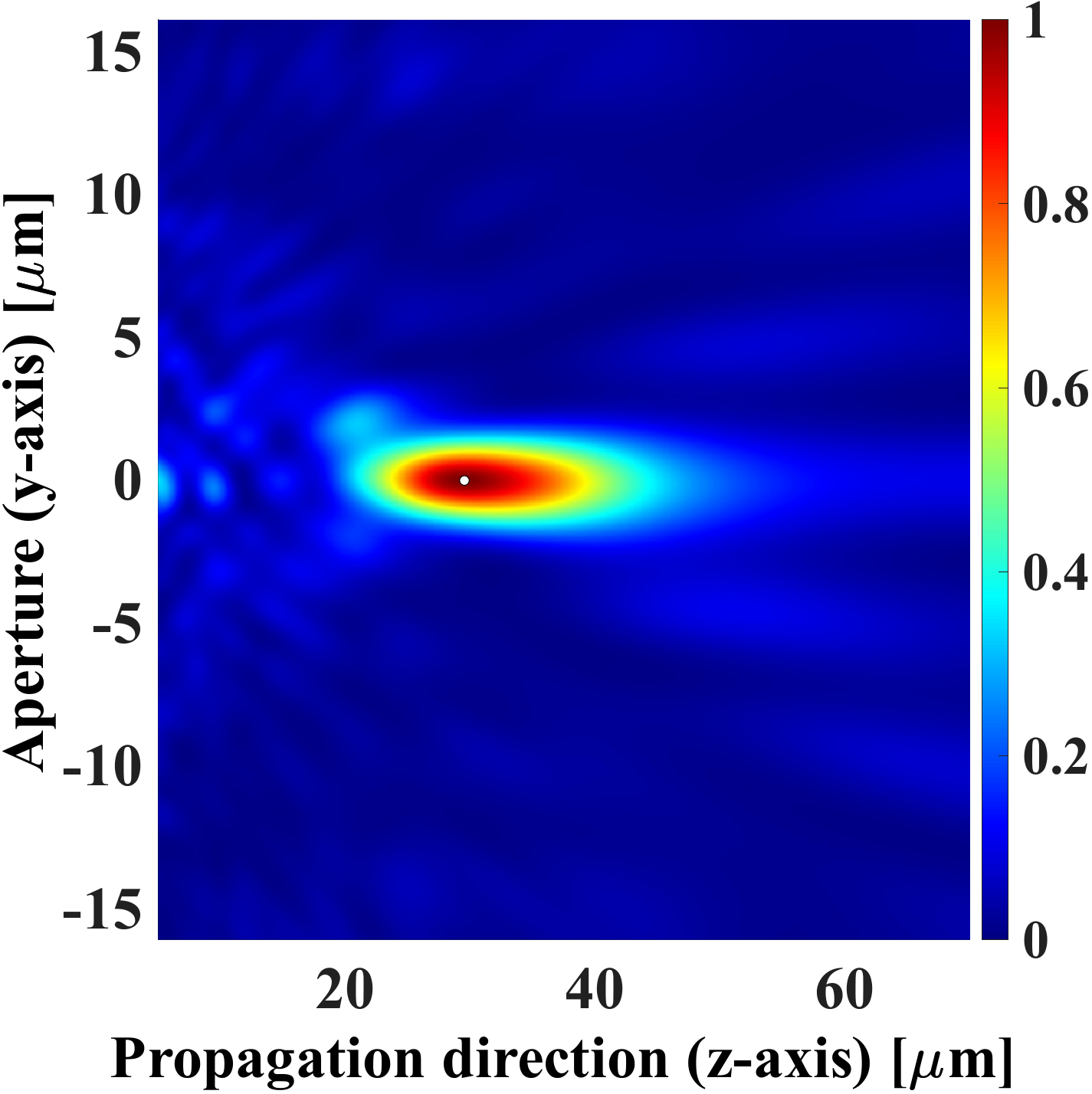}}
        \end{center}
        \vspace*{-5mm}
        \caption{Normalized $E$-field distributions for rotation angles of 60$^\circ$ in the $xy$-, $zx$-, and $yz$-plane under y-LP illumination with CW and CCW rotation. (a-c) CW rotation and (d-f) CWW rotation.}
        \label{Fig: Fig09}
    \end{figure*}

    Furthermore, the focusing performance under LP light along the $x$- and $y$-directions is illustrated in Figs.~\ref{Fig: Fig08} and~\ref{Fig: Fig09} for both CW and CCW rotations, respectively. In these cases, the metalens exhibits focusing characteristics consistent with those observed under circularly polarized excitations, yielding results of comparable quality. The focal lengths along the propagation (axial) direction, as well as $\textit{NA}_{\rm A.}$ and $F_{\rm N_{A.}}$, for all simulated cases are summarized in Table~\ref{Tab: Table02}.
    
    The analytically calculated PF exhibits a linear increase with the rotation angle of Disk $II$. However, the full-wave numerical simulations reveal that PF follows a non-linear trend as the rotation angle of Disk $II$ increases. As noted, this discrepancy arises from the limited ratio of the metalens aperture size to the operating wavelength. This small aperture-to-wavelength ratio increases undesirable diffraction and scattering phenomena, which consequently degrade the metalens’s focusing capability.
    
\begin{figure}
        \centering
        \includegraphics[width=0.6\linewidth,trim={0.0cm 0.0cm 0.0cm 0.0cm},clip=true]{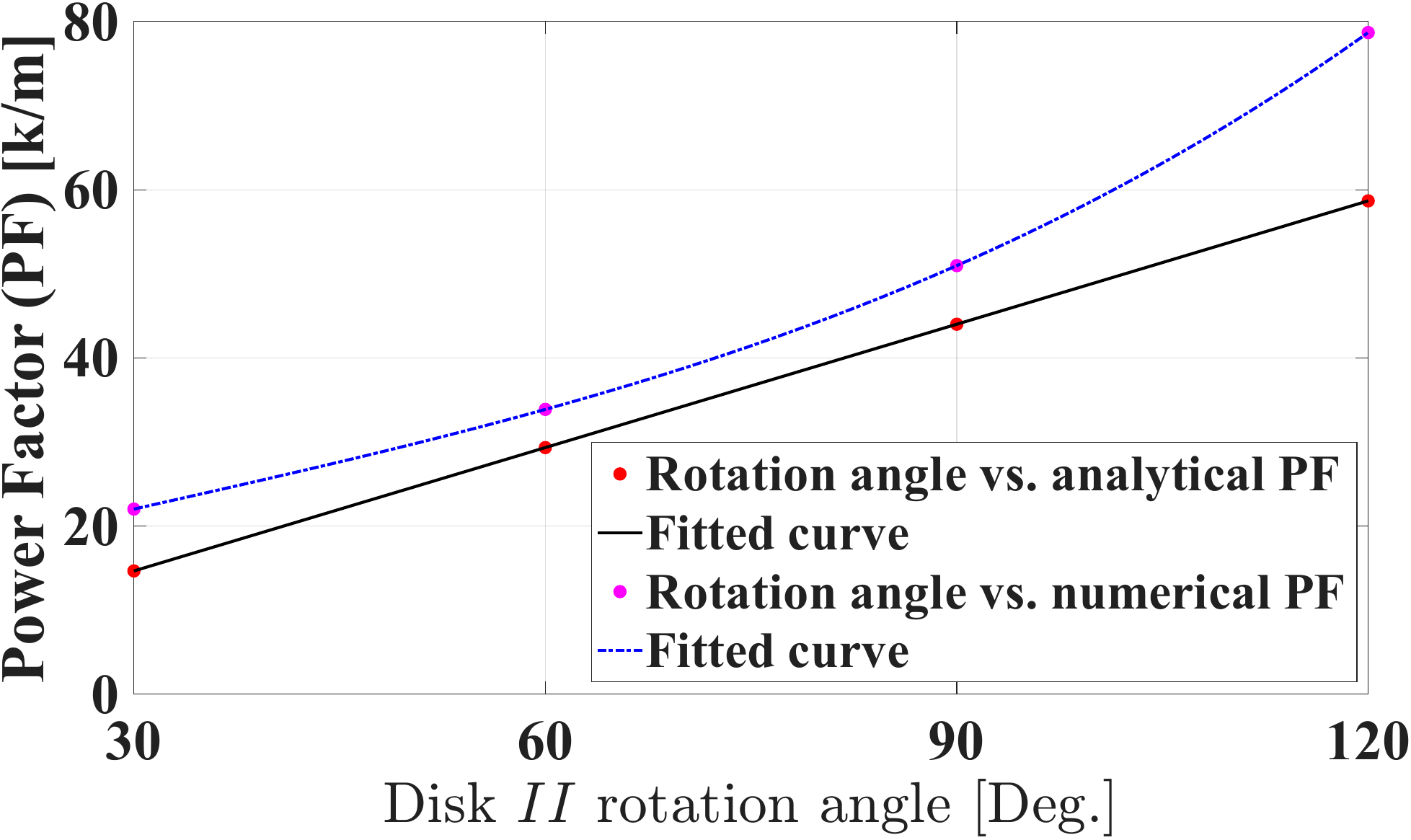}
        \caption{Power factor variation with respect to Disk $II$ rotation angle.}
        \label{Fig: Fig10}
    \end{figure}

    As the rotation angle increases, the Moiré-PB system demonstrates a significant enhancement in effective focusing strength, which is proportional to the twist angle between its constituent disks (Tables~\ref{Tab: Table01} and~\ref{Tab: Table02}). Nevertheless, the positioning accuracy of the focal point gradually degrades as the mutual rotation angle approaches $180^\circ$. This deviation originates from the increased phase step size induced by large angular displacements, which causes spatial undersampling of the phase gradient across the aperture. Respectively, the synthesized phase profile fails to fully utilize the continuous $[0, 2\pi)$ phase range needed to construct well-defined concentric Fresnel zones, leading to phase quantization artifacts and focal spot degradation. Importantly, this performance degradation represents a fundamental physical bound inherent to the Moiré phase synthesis framework and cannot be compensated for without altering the underlying grid sampling density. Consequently, this limitation highlights an unavoidable trade-off in rotational Moiré optics between maximizing the upper numerical aperture limit and preserving fine focal-point positioning accuracy. Overall, the results shown in Fig.~\ref{Fig: Fig10} demonstrate that the device behaves as a mechanically tunable focusing element over the explored angular range with a quasi-linear angular dependence. Minor deviations between the analytical predictions and the numerical results arise from diffraction effects not captured by the analytical model.
    
    The presented analytical and numerical results demonstrate that the Moiré-PB metalens offers a significant advancement in dynamic wavefront control. The system integrates Moiré rotational tunability with PB polarization-sensitive phase control by utilizing two structurally identical disks. Furthermore, this self-complementary, single-mask layout eliminates design complexity and offers a low-cost, scalable manufacturing strategy. In fact, this combination enables creation of compact yet mechanically reconfigurable flat optical devices. The current study demonstrates efficient phase modulation and focal length tuning through the controlled rotation of Disk $II$. However, several areas for future improvement remain.
    
    One avenue for enhancement is to scale the aperture size beyond the current demonstration (i.e., increase the $D/\lambda$ ratio). This would improve focusing efficiency and beam quality. It could mitigate diffraction-limited spot sizes by employing larger meta-atom arrays while maintaining the low-loss total transmission power ($>$83.65\%) observed in CST simulations of the {\tt Ge} nanobars. However, when scaling the aperture size, it is important to ensure that the chosen linear scaling factor $a$ does not exceed the maximum allowable criterion. Exceeding this limit would disrupt the phase profile of the Moiré structure, thereby reducing focusing efficiency for larger rotation angles. Further prospects include achieving broadband operation by engineering meta-atoms with dispersion-controlled {\tt Ge} resonators or by integrating PB elements. This would extend performance across near- to mid-infrared wavelengths.
\section{Understanding the Relationship Between Diffraction-Induced Blurring, Fresnel Number, and Numerical Aperture}
    As the second disk is rotated to smaller angles, the \textit{NA} drops drastically. In this low-\textit{NA} regime, the operational $F_{N}$ also drops significantly. For instance, full-wave numerical simulations reveal that at a 30$^{\circ}$ rotation angle, the retrieved numerical focal length deviates significantly from the analytical prediction. This blurred focal spot, discrepancy, and the accompanying diffuse are direct physical consequences of a low $F_{N}$ and an insufficient aperture-to-wavelength ratio, where dominant diffraction and scattering phenomena overpower the ideal quadratic phase profile. At a rotation angle of 30$^{\circ}$, the $F_{N}$ reaches a value of 1.90, placing the system within the Fresnel (near-field) diffraction regime. In fact, this indicates moderate near-field focusing behavior, as opposed to a strongly localized near-field effect. To maintain a well-defined and sharp focal spot at smaller rotation angles without altering the core meta-atom transmission efficiency, the absolute aperture size must be scaled to significantly increase the $D/\lambda$ ratio. Based on the diffraction constraints, expanding the meta-atom array to achieve a minimum diameter of $D \ge 75\ \mu\text{m}$ ($D/\lambda \ge 37$) would ensure that the Fresnel number stays comfortably above the diffractive blurring threshold ($N_F > 10$), preserving a high focusing intensity and a linear power factor response throughout the varifocal sweep.
    
    To approve the impact of aperture scaling on focusing performance, we employed angular spectrum propagation, which is a Fourier-based technique that transforms the electric field into the spatial-frequency domain via an FFT. However, the FFT inherently imposes periodic boundary conditions. As a result, diffracted fields that reach the edges of the aperture wrap around and reappear on the opposite side. This introduces non-physical interference, distorting the intensity profile and potentially shifting the focal position. Zero padding mitigates this by embedding the physical aperture within a larger zero-filled grid, allowing the field to diffract naturally into empty space and enabling reliable focal spot extraction. The physical aperture is discretized on a Cartesian grid, with the Moiré-PB phase distributions of both disks applied only within the circular region. The grid is then extended to a total size of $N_{\rm tot}=(1+P_{\rm F})N$, where added samples are set to zero. $P_{\rm F}$ represents the padding factor and is set to 4 in this study. This expansion reduces spatial-frequency spacing, enhancing spectral resolution and preventing aliasing, thereby ensuring smooth field propagation and accurate focal distance determination. To evaluate the effect of the $D/\lambda$ ratio on focal accuracy, rotation angles of 20$^\circ$ are simulated for apertures of size $D$, $2D$, and $3D$. For this rotation of Disk $II$, the corresponding focal length is 102.27~$\mu\text{m}$. As demonstrated in Table~\ref{Tab: Table03}, increasing this ratio results in focal spots that are closer to the expected positions.
    \begin{table}[!htbp]
        \centering
        \caption{Predicted focal response ($f_{\rm AS}$) using angular spectrum propagation methodology as a function of the aperture size of Moiré-PB metalens when the rotation angle is 20$^\circ$.}
        \bigskip
        \setlength{\tabcolsep}{3pt}
        \begin{tabular}{c|c|c|c}
            \toprule
            \toprule
             \rowcolor{Gray}
              Aperture size (D) [$\mu\text{m}$] & 32.2 & 64.4 & 96.6 \\
            \hline
              $f_{\rm AS}$ [$\mu\text{m}$] & 63.61 & 93.94 & 95.41 \\
            \toprule
            \toprule
        \end{tabular}
        \begin{minipage}{7.5cm}
        \vspace{0.05cm}
        \end{minipage}
        \label{Tab: Table03}
    \end{table}
\section{Statistical Modeling and Numerical Analysis of Multi-Source Structural Perturbations in Rotational Varifocal Metalens}
\begin{figure}
        \centering
        \includegraphics[width=0.6\linewidth,trim={0.0cm 0.0cm 0.0cm 0.0cm},clip=true]{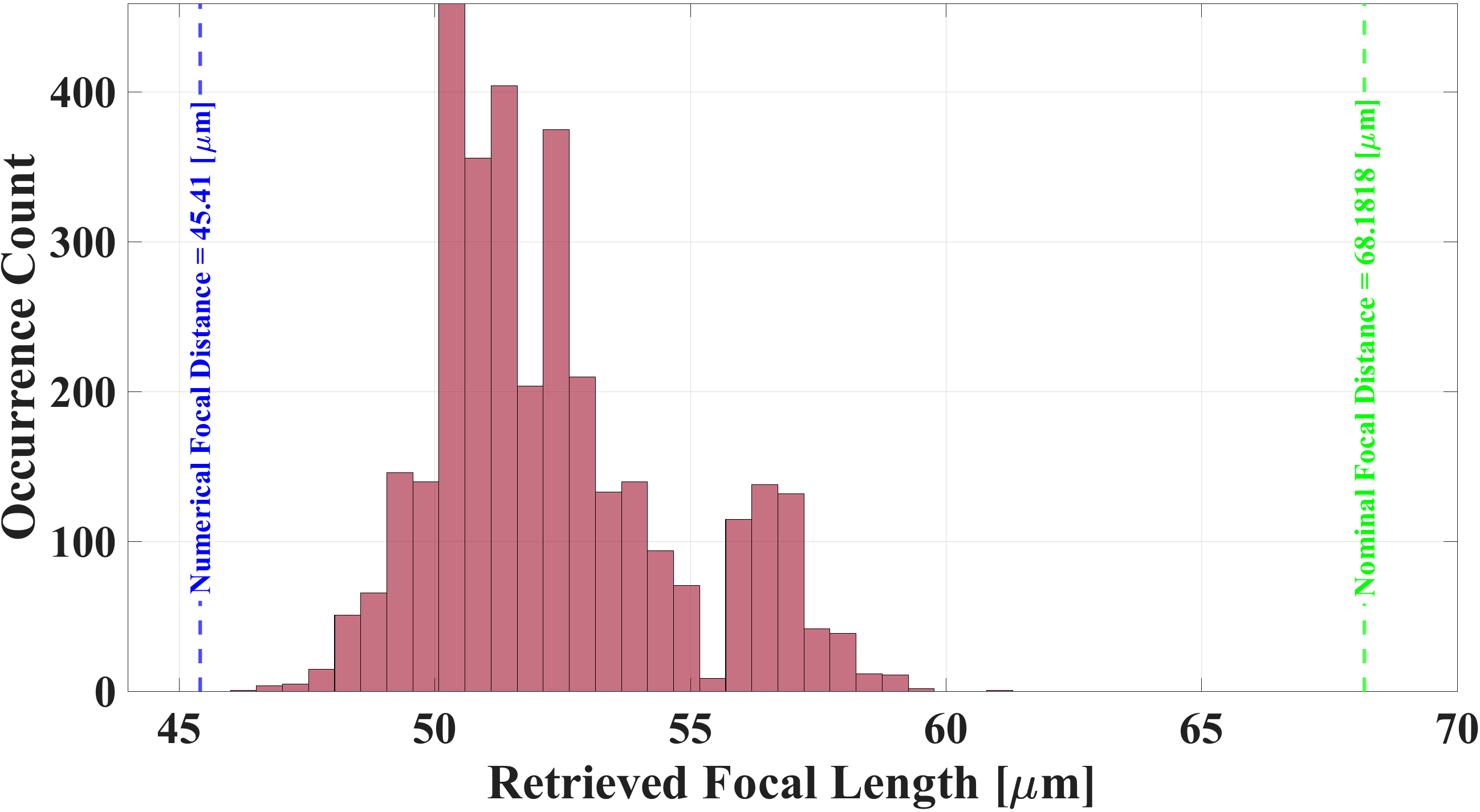}
        \caption{Monte Carlo tolerance analysis of the retrieved focal length under geometric fabrication variations at a rotation angle of 30$^\circ$.}
        \label{Fig: Fig11}
    \end{figure}

    To evaluate the practical resilience and operational bounds of the developed hybrid Moiré-PB metalens under realistic experimental imperfections, a comprehensive statistical tolerance study was performed using 3375 Monte Carlo simulation trials, with a 30$^\circ$ rotation of Disk $II$, Fig.~\ref{Fig: Fig11}. By simultaneously subjecting the system to a 1.8$^\circ$ rotational misalignment ($\sigma_{\Theta}$), a 15~nm axial gap variation ($\sigma_{g}$), a 9$^\circ$ half-wave plate phase deviation ($\sigma_{\rm HWP}$), and a 0.15~rad spatial phase noise layer ($\sigma_{\phi}$), we capture a comprehensive snapshot of coupled structural degradation. The nominal analytical focal length sits at 68.18~$\mu\text{m}$ based on infinite-aperture assumptions. At the same time, the unperturbed numerical configuration naturally shifts to 45.41~$\mu\text{m}$ due to diffraction originating from the finite 32.2~$\mu\text{m}$ aperture size. Upon injecting the collective Gaussian perturbations, the mean analytical focal length over the 3375 trials stabilizes at 52.28~$\mu\text{m}$, once standard Gaussian image smoothing is applied to the raw field cross-sections ($\mathbf{E}_{xz}$ and $\mathbf{E}_{yz}$) to suppress localized spatial noise spikes (see Supplementary Material Section 5). These methodological upward and downward shifts from the nominal numerical and analytical baselines arise from a physical reduction in the lens’s effective numerical aperture. Since the outer structural zones experience accumulated phase mismatch, they lose phase coherence more quickly than the center. Notably, the estimated tight standard deviation, $\sigma_{f}\approx 2.38~\mu\text{m}$, across all metrics indicates that the platform maintains a fair level of positional consistency despite a noticeable focal elongation. In fact, a maximum of 32.05\% focal length misalignment from the nominal focal length is observed at the worst-case scenarios. The occurrence of a secondary minor peak near 56.2~$\mu\text{m}$ in the distribution profile may further reveal subtle bimodal behavior, indicating severe, asymmetric phase mismatch in which the primary beam waist undergoes localized focal splitting. Ultimately, these findings underscore that while the absolute focal plane shifts within structural tolerances, the hybrid flat-optics assembly maintains reliable focusing integrity without catastrophic performance failure, thereby solidifying its manufacturing viability.
    
    To preserve the sharpness and diffraction-limited intensity of the focal spot, spatial variations or positioning inaccuracies across the Fresnel zone boundaries must be strictly constrained~\cite{BornWolf1999}. Based on the local zone width of $\Delta x=x_{\rm n} - x_{\rm n-1}\approx\lambda\sqrt{x_{\rm n}^{\rm 2} + f^{\rm 2}}/(2x_{\rm n})$ and according to the Rayleigh quarter-wavelength criterion ($\Delta \phi \le \pi/2$), maintaining destructive interference suppression and preventing phase degradation requires that the maximum structural misalignment or discretization tolerance $\delta x$ does not exceed at most half of the local zone width, i.e., $|\delta x| \le \frac{1}{2} \Delta x$.  Exceeding this upper bound induces phase decoherence across outer radial regions, leading to severe spot broadening, a reduced Strehl ratio, and an overall deterioration of focal spot sharpness~\cite{BornWolf1999}. The half of the local zone width is equal to 865~nm in the case with a 90$^\circ$ CW/CCW rotation angle of Disk $II$ and $x_{\rm n}=D/2=16.1~\mu\text{m}$. As a consequence, to complete the tolerance analysis, eight numerical case studies are performed to assess the effects of structural center misalignment between the cascaded disks. This analysis focuses on the scenario with a 90$^\circ$ CW rotation angle of Disk $II$ under RCP illumination, as summarized in Table~\ref{Tab: Table04}. We individually simulated center offsets of 1000~nm and 2000~nm for Disk $II$ along different azimuth angles to quantify the resulting focal lengths and their deviations from the corresponding ideally aligned configurations. The calculated focal-length deviations indicate that the device maintains reliable performance under the Rayleigh quarter-wavelength criterion center misalignment, demonstrating a practical tolerance for assembly inaccuracies.
    \begin{table}[!htbp]
        \centering
        \caption{Focal length ([$\mu\text{m}$]) under center misalignment between cascaded disks and its deviation error from numerical studies.}
        \bigskip
        \setlength{\tabcolsep}{3pt}
        \begin{tabular}{c|c|c|c|c}
            \toprule
            \toprule
             \rowcolor{Gray}
              Misalignment along  &  &  &  &  \\
             \rowcolor{Gray}
              polar direction~[$^\circ$] & \multirow{-2}{*}{0} & \multirow{-2}{*}{45} & \multirow{-2}{*}{90} & \multirow{-2}{*}{135} \\
            \hline
              1000~nm &  &  &  &  \\
              Misalignment & \multirow{-2}{*}{20.08} & \multirow{-2}{*}{20.31} & \multirow{-2}{*}{19.62} & \multirow{-2}{*}{20.08} \\
            \hline
              Bifocality & \xmark & \xmark & \xmark & \xmark \\
            \hline
              Focal sharpness & \checkmark & \checkmark & \checkmark & \checkmark \\
            \hline
              Error~[\%] & 2.35 & 3.52 & 0.002 & 2.35 \\
            \toprule
             \rowcolor{Gray}
              2000~nm &  &  &  &  \\
             \rowcolor{Gray}
              Misalignment & \multirow{-2}{*}{18.93} & \multirow{-2}{*}{19.62} & \multirow{-2}{*}{19.16} & \multirow{-2}{*}{18.58} \\
            \hline
             \rowcolor{Gray}
              Bifocality & \checkmark & \xmark & \xmark & \checkmark \\
            \hline
             \rowcolor{Gray}
              Focal sharpness & \xmark & \checkmark & \xmark & \xmark \\
            \hline
             \rowcolor{Gray}
              Error~[\%] & 3.52 & 0.002 & 2.35 & 5.28 \\
            \toprule
            \toprule
        \end{tabular}
        \begin{minipage}{7.5cm}
        \vspace{0.05cm}
        \end{minipage}
        \label{Tab: Table04}
    \end{table}

\section{Discussion of Results and Conclusions}
    
    \begin{table*}[!htbp]
        \tiny
        \centering
        \caption{Comparison with state-of-the-art hybrid metalens/metasurface.}
        \vspace*{2mm}
        \setlength{\tabcolsep}{3pt}
        \begin{tabular}{c|c|c|c|c|c|c}
            \toprule
            \toprule
            \rowcolor{Gray}
              & Tunability & Employed & Experimental-based & Structural & Operating &  \\
            \rowcolor{Gray}
             \multirow{-2}{*}{Refs.} & mechanism & methods & study & complexity & band & \multirow{-2}{*}{Application} \\
            \toprule
              & Mechanical & Propagation &  & Two stretchable &  & Imaging \\
             \cite{liuReconfigurableMetalensDuallinearfocus2023} & stretching \& Polarization & \& & \xmark & layers & \multirow{-2}{*}{1} & Security inspection \\
              & switching & PB-phase &  & (Complex) & \multirow{-2}{*}{[THz]} & Non-destructive testing \\
            \hline
            \rowcolor{Gray}
              & Mechanical mutual & Moiré &  & Three different &  & Wireless communication \\
            \rowcolor{Gray}
             \cite{XUTerahertz4DBessel2025} & rotation \& Polarization & \& & \checkmark & layers & \multirow{-2}{*}{291} & THz radar \\
            \rowcolor{Gray}
              & handedness & PB-phase &  & (Complex) & \multirow{-2}{*}{[GHz]} & Super-resolution imaging \\
            \hline
              & Relative rotation & Propagation &  & Two different &  & Optical communications \\
             \cite{nanGenerationBesselBeams2025} & \& & \& & \checkmark & layers & \multirow{-2}{*}{0.9} & Super-resolution imaging \\
              & Overall rotation & Moiré phase &  & (Complex) & \multirow{-2}{*}{[THz]} & Medical photonics \\
            \hline
            \rowcolor{Gray}
              & Relative in-plane & Propagation &  & Two different &  & Polarimetric imaging \\
            \rowcolor{Gray}
             \cite{yeUltraBroadbandMoirePBDoublet2024} & rotation of decentered & \& & \checkmark & layers & \multirow{-2}{*}{400-780} & Biological sample imaging \\
            \rowcolor{Gray}
              & liquid crystal PB lenses & PB-phase &  & (Complex) & \multirow{-2}{*}{[nm]} & Resolution-tunable edge detection \\
            \hline
              & In-plane layer & Propagation &  & Two different &  & Reconfigurable THz imaging \\
             \cite{Li_2026_1} & rotation \& Spin-decoupled & \& & \checkmark & layers & \multirow{-2}{*}{0.5} & Microscale spectral analysis \\
              & polarization switching & PB-phase &  & (Complex) & \multirow{-2}{*}{[THz]} & 6G wireless communication \\
            \hline
            \rowcolor{Gray}
              & Relative in-plane & Moiré &  & Two different &  & Non-destructive industrial testing \\
            \rowcolor{Gray}
             \cite{Li_2026_2} & rotation angle between & \& & \checkmark & layers & \multirow{-2}{*}{0.5} & Biomedical diagnostics \\
            \rowcolor{Gray}
              & cascaded metasurface layers & PB-phase &  & (Complex) & \multirow{-2}{*}{[THz]} & Portable THz polarization optics \\
            \hline
               & Mechanical & Moiré &  & Two identical &  &  \\
              \multirow{-2}{*}{This} & mutual & \& & \xmark & layers & \multirow{-2}{*}{144.90-173.40} & \multirow{-2}{*}{Polarization independent focusing} \\
              \multirow{-2}{*}{work} & rotation & PB-phase &  & (Simple) & \multirow{-2}{*}{[THz]} & \multirow{-2}{*}{Adaptive Imaging} \\
            \toprule
            \toprule
        \end{tabular}
        \label{Tab: Table05}
    \end{table*}
    
    This work has successfully demonstrated a transmissive mid-infrared metalens. The device integrates Moiré and Pancharatnam-Berry phases for rotational wavefront control using a simple design with two identical layers. To achieve this aim, {\tt Ge} meta-atoms on {\tt CaF$_{\rm 2}$} substrates have been optimized as half-wave plates. They have achieved $>99.95\%$ PCR and $>83.65\%$ total transmission power at $2~\mu\text{m}$ wavelength. The anisotropic nanobar geometry has provided a phase retardation of $180.77^\circ$, closely matching the theoretical half-wave condition. By rotating the second disk from $30^\circ$ to $120^\circ$, the focal length can be continuously tuned from about $45~\mu\text{m}$ down to $12~\mu\text{m}$. This tuning corresponds to numerical apertures ranging from 22.99\% to 68.67\%. Such mechanical rotation mechanism eliminates the need for axial displacement, offering a compact varifocal solution. Additionally, the metalens has exhibited expected robust focusing performance under RCP, LCP, EP, $x$-LP, and $y$-LP illuminations. For LP and EP excitations, both clockwise and counterclockwise rotations produce equivalent focal lengths. Full-wave simulations confirmed the predicted focal length tuning, with percentage errors of approximately $14\%$ between the analytical and numerical results. The metalens power factor has exhibited a quasi-linear dependence on rotation angle, enabling predictable focal length adjustment.
    
    Table~\ref{Tab: Table05} compares state-of-the-art hybrid metasurfaces across tunability, methodology, structural complexity, operating band, and application. Unlike prior works employing distinct multi-layer configurations with high structural complexity, our design realizes reconfigurable wavefront control using a simple design with two identical layers. The proposed metalens enables focusing for adaptive imaging across the 144.90-173.40~THz band utilizing hybrid Moiré and PB phases via mechanical mutual rotation.
    
    As a possible future direction of research, integration of Moiré-PB metalens with photodetectors and optoelectronic sensors (such as $III-V$ detectors) or nonlinear crystals might be studied, with potential for developing compact spectrometers or tunable second-harmonic generators. This integration would use {\tt CaF$_{\rm 2}$} substrates to enable low-loss propagation. To ensure cost-effective scalability, fabrication tolerances could be reduced through techniques such as electron-beam lithography or nanoimprint lithography. Future efforts should also focus on correcting aberrations using aspheric Moiré profiles.
    
    Ultimately, these developments could potentially position Moiré-PB metalenses as flexible platforms for endoscopy and adaptive microscopy. To facilitate real-time and feedback-controlled focal strength tuning, the disks could be mechanically operated via precision microactuators. Suitable possibilities lie in using MEMS rotors or piezoelectric stacks.
    
    Thus, we may conclude that the Moiré-PB-phase integration concept developed in this article may establish a universal platform for adaptive imaging and beam control since it uniquely combines continuous mechanical tunability, polarization control, and broadband potential within a simple and single compact optical device. We believe that this work offers a practical pathway toward reconfigurable flat optics for the IR imaging, spectroscopy and other applications.
\section*{Acknowledgement}
    K. Kaboutari acknowledges co-financing by Funda\c{c}\~{a}o para a Ci\^{e}ncia e a Tecnologia (Portuguese Foundation for Science and Technology) through the Carnegie Mellon Portugal Program under the fellowship PRT/BD/154201/2022 and DOI identifier \href{https://doi.org/10.54499/PRT/BD/154201/2022}{10.54499/PRT/BD/154201/2022}. S. Maslovski and P. Pinho acknowledge funding by national funds through FCT – Fundação para a Ciência e a Tecnologia, I.P., and, when eligible, by EU funds under the project UID/50008/2025 – Instituto de Telecomunicações, with DOI identifier - \href{https://doi.org/10.54499/UID/50008/2025}{10.54499/UID/50008/2025}.
\bibliographystyle{elsarticle-num}
\bibliography{References}
\end{document}